\documentclass{aa}  

\usepackage{graphicx}
\usepackage{txfonts}
\usepackage{natbib}
\usepackage{orcidlink}
\usepackage{enumitem}
\usepackage{lscape} 
\begin{document} 

\title{The GAPS programme at TNG}

\subtitle{LXXIV. A reanalysis of the planetary systems TOI-1272 and TOI-1694 with HARPS-N and retraction of the planetary interpretation of TOI-1272\,c}
   \titlerunning{Two Neptunes on the ridge}
   \authorrunning{L. Mancini et al.}

\author{
L. Mancini\orcidlink{0000-0002-9428-8732}\inst{1,2}\fnmsep\thanks{lmancini@roma2.infn.it}
\and
L. Naponiello\orcidlink{0000-0001-9390-0988}\inst{2}
\and
M. Damasso\orcidlink{0000-0001-9984-4278}\inst{2}
\and
K. Biazzo\orcidlink{0000-0002-1892-2180}\inst{3}
\and
A.~S.~Bonomo\orcidlink{0000-0002-6177-198X}\inst{2}
\and
A.~F.~Lanza\orcidlink{0000-0001-5928-7251}\inst{4}
\and
J. Lillo-Box\orcidlink{0000-0003-3742-1987}\inst{5}
\and
M. Pinamonti\orcidlink{0000-0002-4445-1845}\inst{2}
\and
R. Cosentino\orcidlink{0000-0003-1784-1431}\inst{6}
\and
A. Bignamini\orcidlink{0000-0002-5606-6354}\inst{7}
\and
W. Boschin\orcidlink{0000-0001-9978-9109}\inst{6}
\and
A. Fiorenzano\inst{6}
\and
P. Giacobbe\orcidlink{0000-0001-7034-7024)}\inst{2}
\and
F. Manni\orcidlink{0009-0003-1841-7381}\inst{1,2}
\and
M. Rainer\orcidlink{0000-0002-8786-2572}\inst{8}
\and
G. Scandariato\inst{9}
\and
A. Sozzetti\orcidlink{0000-0002-7504-365X}\inst{2}
}

\institute{
Department of Physics, University of Rome ``Tor Vergata'', Via della Ricerca Scientifica 1, 00133 Rome, Italy
\and
INAF -- Turin Astrophysical Observatory, via Osservatorio 20, 10025 Pino Torinese, Italy
\and
INAF -- Astronomical Observatory of Rome, Via Frascati 33, 00178 Monte Porzio Catone, Italy
\and
INAF -- Astrophysical Observatory of Catania, Via S. Sofia 78, 95123 Catania, Italy
\and
Centro de Astrobiolog\'ia (CAB), CSIC-INTA, Camino Bajo del Castillo s/n, 28692, Villanueva de la Ca\~nada, Madrid, Spain 
\and
Fundacion Galileo Galilei, Rambla J.\,A. F\'{e}rnandez P\'{e}rez 7, 38712 Bre\~{n}a Baja, La Palma, Santa Cruz de Tenerife, Spain
\and
INAF -- Astronomical Observatory of Trieste, via Tiepolo 11, 34143 Trieste, Italy
\and
INAF -- Brera Astronomical Observatory, Via E. Bianchi 46, 23807 Merate, Italy
\and
INAF -- Astrophysical Observatory of Catania, Via S. Sofia 78, 95123 Catania, Italy
}
   \date{Received 24 March 2026 / Accepted dd May 2026}

 
  \abstract
{Hot Neptunes are close-in exoplanets that occupy a sparsely populated region of parameter space known as the ``hot-Neptune desert''. Their presence in this extreme environment is puzzling as it implies a complex history involving intense stellar radiation, atmospheric loss, and unique migration patterns that differ from more conventional Neptunes at larger orbital periods.}
{We are running an observational programme aimed at enlarging the number of close-in Neptune-sized planets, with well-measured physical and orbital parameters. It contributes to the compilation of a statistically significant sample that will aid in clarifying the formation and migration pathways characterising this class of exoplanets.}
{We used currently available TESS photometry, along with new (HARPS-N) and archival (HIRES) high-precision radial-velocity measurements, to review the main properties of the planetary systems TOI-1272 and TOI-1694 by means of joint-fit analyses that, in the case of TOI-1272, included Gaussian-process regressions for carefully modelling stellar activity.}
{Our final estimates of the parameters of the two systems are consistent with previous measurements but have smaller uncertainties. We identified the radial-velocity variation of TOI-1272 found in the HIRES data as stellar activity rather than planetary in nature and, therefore, we rejected the non-transiting planet TOI-1272\,c. This opens up the possibility that TOI-1272\,b's eccentricity is the result of high-eccentricity migration. 
The greater amount of data at our disposal also allowed us to point out that both TOI-1694\,b and TOI-1694\,c move on slightly eccentric orbits. 
The current orbital architecture of TOI-1694 suggests a history of migration driven by both disc and dynamical interactions.}
{}


   \maketitle
%

\section{Introduction}
\label{sec:introduction}
\begin{figure}
\centering
\includegraphics[width=1\linewidth]{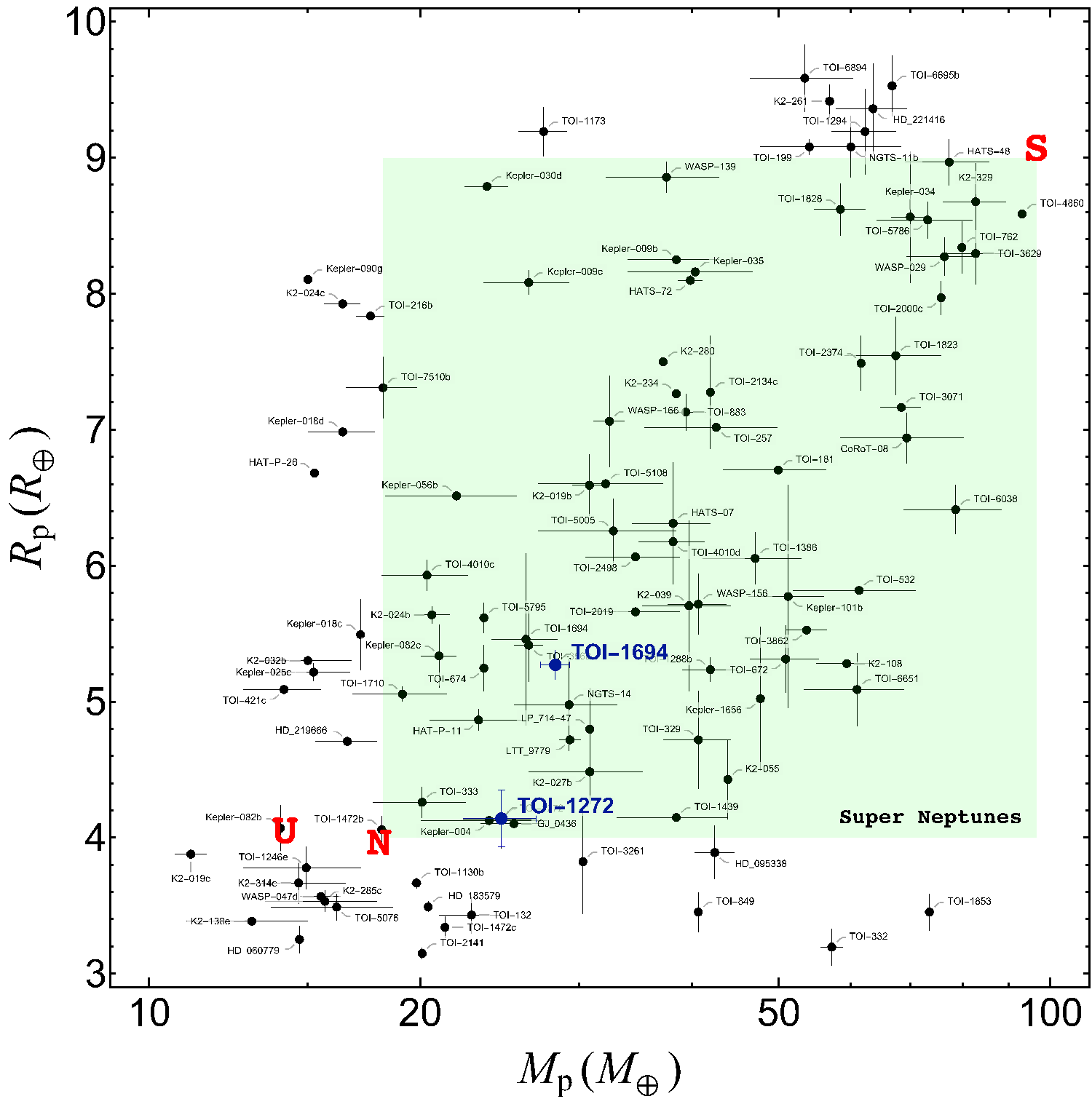}
\caption{Section of the $R_{\rm p}$ versus $M_{\rm p}$ linear-log diagram of known transiting exoplanets with a mass between 10 and 100\,$M_{\oplus}$ (measured with an accuracy to within $20\%$) and a radius between 3 and 10\,$R_{\oplus}$. Data taken from {\tt TEPCat}. The positions of TOI-1274\,b and TOI-1694\,b are highlighted (this work). Exoplanets inside the green zone are giant planets with sizes and masses between those of Neptune and Saturn. The position of Uranus, Neptune, and Saturn are also marked with red capital letters.} 
\label{fig:diagramMpRp}
\end{figure}

Giant planets with sizes and masses between those of Neptune and Saturn bridge the gap between ice giants and gas giants, and they are often termed super-Neptunes, sub-Saturns, or more generically intermediate-mass giants \citep[see, e.g.,][]{bonomo2014,bakos2015,bayliss2015,knudstrup2023,castro2024b}. These planets populate a wide-ranging region of the parameter space with no representation in the Solar System (Fig.~\ref{fig:diagramMpRp}), further proving that our immediate celestial neighbourhood is not the archetype of planetary systems in the Milky Way.
In particular, exoplanets with radii between roughly 3 and 8.5\,$R_{\oplus}$, masses between 10 and 100\,$M_{\oplus}$, and orbital periods shorter than approximately three days occupy a critical and still enigmatic regime in the exoplanet-parameter space. They reside within the so-called ``hot Neptune desert'', which is a region of dramatically low occurrence, which was identified in the 2010s \citep[e.g.][]{szabo2011} and still challenges the theoretical models of planet formation, migration, and atmospheric evolution. This deficit is particularly striking because although intermediate-sized planets are generally easier to detect at short periods, they remain extremely rare compared to smaller super-Earths or larger gas giants.

Despite early data from the {\it Kepler} mission indicating that this region was virtually uninhabited \citep{latham2011}, recent discoveries from the TESS mission \citep{ricker2015} have identified several ``desert dwellers'' revealing that the desert is not entirely empty, but rather a complex landscape of rare and unique planets (Fig.~\ref{fig:desert}).
The Neptune desert is currently attributed to a combination of atmospheric mass loss and dynamical migration. The lower edge of the desert is believed to have been shaped by photoevaporation caused by intense X-ray and extreme ultraviolet irradiation from the parent stars, which strips away the gaseous envelopes of low-to-intermediate-mass planets, leaving behind exposed rocky cores \citep{ionov2018,owen2018,koskinen2022,thorngren2023}. In contrast, the upper boundary is likely sculpted by high-eccentricity migration and tidal disruption \citep{matsakos2016,owen2018}. Recent characterisation of desert dwellers revealed a remarkable diversity in density that challenges conventional formation theories. Despite having low core masses, some ultra-hot Neptunes have managed to retain extensive hydrogen-helium envelopes, resisting complete atmospheric escape \citep{jenkins2020,nabbie2024}. Conversely, other planets in the Neptune desert are exceptionally dense (reaching 10\,g\,cm$^{-3}$), with core masses exceeding $30\,M_{\oplus}$ \citep{armstrong2020,osborn2023}. Such massive, gas-poor planets are often interpreted as the exposed cores of giant planets (e.g., \citealt{hallatt2026}), but extreme cases required unconventional hypotheses, such as catastrophic formation scenarios (e.g., multiple planetary collisions; \citealt{naponiello2023}), to explain their existence.

\begin{figure}
\centering
\includegraphics[width=0.9\linewidth]{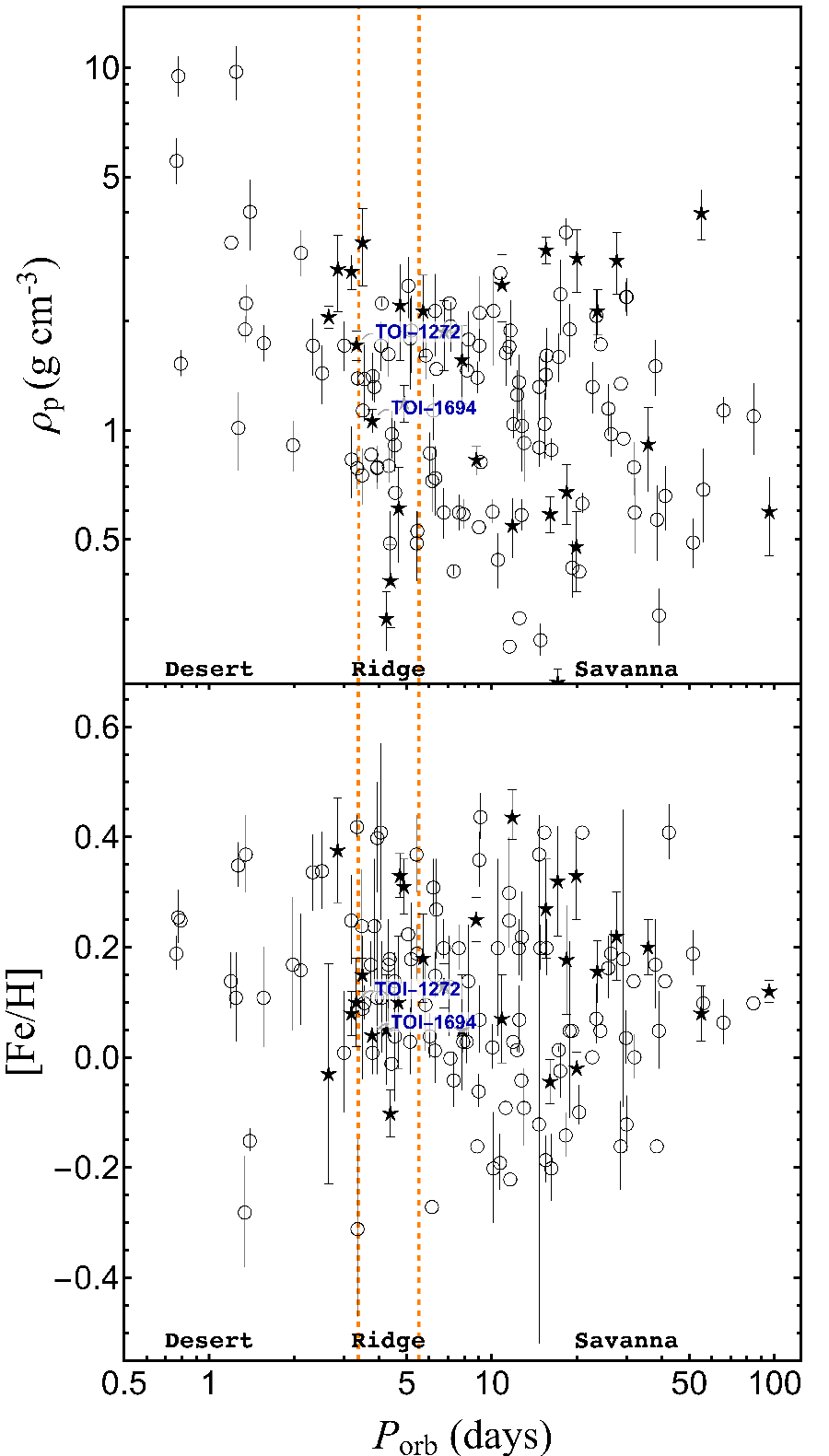}
\caption{{\it Top panel}: Log-log diagram of the planetary density versus orbital period of known transiting exoplanets with (3\,$R_{\oplus}$\,<\,$R_{\rm p}$\,<\,7\,$R_{\oplus}$) and mean density measured with an accuracy to within $30\%$. {\it Bottom panel}: Linear-log diagram of the parent-star metallicity versus planets' orbital period. Circles indicate planets with eccentricity $e \leq 0.1$, while five-pointed stars indicate those with $e > 0.1$. Data taken from {\tt TEPCat}. Horizontal error bars have been suppressed for clarity. The positions of TOI-1274\,b and TOI-1694\,b are highlighted. The orange-dashed lines delimit the three regions recognised by \citet{castro2024a}.
} 
\label{fig:diagramRhoPorb}
\end{figure}

As pointed out by \citet{castro2024a}, beyond this desert, planetary distribution transitions into the ``Neptunian ridge'', a peak in occurrence at orbital period ($P_{\rm orb}$) between 3.2\,days and 5.7\,days, and the ``Neptunian savanna'' at longer orbital periods (5.7\,days\,<\,$P_{\rm orb}$\,<\,100\,days), each exhibiting distinct physical and orbital characteristics (Fig.~\ref{fig:desert}).
%
%
%
%
The parent stars of Neptune-sized exoplanets in both the desert and the ridge are generally more metal rich than those of the savanna \citep{dong2018,vissapragada2025,doyle2025}. This is evident from Fig.~\ref{fig:diagramRhoPorb} (bottom panel) and Fig.~\ref{fig:CDF} (left-hand panel), in which we compared the cumulative distribution function (CDF) related to the host-star metallicities of the Neptune-desert sample with those of the ridge and savanna samples. Here, we considered all planets with a measured radius between 3 and 8.5\,$R_{\oplus}$ orbiting stars of known metallicity. The data were taken from the Transiting Extrasolar Planet Catalogue ({\tt TEPCat}) \citep{southworth2011}.
The null hypothesis that the Neptune-desert and Neptune-savanna datasets have the same distribution is rejected at the $5\%$ level based on the Kolmogorov-Smirnov test ($p=2.6\times10^{-4}$), whereas it is not rejected for the Neptune-desert and Neptune-ridge datasets ($p=0.7$). 

It is also recognisable that the Neptune desert is populated by planets that are denser than those of the ridge and savanna, see Fig.~\ref{fig:diagramRhoPorb} (top panel), with several extreme examples, as we already mentioned.
In the right-hand panel of Fig.~\ref{fig:CDF}, we compared the CDF related to the planetary density of the Neptune-desert sample with those of the ridge and savanna samples (we only considered planets with a mean density measured with an accuracy of within $30\%$). Data were taken from {\tt TEPCat}. Here, the null hypothesis that the Neptune-desert dataset has the same distribution as the ridge and savanna datasets is rejected at the $5\%$ level based on the Kolmogorov-Smirnov test ($p=5.7\times10^{-4}$ and $p=1.5\times10^{-3}$, respectively). 

The evolutionary frameworks for planets in the Neptune desert, ridge, and savanna appear, therefore, to be distinguished by different migration histories, atmospheric loss mechanisms, and host star properties. Specific frameworks have been proposed as summarised by \citet{vissapragada2025}
%
%
%
%
and distinguishing between such formation pathways requires a large statistical sample and precise and accurate measurements of the physical parameters of the planetary systems populating these three regions. It is also essential to correctly know the orbital parameters (eccentricity and spin-orbit obliquity) as they provide critical insight into their dynamical history. 
In this context, we started an observational program within the long-term GAPS (Global Architecture of Planetary Systems; e.g., \citealt{damasso2015,esposito2017}) project to confirm and characterise planets having intermediate masses and radii between super-Earths and sub-Saturns \citep{naponiello2022,naponiello2023,naponiello2025a,naponiello2026b}. This was followed by the more focussed HONEI (Hot Neptune Initiative) observational program \citep{naponiello2025b,manni2025}. Both programs are mainly based on data collected with the HARPS-N (High Accuracy Radial velocity Planet Searcher for the Northern hemisphere; \citealt{cosentino2012}) spectrograph at the Telescopio Nazionale Galileo.

In this new paper of our series, we present new TESS photometric data and HARPS-N radial velocity (RV) measurements that allowed us to review the physical and orbital parameters of two known planetary systems, TOI-1272 and TOI-1694 (see Sect.~\ref{sec:targets}), which are almost the same distance from us, 137 and 124 pc, respectively. The two parent stars have approximately the same effective temperature ($T_{\rm eff}$\,$\sim$\,5000\,K) and are known to host two planets each, including a transiting Neptune-sized planet and a non-transiting planet. The two transiting exoplanets, TOI-1272\,b and TOI-1694\,b, both have an eccentric orbit with a period of $\sim$\,3.3\,days, which places them in the ridge. The two non-transiting exoplanets, TOI-1272\,c and TOI-1694\,c have an orbital period of $\sim$\,8.7 and $\sim$\,389\,days, respectively. The existence of the non-transiting planet, TOI-1272\,c, is seriously questioned by our analysis. 

The paper is organised as follows. 
In Sect.~\ref{sec:targets}, we summarise the properties of the two planetary systems that are the subject of this study. 
In Sect.~\ref{sec:observation}, we describe the TESS photometry that we used and present new times series of HARPS-N data.
%
Sect.~\ref{sec:host_stars} provides a characterisation of the two parent stars, while that of all their planets is described in Sect~\ref{sec:analysis_characterisation}. 
Our analysis of the TOI-1272 system (Sect.~\ref{sec:analysis_TOI-1272}) highlights how stellar activity has been previously misinterpreted as a planetary signal and, consequently, that the planet TOI-1272\,c, claimed in the discovery paper, should be removed from the confirmed planet catalogues. 
Our conclusions are summarised in Sect.~\ref{sec:conclusions}.

\section{Target properties}
\label{sec:targets}
In this section, we summarise the main properties, based on previous studies, of the two planetary systems that are the subject of this study. 

\subsection{TOI-1272}
\label{sec:TOI-1272}
The discovery of the TOI-1272 planetary system was reported by \citet{macDougall2022}. It is composed of a main-sequence dwarf star ($V$\,=\,11.76\,mag, $M_{\star}$\,$\sim$\,0.8\,$M_{\sun}$, $R_{\star}$\,$\sim$\,0.8\,$R_{\sun}$, $T_{\rm eﬀ}$\,$\sim$\,5000\,K) and an eccentric transiting hot super-Neptune, TOI-1272\,b ($M_{\rm p}$\,$\sim$\, 25\,$M_{\oplus}$, $R_{\rm p}$\,$\sim$\,4.1\,$R_{\oplus}$, $e$\,$\sim$\,0.3), with an orbital period of 3.316\,days. The values of these parameters, which allowed the authors to confirm the existence of the planet, are based on the analysis of TESS photometry (sectors 15, 16, 22), 62 spectra obtained with the HIRES instrument \citep{vogt1994} at the Keck Observatory, and additional ground-based time-series photometry. Referring to {\it Gaia} Data Release 2 (DR2), \citet{macDougall2022} noticed a neighbour star within $30^{\prime \prime}$ that dilutes the TOI-1272 light curve by less than $1\%$. A variability signal at $28.3 \pm 0.6$ days was noted in the TESS data, which was associated with stellar rotation. The analysis of HIRES RV measurements allowed \citet{macDougall2022} to also identify a non-transiting outer companion, TOI-1272\,c, on an 8.7 day orbit with a mass of $M_{\rm p}\sin{i}=26.7\pm3.1M_{\oplus}$ and $e \lesssim 0.35$. However, their transit-timing-variation (TTV) analysis, whose search was extended over a photometric baseline of $\sim$\,600\,days, did not show evidence in favour of TTVs.

The parameters of the TOI-1272 system were revised by \citet{macDougall2023} in a study that included 85 TESS target stars from the TESS-Keck Survey sample \citep{chontos2022}. 
TOI-1272 was also included in a subsequent uniform analysis of the TESS-Keck Survey data performed by \citet{polanski2024}, who found no evidence of likely false positives among their entire sample. 
Based on their new analysis, they measured $M_{\rm p}\sin{i}=21.7\pm3.6\,M_{\oplus}$ and $e \lesssim 0.12$ for TOI-1272\,c. A comparison of the results of all these studies is reported in Tables~\ref{tab:planet_TOI-1272} and \ref{tab:star_TOI-1272}. We have excluded the parameters of TOI-1272\,c from Table~\ref{tab:planet_TOI-1272}.
%

\subsection{TOI-1694}
\label{sec:TOI-1694}
The discovery of the TOI-1694 planetary system was announced by \citet{VanZandt2023}, who reported that the parent star is an early K dwarf ($V=11.4$\,mag, $M_{\star}\sim 0.8\,M_{\sun}$, $R_{\star}\sim 0.8\,R_{\sun}$, $T_{\rm eﬀ} \sim 5000$\,K) hosting two planets, namely TOI-1694\,b and TOI-1694\,c. They were confirmed thanks to TESS photometry (sectors 19 and 20) and 20 HIRES spectra. The inner planet in this system, TOI-1694\,b, is a transiting hot super-Neptune ($M_{\rm p}\sim 26\,M_{\oplus}$, $R_{\rm p}\sim 5.4\,R_{\oplus}$) with an orbital period of 3.377\,days; the outer planet, TOI-1694\,c, is a non-transiting Jupiter analogue ($M_{\rm p}\sin{i}\sim 1 \,M_{\rm Jup}$) with an orbital period of $\sim 390$ days and low eccentricity ($e<0.2$).

\citet{mistry2023} presented an accurate study of the transit parameters of TOI-1694\,b in a paper related to the validation of 11 TESS planet candidates by examining their light curves and computing false-positive probabilities using statistical validation tools. High-angular-resolution imaging, performed by using adaptive optics and speckle imaging techniques, did not detect any nearby sources. Ground-based photometry was also used by \citet{mistry2023}.

TOI-1694 was also included in the above-mentioned study by \citet{polanski2024}, who reported slightly different values of several parameters (Tables~\ref{tab:planet_TOI-1694} and \ref{tab:star_TOI-1694}). 
In particular, they mistakenly reported a value of the semi-major axis of the TOI-1694\,c's orbit ($0.10\pm0.02$\,au), which is an order of magnitude lower than the correct value (see Fig.~6 from \citealt{VanZandt2023}). This incorrect value, which was automatically included in the exoplanet archives, has had a negative impact on statistical studies of cold Jupiters, mainly in that this planet has remained out of the scope of investigations by means of {\it Gaia} astrometric data.

Finally, \citet{handley2025} measured the Rossiter–McLaughlin effect by observing TOI-1694 during a transit of TOI-1694\,b with the KPF spectrograph \citep{gibson2024}. The sky-projected obliquity was constrained as $\lambda=9^{\circ}$$^{+22^{\circ}}_{-18^{\circ}}$, a value that suggests a nearly aligned orbit.


\begin{table}
\centering %
\caption{Stellar parameters of TOI-1272.} %
\label{tab:star_TOI-1272} %
\resizebox{\hsize}{!}{
\begin{tabular}{lccc|cc}
\hline %
\hline  \\[-8pt]
Parameter & Unit & This work & Source & MacDougall & MacDougall \\
 &  &  &  & et al. (2022) & et al. (2023) \\
\hline  \\[-6pt] %
\multicolumn{6}{l}{\large{\bf Cross-identifications}} \\ [2pt] %
TOI \dotfill & \dotfill & 1272 & TOI catalog & \dots & \dots \\
TIC \dotfill & \dotfill & 417948359 & Tycho-2 & \dots & \dots \\
2MASS \dotfill & \dotfill& J13164717+4951399 & 2MASS & \dots & \dots \\
{\it Gaia} \dotfill & \dotfill& 1556242405699527424 & {\it Gaia}~DR3 & \dots & \dots \\ [6pt] %
\multicolumn{6}{l}{\large{\bf Astrometric properties}} \\ [2pt] %
$\alpha$\,(J2015.5) \dotfill & h:m:s & \,\,\,\,13:16:47.09 & {\it Gaia}~DR2 & \dots & \dots \\
$\delta$\,(J2015.5)  \dotfill & $^{\circ}$:$^{\prime}$:$^{\prime \prime}$ & +49:51:39.8\, & {\it Gaia}~DR2 & \dots & \dots \\
$\pi$ \dotfill & mas & $7.26 \pm 0.01$ & {\it Gaia}~DR3 & $7.24 \pm 0.021$ & \dots \\
$\mu_\alpha \cos{\delta}$ \dotfill & mas\,yr$^{-1}$ & $-63.24 \pm 0.01$~~~~~ & {\it Gaia}~DR3 & \dots & \dots \\
$\mu_\delta$ \dotfill & mas\,yr$^{-1}$  & $4.75 \pm 0.01$ & {\it Gaia}~DR3 & \dots & \dots \\ [6pt] %
\multicolumn{6}{l}{\large{\bf Photometric properties}} \\ [2pt] %
$B$ \dotfill & mag & $12.846 \pm 0.024$ & APASS Johnson$^{(a)}$ & \dots & \dots \\  
$G_{\rm BP}$ \dotfill & mag & $12.0694 \pm 0.0008$ & {\it Gaia}~DR3 & \dots & \dots \\
$V$ \dotfill & mag & $11.878 \pm 0.011$ & APASS Johnson$^{(a)}$ & \dots & \dots \\ 
$g'$ \dotfill & mag & $12.320 \pm 0.030$ & APASS Sloan$^{(a)}$ & \dots & \dots \\  
$r'$ \dotfill & mag & $11.566 \pm 0.006$ & APASS Sloan$^{(a)}$ & \dots & \dots \\ 
$G$ \dotfill & mag & $11.5948 \pm 0.0003$ & {\it Gaia}~DR3 & \dots & \dots \\
$i'$ \dotfill & mag & $11.311 \pm 0.025$ & APASS Sloan$^{(a)}$ & \dots & \dots \\ 
$T$ \dotfill & mag & $11.0244 \pm 0.0061$ & {\it Gaia}~DR3 & \dots & \dots \\
$G_{\rm RP}$ \dotfill & mag & $10.9633 \pm 0.0005$ & {\it Gaia}~DR3 & \dots & \dots \\
$J$ \dotfill & mag & $10.218 \pm 0.019$ & 2MASS$^{(b)}$ & \dots & \dots \\
$H$ \dotfill & mag & $\,\,\,9.793 \pm 0.029$ & 2MASS$^{(b)}$  & \dots & \dots \\
$K_{\rm s}$ \dotfill  & mag & $\,\,\,9.701 \pm 0.020$ & 2MASS$^{(b)}$  & \dots & \dots \\
$W1$\,(3.4\,$\mu$m) \dotfill & mag & $\,\,\,9.632 \pm 0.023$ & AllWISE$^{(c)}$ & \dots & \dots \\
$W2$\,(4.6\,$\mu$m) \dotfill & mag & $\,\,\,9.698 \pm 0.020$ & AllWISE$^{(c)}$ & \dots & \dots \\
$W3$\,(12\,$\mu$m)  \dotfill & mag & $\,\,\,9.613 \pm 0.041$ & AllWISE$^{(c)}$ & \dots & \dots \\ [6pt] %
\multicolumn{6}{l}{\large{\bf Spectroscopic properties}} \\ [2pt] %
Spectral type \dotfill & & $\rm K2\,V-K3\,V$$^{(d)}$ & This work & \dots & \dots \\
$T_{\rm eff}$$^{(e)}$ \dotfill & K & $4985\pm40$ & This work & $4985 \pm 121$ & $5065^{+52}_{-50}$ \\
$v\sin{i_{\star}}$ \dotfill & km\,s$^{-1}$ & $1.2\pm0.9$ & This work & \dots & \dots \\
$v_{\rm micro}$ \dotfill & km\,s$^{-1}$& $0.51 \pm 0.32$ & This work & \dots & \dots \\%
$\rm{[Fe/H]}$ \dotfill & dex & $0.10 \pm 0.08$ & This work & $0.17 \pm 0.06$ & $0.18^{+0.05}_{-0.06}$ \\%
[6pt] %
\multicolumn{6}{l}{\large{\bf Derived parameters}} \\ [2pt] %
$L_{\star}$ \dotfill & $L_{\sun}$ & $0.3424_{-0.047}^{+0.048}$ & This work & \dots & \dots \\ [2pt] %
$M_{\star}$ \dotfill & $M_{\sun}$ & $0.828_{-0.032}^{+0.037}$ & This work & $0.851 \pm 0.049$ & $0.88_{-0.02}^{+0.01}$ \\ [2pt]  %
$R_{\star}$ \dotfill & $R_{\sun}$ & $0.800_{-0.013}^{+0.011}$ & This work & $0.788 \pm 0.033$ & $0.79_{-0.01}^{+0.01}$ \\ [2pt] %
$\log g_{\star}$ \dotfill & cgs & $4.551\pm0.022$ & This work & $4.55 \pm 0.10$ & \dots \\ [2pt] %
$\rho_{\star}$\dotfill & g\,cm$^{-3}$ & $2.29_{-0.13}^{+0.15}$ & This work & $2.453 \pm 0.343$ & $2.46_{-0.09}^{+0.08}$ \\ [2pt] %
$\log R^{\prime}_{\rm HK}$$^{(f)}$\dotfill & dex &  $-4.9315\pm0.0007$ & This work & $-4.705$ & \dots \\ [2pt] %
Age\dotfill & Gyr & $7.0^{+4.5}_{-4.3}$ & This work & $3.65^{+4.17}_{-0.98}$ & $1.1_{-0.8}^{+1.6}$ \\ [2pt] %
$A_V$ \dotfill & mag & $< 0.022$ & This work & \dots & \dots \\ [2pt] %
$d$ \dotfill & pc  & $137.73_{-0.20}^{+0.21}$ & This work & \dots & \dots  \\ [2pt] %
\hline %
\end{tabular}
}
\tablefoot{
$^{(a)}$Values taken from the AAVSO Photometric All Sky Survey (APASS) project \citep{henden2016}. $^{(b)}$Values taken from the Two Micron All Sky Survey (2MASS) project \citep{skrutskie2006}. $^{(c)}$Values taken from the Wide-field Infrared Survey Explorer (WISE) mission \citep{cutri2021}.
$^{(d)}$The spectral type was obtained from \citet{pecaut2013} (Table 5, Version 2022.04.16). $^{(e)}$The effective temperature was obtained using the {\tt EXOFASTv2} tool (Sect.~\ref{sec:stellar_parameters}) considering both the colour index and the effective temperature.
$^{(f)}$This is the weighted mean from the values reported in Table~\ref{tab:RV_TOI-1272}.
}
\end{table}

\begin{table}
\centering %
\caption{Stellar parameters of TOI-1694.} %
\label{tab:star_TOI-1694} %
\resizebox{\hsize}{!}{
\begin{tabular}{lccc|ccc}
\hline %
\hline  \\[-8pt]
Parameter & Unit & This work & Source & Van Zandt & Mistry & MacDougall  \\
          &      &           &        & et al. (2023) & et al. (2023) & et al. (2023)  \\
\hline  \\[-6pt] %
\multicolumn{7}{l}{\large{\bf Cross-identifications}} \\ [2pt] %
TOI \dotfill & \dotfill & 1694 & TOI catalog & \dots & \dots & \dots \\
TIC \dotfill & \dotfill & 396740648 & TIC & \dots & \dots & \dots \\
TYC \dotfill & \dotfill & 4108-01434-1 & Tycho-2 & \dots & \dots & \dots \\
2MASS \dotfill & \dotfill& J06305955+6621384 & 2MASS & \dots & \dots & \dots \\
{\it Gaia} \dotfill & \dotfill& 1104269596843209472 & {\it Gaia}~DR3 & \dots & \dots & \dots \\ [6pt] %
\multicolumn{7}{l}{\large{\bf Astrometric properties}} \\ [2pt] %
$\alpha$\,(J2015.5) \dotfill & h:m:s & \,\,\,\,06:30:59.69 & {\it Gaia}~DR2 & \dots & \dots & \dots \\
$\delta$\,(J2015.5)  \dotfill & $^{\circ}$:$^{\prime}$:$^{\prime \prime}$ & +66:21:38.1\, & {\it Gaia}~DR2 & \dots & \dots & \dots \\
$\pi$ \dotfill & mas & $8.04 \pm 0.03$ & {\it Gaia}~DR3 & \dots & \dots & $7.99$ \\
$\mu_\alpha \cos{\delta}$ \dotfill & mas\,yr$^{-1}$  & ~~~$46.35 \pm 0.02$~~~~~ & {\it Gaia}~DR3 & \dots & \dots & \dots \\
$\mu_\delta$ \dotfill & mas\,yr$^{-1}$  & $-20.25 \pm 0.02$~~~~ & {\it Gaia}~DR3 & \dots & \dots & \dots \\ [6pt] %
\multicolumn{7}{l}{\large{\bf Photometric properties}} \\ [2pt] %
$B$ \dotfill & mag & $12.477 \pm 0.081$ & APASS Johnson$^{(a)}$ & \dots & \dots & \dots \\ 
$G_{\rm BP}$ \dotfill & mag & $11.7737 \pm 0.0006$ & {\it Gaia}~DR3 & \dots & \dots & \dots \\
$V$ \dotfill & mag & $11.520 \pm 0.046$ & APASS Johnson$^{(a)}$ & \dots & \dots & \dots \\ 
$g'$ \dotfill & mag & $12.044 \pm 0.088$ & APASS Sloan$^{(a)}$ & \dots & \dots & \dots \\  
$r'$ \dotfill & mag & $11.229 \pm 0.067$ & APASS Sloan$^{(a)}$ & \dots & \dots & \dots \\ 
$G$ \dotfill & mag & $11.3084 \pm 0.0003$ & {\it Gaia}~DR3 & \dots & \dots & \dots \\
$i'$ \dotfill & mag & $10.980 \pm 0.065$ & APASS Sloan$^{(a)}$ & \dots & \dots & \dots \\  
$T$ \dotfill & mag & $10.7424 \pm 0.0061$ & TESS & \dots & \dots & \dots \\
$G_{\rm RP}$ \dotfill & mag & $10.6847 \pm 0.0003$ & {\it Gaia}~DR3 & \dots & \dots & \dots \\
$J$ \dotfill & mag & $\,\,\,9.957 \pm 0.029$ & 2MASS$^{(b)}$ & \dots & \dots & \dots \\
$H$ \dotfill & mag & $\,\,\,9.473 \pm 0.030$ & 2MASS$^{(b)}$ & \dots & \dots & \dots \\
$K_{\rm s}$ \dotfill & mag & $\,\,\,9.425 \pm 0.020$ & 2MASS$^{(b)}$ & \dots & \dots & \dots \\
$W1$\,(3.4\,$\mu$m) \dotfill & mag & $\,\,\,9.364 \pm 0.023$ & AllWISE$^{(c)}$ & \dots & \dots & \dots \\
$W2$\,(4.6\,$\mu$m) \dotfill & mag & $\,\,\,9.431 \pm 0.020$ & AllWISE$^{(c)}$ & \dots & \dots & \dots \\
$W3$\,(12\,$\mu$m)  \dotfill & mag & $\,\,\,9.363 \pm 0.041$ & AllWISE$^{(c)}$ & \dots & \dots & \dots \\ [6pt] %
\multicolumn{7}{l}{\large{\bf Spectroscopic properties}} \\ [2pt] %
Spectral type \dotfill & & $\rm K2\,V - K3\,V$$^{(d)}$ & This work & \dots & \dots & \dots \\
$T_{\rm eff}$$^{(e)}$ \dotfill & K & $5005 \pm 65 $ & This work & $5066 \pm 100$ & $5135\pm50$ & $5058_{-55}^{+60}$\\ [2pt] %
$v\sin{i_{\star}}$ \dotfill & km\,s$^{-1}$ & $1.2\pm0.8$ & This work & $1.2 \pm 1.0$ & \dots  & \dots  \\
$v_{\rm micro}$ \dotfill & km\,s$^{-1}$& $0.41 \pm 0.31$ & This work & \dots  & \dots  & \dots \\%
$\rm{[Fe/H]}$ \dotfill & dex & $+0.04 \pm 0.08$ & This work & $0.12 \pm 0.06$ & $+0.06\pm 0.08$ & $+0.13_{-0.06}^{+0.05}$ \\%
[6pt] %
\multicolumn{7}{l}{\large{\bf Derived parameters}} \\ [2pt] %
$L_{\star}$ \dotfill & $L_{\sun}$ & $0.381_{-0.017}^{+0.022}$ & This work & \dots & \dots& \dots \\ [2pt] %
$M_{\star}$ \dotfill & $M_{\sun}$ & $0.828_{-0.032}^{+0.039}$ & This work & $0.84 \pm 0.03$ & $0.85\pm0.11$ & $0.85_{-0.02}^{+0.02}$ \\ [2pt] %
$R_{\star}$ \dotfill & $R_{\sun}$ & $0.812_{-0.022}^{+0.023}$ & This work & \dots & $0.818\pm0.048$ & $0.80_{-0.01}^{+0.01}$ \\ [2pt] %
$\log g_{\star}$ \dotfill & cgs & $4.536_{-0.026}^{+0.028}$ & This work & $4.53 \pm 0.10$ & $4.66\pm0.10$ & \dots \\ [2pt] %
$\rho_{\star}$\dotfill & g\,cm$^{-3}$ & $2.18_{-0.18}^{+0.20}$ & This work & \dots & \dots & $2.3_{-0.15}^{+0.14}$ \\  [2pt] %
$\log R^{\prime}_{\rm HK}$$^{(f)}$\dotfill & dex &  $-4.8664\pm0.0016$ & This work & <-4.7 & \dots & \dots \\ [2pt] %
Age\dotfill & Gyr & $7.8^{+4.1}_{-4.6}$ & This work & \dots & \dots & $4.3_{-2.6}^{+3.5}$ \\ [2pt] %
$A_V$ \dotfill & mag & $<0.065$ & This work & \dots & \dots & \dots \\  [2pt] %
$d$ \dotfill & pc  & $124.36_{-0.43}^{+0.42}$ & This work & \dots & \dots & \dots \\  [2pt] %
\hline %
\end{tabular}
}
\tablefoot{Notes are the same as Table~\ref{tab:star_TOI-1272}.
}
\end{table}

\section{Observations and data reduction}
\label{sec:observation}
The current approved procedure for characterising transiting-exoplanet systems involves the joint modelling of transit photometric data and RV measurements. 
High-angular-resolution imaging is also part of the standard process to assess possible contamination by background stars \citep[e.g.][]{lillo-box12,evans2016}. The datasets used in this study are described in the following. 

\subsection{TESS and ASAS-SN photometry of TOI-1272}
\label{sec:photometry1272}
TOI-1272 was observed by TESS while monitoring sectors 15, 16, and 22 (those used by \citealt{macDougall2022}) as well as, more recently, sectors 49 and 76. In particular, we adopted the Presearch Data Conditioning Simple Aperture Photometry (PDCSAP; \citealt{Stumpe2012, Stumpe2014,Smith2012}) short-cadence light curve (with exposure times of 120\,s), which is provided by the TESS Science Processing Operations Center (SPOC) pipeline and retrieved via the Python package \texttt{lightkurve} \citep{lightkurve} from the Mikulski Archive for Space Telescopes Portal. No significant TTVs were reported for TOI-1272\,b by \citet{naponiello2026a} based on the same dataset.

We retrieved archival $V$-band photometry from the All-Sky Automated Survey for Supernovae (ASAS-SN; \citealt{Kochanek2017}), which provides a longer time baseline but at a lower cadence. For the ASAS-SN data, we removed a long-term trend by fitting a low-order polynomial and computed a generalised Lomb-Scargle (GLS; \citealt{zechmeister2009}) periodogram. The main peak is found at $P$\,=\,24.3\,d, with a false alarm probability (FAP) slightly below 1\%. Similarly, the GLS computed on the complete TESS dataset (all five sectors combined) shows a dominant peak at $P$\,=\,25.3\,d, consistent with the ASAS-SN result. We also applied the TESS Systematics-Insensitive Periodogram (TESS-SIP), following \citet{macDougall2022}, finding that the highest peak occurs at $P$\,$\sim$\,26\,d.  When splitting the data into four segments (S15–16, S22, S49, and S76), the dominant periods are $25.5$\,d, $28.8$\,d, $23.6$\,d, and $17.6$\,d, respectively. The dispersion of these values and, in particular, the shorter period detected in the most recent sector suggest evolving active regions and a non-stationary starspot pattern.

Interestingly, several of the detected periods are close to integer multiples of $8.7$\,d, since $2\times8.7=17.4$ ($\approx$\,17.6\,d) and $3\times8.7=26.1$ ($\approx$\,25.5\,d). The peaks at $17-18$\,d and $25-26$\,d are therefore consistent with a scenario in which the $8.7$\,d signal, the period of the supposed planet TOI-1272\,c, is related to stellar activity and its harmonics. Overall, the photometry supports a stellar rotation period of $P_{\rm rot}$\,=\,$25-26$\,d, with variability in the recovered peak likely driven by the temporal evolution of magnetic activity. 
However, it is not possible to establish a more accurate constraint on $P_{\rm rot}$ based solely on all available photometry (TESS and ASAS-SN).

We also retrieved an archival light curve of roughly 5000 points collected by the SuperWASP survey \citep{butters2010} over $\sim$\,120 nights. The Lomb-Scargle periodogram for this dataset does not show any clear periodicity.

\subsection{TESS photometry of TOI-1694}
\label{sec:photometry1694}
TOI-1694 was observed by TESS while scanning sectors 19, 20 (those analysed by \citealt{VanZandt2023}), and the most recent sector 73 (included in \citealt{handley2025}). For the data in the first two sectors, we adopted the short-cadence 120\,s PDCSAP light curve provided by the TESS SPOC pipeline; whereas for the data from sector 73, we used the same light curve that was obtained by \citet{handley2025} after a detrending and removal of systematics (see their Fig.\,1) since TOI-1694 was only partially captured on the outermost pixel of the TESS imaging area. This light curve was kindly provided to us by the team of authors. 
No significant TTVs were reported for TOI-1694\,b by \citet{naponiello2026a} using the same dataset. Furthermore, our analysis, carried out on the basis of a GLS, did not indicate any significant signals related to stellar activity.

\subsection{HARPS-N spectra}
\label{sec:HARPS-N_spectra}
We acquired times-series spectroscopic data of TOI-1272 and TOI-1694 with HARPS-N, which is a high-resolution ($R$\,=\,115\,000), visible-light (383\,nm\,$\leq$\,$\lambda$\,$\leq$\,690\,nm), fibre-fed echelle spectrograph mounted on the 3.58-m Italian Telescopio Nazionale Galileo (TNG), located at the Observatorio del Roque de los Muchachos in the island of La Palma, Spain. Due to its long-term stability and simultaneous wavelength calibration, HARPS-N can achieve RV measurements with an accuracy of $\sim$\,1\,m\,s$^{-1}$ \citep{cosentino2012}.

HARPS-N high-resolution spectra were obtained within the GAPS-Neptune project \citep{naponiello2022}. Data collection began when the two targets were still considered exoplanet-system candidates. In total, we collected 94 spectra for TOI-1272 between 17 May 2020 and 13 August 2023 (Table~\ref{tab:RV_TOI-1272}) and 89 spectra for TOI-1694 between 3 October 2020 and 7 November 2022 (Table~\ref{tab:RV_TOI-1694}). RVs and activity indices were determined from HARPS-N spectra, which were reduced using version 3.2.0 of the HARPS-N Data Reduction Software ({\tt DRS}; \citealt{dumusque2021}). The activity indices were extracted manually after reduction, using a script that is not officially part of the DRS but that we adapted to make it compatible with the HARPS-N reduced data.

\subsection{High-angular-resolution imaging}
We observed TOI-1694 with the AstraLux lucky-imaging camera \citep{hormuth08} mounted at the 2.2\,m telescope at the Observatorio de Calar Alto on February 26, 2020, at an airmass of 1.16 and using a $z$ filter, obtaining 42\,868 frames with a 10\,ms exposure time per frame. The instrument pipeline reduced the data and selected the best 10\% of the frames to finally stack them in a composite image with an effective exposure time of 43\,s. No companions were found in this image. 
We determined the sensitivity limits following the procedure explained by \cite{lillo-box12,lillo-box14} and used them to compute the blended source confidence, which measures the probability that the target has a companion not detected by our high-resolution image that could potentially mimic the depth of the transit signal. In this case, this probability is as low as 0.12\% (see also \citealt{lillo-box24} for further details). 

Unpublished adaptive optic and speckle images of TOI-1272 were obtained with the NIRC2@KECK\,II \citep{wizinowich2000} and the $^{\backprime}$Alopeke@GEMINI\,North \citep{scott2021}, respectively, and are available in the ExoFOP archive. We did not use them.

\section{Host-star characterisation}
\label{sec:host_stars}
To estimate the photospheric parameters of the host stars, we produced co-added spectra from the available HARPS-N data (see Sect.~\ref{sec:HARPS-N_spectra}). The signal-to-noise ratio (S/N) of the co-added spectra reached typical values of $\sim$\,150\,--\,200 at $\lambda$\,$\sim$\,6000\,\AA. 

\subsection{Atmospheric parameters and iron abundance}
\label{star_atmos_param}
We derived the spectroscopic atmospheric parameters, i.e. the effective temperature ($T_{\rm eff}$), the surface gravity ($\log g$), the microturbulence velocity ($\xi$) and iron abundance ([Fe/H]), by measuring the equivalent widths (EWs) of iron lines. The EWs were measured using the \texttt{ARESv2} code \citep{Sousaetal2015} taking into account the line list by \cite{Biazzoetal2022}. 

The analysis was performed using the 2019 versions of the \texttt{MOOG} radiative transfer code \citep{Sneden1973}, adopting the \citet{CastelliKurucz2003} grid of model atmospheres with new opacities (ODFNEW). We followed the standard methodology described by \citet{Biazzoetal2022}: $(i)$ $T_{\rm eff}$ was derived by imposing the independence of the iron abundance on the line excitation potentials (excitation equilibrium); $(ii)$ $\log{g}$ was determined via the ionisation equilibrium between \ion{Fe}{i} and \ion{Fe}{ii} lines; $(iii)$ $\xi$ was obtained by minimising the trend between the reduced EW (${\rm EW}/\lambda$) of \ion{Fe}{i} lines and the corresponding iron abundance. The final values of the stellar parameters are listed in Tables~\ref{tab:star_TOI-1272} and \ref{tab:star_TOI-1694} for TOI-1272 and TOI-1694, respectively.

\subsection{Projected rotational velocity}
Once the atmospheric parameters were fixed, we derived the projected rotational velocity, $v \sin{i}$, for both stars using the spectral-synthesis method and the {\tt synth} driver of \texttt{MOOG}. We targeted specific spectral regions around $540$, $620$ and $670$\,nm. Following \cite{Breweretal2016}, we assumed a macroturbulence velocity $v_{\rm macro}$\,=\,1.8\,km\,s$^{-1}$ for both targets and found $v\sin{i} = 1.8 \pm 0.9$\,km\,s$^{-1}$ and $1.2\pm0.8$\,km\,s$^{-1}$ for TOI-1272 and TOI-1694, respectively. 

\subsection{Stellar parameters}
\label{sec:stellar_parameters}
To determine the mass, radius, and age of both host stars, we used the {\tt EXOFASTv2} tool (\citealt{2017ascl.soft10003E, Eastman2019}; see also \citealt{naponiello2025a} for more details), which allows simultaneous modelling of the stellar spectral energy distribution (SED) and the MIST evolutionary tracks (e.g., \citealt{Paxton2015}) in a differential evolution Markov chain Monte Carlo Bayesian framework. We used the APASS Johnson $B$ and $V$, Sloan $g'$, $r'$, and $i'$ magnitudes, the 2MASS near-infrared $J$, $H$, and $K_{\rm s}$ magnitudes, and the WISE $W1$, $W2$, and $W3$ infrared magnitudes to model the SED of both stars (Tables~\ref{tab:star_TOI-1272} and \ref{tab:star_TOI-1694}). Moreover, we imposed Gaussian priors on the {\it Gaia} DR3 parallax as well as on the $T_{\rm eff}$ and [Fe/H] derived in Sect.~\ref{star_atmos_param}. The stellar SED and its best fit are shown for both stars in Fig.~\ref{fig:stellarSEDs}. The stellar masses, radii, ages and their $1\sigma$ uncertainties were derived from the medians and 15.86\%\,$-$\,84.14\% percentiles of the posteriors, respectively. These values are given in Tables~\ref{tab:star_TOI-1272} and \ref{tab:star_TOI-1694} for TOI-1272 and TOI-1694, respectively.

\subsection{Age estimate for TOI-1272 via gyrochronology}
We also inferred the age of TOI-1272 thanks to the thanks to the gyrochronology method. Taking into account Fig.~4 of \citet{Gruneretal23} and a {\it Gaia} colour index $G_{\rm BP} - G_{\rm RP} = 1.11$\,mag, and assuming a rotation period of $\sim$\,$26-28$\,d, we estimated an age of about 4\,Gyr for TOI-1272 with an uncertainty of at least 25\,$-$\,30\%, compatible with that of the previous section. 

The chromospheric index, $\log R^{\prime}_{\rm HK}$$\sim$\,$-4.705$, as found by \citet{macDougall2022}, points to an age of 2\,--\,3\,Gyr according to the relations of \citet{Lorenzo-Oliveiraetal18} and \citet{Carvalho-Silvaetal25}. Nevertheless, such an age estimate can be affected by a  modulation of the chromospheric index associated with a stellar activity cycle. Considering that the cycle amplitude can be similar to that of the Sun ($\Delta \log R^{\prime}_{\rm HK}$\,$\sim$\,0.25) or greater, TOI-1272 could have a mean chromospheric index lower than that reported by \citet{macDougall2022} and, therefore, compatible with an age of $\sim$\,4~Gyr.

\begin{table}
\caption{Orbital and physical parameters for TOI-1272\,b.} %
\label{tab:planet_TOI-1272} %
\centering %
\resizebox{\hsize}{!}{
\begin{tabular}{lc|ccc}
\hline\hline \\ [-8pt] %
~~~~~~~Parameter &  This work & MacDougall & MacDougall & Polanski \\ %
&           & et al. (2022)    & et al. (2023)    & et al. (2024)  \\ [2pt] %
\hline \\[-8pt] %
\multicolumn{5}{l}{\bf Transit parameters} \\ %
$P_{\rm orb}$\,(d)         \dotfill & 3.3159765\,(15) & 3.315990\,(20) & 3.3159790\,(60) & 3.315990\,(18) \\ [2pt] %
$T_{\rm 0}$\,(BJD$-2458000$) \dotfill & 713.03098\,(40) & 713.0253\,(60) & 713.03150\,(60) & 713.02552\, (36) \\ [2pt] %
$T_{\rm 14}$\,(h)           \dotfill & $1.521^{+0.021}_{-0.019}$ & \dots & $1.57^{+0.06}_{-0.05}$ & \dots   \\ [2pt] %
$R_{\rm p}/R_{\star}$        \dotfill & $0.04881^{+0.00085}_{-0.00094}$ & \dots & $0.0477_{-0.0017}^{+0.0025}$ & \dots  \\ [2pt]
$b$                          \dotfill & $0.53^{+0.07}_{-0.12}$ & $0.45^{+0.15}_{-0.21}$ & \dots & \dots \\  [2pt]
$i$\,(deg)                  \dotfill & $86.00^{+0.67}_{-0.39}$ & \dots & \dots & \dots  \\  [2pt]
$a/R_{\star}$                \dotfill & $11.03\pm0.23$ & \dots & \dots & \dots  \\  [2pt]
$q_1$,\,\textsc{tess}$^{(a)}$\dotfill & $0.17^{+0.18}_{-0.10}$ & \dots & \dots & \dots  \\  [2pt]
$q_2$,\,\textsc{tess}$^{(a)}$\dotfill & $0.35^{+0.37}_{-0.25}$ & \dots & \dots & \dots  \\  [6pt]
\multicolumn{5}{l}{\bf RV parameters} \\ [2pt] %
$K$\,(m\,s$^{-1}$)           \dotfill & $12.5^{+0.68}_{-0.66}$ & $12.6 \pm 1.1$ & \dots & $13.5\pm 1.3$ \\ [2pt] %
$\sqrt{e}\sin\omega$         \dotfill & \,$0.495\pm0.060$ & \dots & \dots & \dots \\ [2pt] %
$\sqrt{e}\cos\omega$         \dotfill & $-0.310\pm0.060$\,\, & \dots & \dots & \dots \\ [2pt] %
$e^{(b)}$                    \dotfill & $0.345^{+0.046}_{-0.041}$ & $0.338_{-0.062}^{+0.056}$ & $0.350 \,(55)$ & $0.350 \,(55)$ \\ [2pt] %
$\omega$\,(deg)              \dotfill & $122\pm7$ & $123.6 \pm 11.5$ & \dots & $141 \pm 15$ \\   [6pt] %
\multicolumn{5}{l}{\bf Instrumental parameters} \\
$\sigma_{\rm{TESS}}$$^{(c)}$\,(ppm)           \dotfill & $0.26^{+8.73}_{-0.25}$ & \dots & \dots & \dots  \\  [2pt]
$\sigma_{\rm{HARPS-N}}$$^{(c)}$\,(m\,s$^{-1}$)\dotfill & $1.74^{+1.17}_{-1.11}$ & \dots & \dots & \dots  \\  [2pt]
$\mu_{\rm{HARPS-N}}$$^{(d)}$\,(m\,s$^{-1}$)   \dotfill & $1901.5\pm1.3$ & \dots & \dots & \dots  \\  [2pt]
$\sigma_{\rm{HIRES}}$\,(m\,s$^{-1}$)  \dotfill & $1.43^{+0.96}_{-0.93}$ & \dots & \dots & \dots  \\  [2pt]
$\mu_{\rm{HIRES}}$\,(m\,s$^{-1}$)     \dotfill & $0.09\pm1.23$ & \dots & \dots & \dots  \\  [6pt] %
\multicolumn{5}{l}{\bf Derived parameters} \\ [2pt] %
$M_{\rm p}$\,($M_{\oplus}$)   \dotfill & $24.1\pm1.4$ & $24.6 \pm 2.3$ & \dots & $27.0^{+2.7}_{-2.5}$ \\ [2pt] %
$R_{\rm p}$\,($R_{\oplus}$)   \dotfill & $4.26\pm0.10$ & $4.14\pm0.21$ & $4.13^{+0.23}_{-0.15}$ & \dots \\ [2pt] %
$\rho_{\rm p}$\,(g\,cm$^{-3}$)\dotfill & $1.72^{+0.16}_{-0.15}$ & $1.9\pm0.3$ & \dots & $2.09^{+0.44}_{-0.59}$ \\ [2pt] %
$g_{\rm p}$\,(m\,s$^{-2}$)    \dotfill & $13.03^{+0.98}_{-0.94}$ & \dots & \dots & \dots  \\  [2pt]
$a$\,(au)                     \dotfill & $0.0410\,(11)$ & $0.0412\,(08)$ & $0.0417^{+0.0002}_{-0.0003}$ & $0.0417 \,(07)$  \\ 
$T_{\rm eq}$$^{(e)}$\,(K)       \dotfill & $1051^{+15}_{-14}$ & $961\pm32$ & \dots & \dots  \\  [2pt] %
TSM$^{(f)}$                   \dotfill & $54.8^{+4.7}_{-4.4}$ & \dots & \dots & \dots  \\  [6pt] %
\hline
\end{tabular}
}
\tablefoot{The median values of the best-fit parameters for TOI-1272\,b, along with their upper and lower $68\%$ credibility intervals as uncertainties. These values, both fitted and derived, were obtained from the posterior distributions of the corresponding models (this work). Values from literature are also reported for comparison. The numbers in brackets represent the uncertainties in the preceding digits.
$^{(a)}$$q_1 \equiv (u_1+u_2)^2$ and $q_2 \equiv (u_1/2)(u_1+u_2)^{-1}$, where $u_1$ and $u_2$ are the limb-darkening coefficients of the quadratic law \citep{Kipping2013}.
$^{(b)}$The $95\%$ confidence upper limit on the eccentricity determined when $\sqrt{e} \cos{\omega}$ and $\sqrt{e} \sin{\omega}$ are allowed to vary in the fit.
$^{(c)}$$\sigma_{\rm{TESS}}$ and $\sigma_{\rm{HARPS-N}}$ are jitters added in quadrature to the errorbars of TESS and HARPS-N, respectively.
$^{(d)}$This is the systemic RV for HARPS-N.
$^{(e)}$This represents the equilibrium temperature assuming a Bond albedo of zero and an uniform redistribution of heat to the night side.
$^{(f)}$Transmission spectroscopy metric (TSM; \citealt{Kempton2018}).
}
\end{table}

\begin{table*}
\caption{Orbital and physical parameters for TOI-1694\,b and TOI-1694\,c.} %
\label{tab:planet_TOI-1694} %
\centering %
\resizebox{\hsize}{!}{
\begin{tabular}{lcccccc|ccc}
\hline\hline \\ [-8pt] %
 & & & TOI-1694\,b & & & & & TOI-1694\,c  \\ [2pt] %
 \hline \\ [-8pt]%
Parameter & This work & Van Zandt et & Mistry et & MacDougall & Polanski et & Handley et &  This work & Van Zandt et & Polanski et \\ %
 &  & al. (2023) & al. (2023) & et al. (2023) & al. (2024) & al. (2025) &  & al. (2023) & al. (2024) \\ [2pt] %
\hline \\[-8pt] %
\multicolumn{5}{l}{Transit parameters} \\ %
$P_{\rm orb}$\,(d)\dotfill & $3.7701389\,(17)$ & $3.7701379\,(33)$ & $3.770179\,(60)$ & $3.770137\,(89)$ & $3.770107\,(85)$ & $3.77015\,(10)$ & $386.25_{-2.13}^{+2.08}$ & $389.2 \pm 3.9$ & $393.1 \pm 4.7$\\ [2pt] %
$T_{\rm 0}$\,(BJD$-2458000$)\dotfill & $843.65726\,(40)$ & \dots & $817.26620\,(70)$ & $817.26640\,(60)$ & $817.26629\,(61)$ & $817.26640\,(40)$ & $1174.84_{-2.81}^{+2.97}~~$ & \dots & $170.5\pm4.3$\\[2pt] %
$T_{\rm 14}$\,(h)\dotfill & $2.178 \pm 0.043$ & \dots & \dots & \dots & $2.869626_{-0.032}^{+0.035}$ & \dots & \dots & \dots & \dots \\ [2pt]
$R_{\rm p}/R_{\star}$\dotfill & $0.05947_{-0.00082}^{+0.00080}$  & \dots & $0.0610^{+0.0017}_{-0.0013}$ & $0.0609_{-0.0010}^{+0.0014}$ & $0.0609^{+0.0013}_{-0.0010}$& $0.061^{+0.02}_{-0.01}$ & \dots & \dots & \dots \\ [2pt]
$b$\dotfill & $0.17^{+0.12}_{-0.11}$  & \dots & $0.33^{+0.17}_{-0.20}$ & \dots & $0.26^{+0.20}_{-0.18}$ & $0.35^{+0.17}_{-0.19}$ & \dots & \dots & \dots \\  [2pt]
$i$\,(deg)\dotfill & $89.25^{+0.49}_{-0.51}$  & \dots & $88.17_{-1.19}^{+1.15}$ & \dots & \dots & $87.61^{+1.07}_{-0.99}$ & \dots & \dots & \dots \\  [2pt]
$a/R_{\star}$\dotfill & $11.79^{+0.28}_{-0.30}$ & \dots & $10.21_{-0.79}^{+0.47}$ & \dots & \dots & $10.11^{+0.52}_{-0.86}$ & $258.09^{+6.11}_{-6.54}$ & \dots & \dots \\ [2pt]
$q_1$,\,\textsc{tess}$^{(a)}$\dotfill & $0.26^{+0.11}_{-0.06}$ & \dots & \dots & \dots & \dots & $0.38^{+0.15}_{-0.11}$ & \dots & \dots & \dots \\ [2pt]
$q_2$,\,\textsc{tess}$^{(a)}$\dotfill & $0.80^{+0.14}_{-0.21}$ & \dots & \dots & \dots & \dots & $0.57^{+0.20}_{-0.19}$ & \dots & \dots & \dots \\ [6pt]
\multicolumn{5}{l}{RV parameters} \\ [2pt] %
$K$\,(m\,s$^{-1}$)\dotfill & $13.26^{+0.45}_{-0.46}$ & $12.06 \pm 0.96$ & $11.81$ & \dots & $14.3\pm1.1$ & \dots & $31.71^{+0.75}_{-0.74}$ & $33.4 \pm 1.6$ & $28.84 \pm 0.98$\\ [2pt] %
$\sqrt{e}\sin\omega$\dotfill & $-0.325^{+0.045}_{-0.040}$ & \dots & \dots & \dots & \dots & \dots & $-0.315^{+0.047}_{-0.039}$ & \dots & \dots \\ [2pt] %
$\sqrt{e}\cos\omega$\dotfill & $-0.125^{+0.036}_{-0.037}$ & \dots & \dots & \dots & \dots & \dots & $-0.099^{+0.051}_{-0.049}$ & \dots & \dots\\ [2pt] %
$e^{(b)}$\dotfill & $0.122^{+0.025}_{-0.024}$ & $0$ & \dots & \dots & 0 & 0 & $0.111 \pm 0.025$ & $0.18 \pm 0.05$ & 0 \\ [2pt] %
$\omega$\,(deg) \dotfill & $-110.95^{+6.60}_{-7.73}$ & \dots & \dots & \dots & 90 & \dots & $-107.40^{+8.92}_{-9.66}$ & \dots & 90 \\   [6pt] %
\multicolumn{5}{l}{Instrumental parameters} \\ [2pt] %
$\sigma_{\rm{TESS}}$\,(ppm)\dotfill & $0.52^{+6.21}_{-0.49}$  & \dots & \dots & \dots & \dots & \dots & \dots & \dots & \dots \\ [2pt] %
$\sigma_{\rm{HARPS-N}}$\,(m\,s$^{-1}$)\dotfill & $1.53^{+0.60}_{-0.65}$ & \dots & \dots & \dots & \dots & \dots & \dots & \dots & \dots \\ [2pt] 
$\mu_{\rm{HARPS-N}}$\,(m\,s$^{-1}$)\dotfill & $-22\,545.73^{+0.60}_{-0.61}$ & \dots & \dots & \dots & \dots & \dots & \dots & \dots & \dots \\ [6pt] %
$\sigma_{\rm{HIRES}}$\,(m\,s$^{-1}$)\dotfill & $1.77^{+0.89}_{-0.86}$ & \dots & \dots & \dots & \dots & \dots & \dots & \dots & \dots \\ [2pt] 
$\mu_{\rm{HIRES}}$\,(m\,s$^{-1}$)\dotfill & $-1.07^{+0.73}_{-0.71}$ & \dots & \dots & \dots & \dots & \dots & \dots & \dots & \dots \\ [6pt] %
\multicolumn{5}{l}{Derived parameters} \\ [2pt] %
$M_{\rm p}$\,($M_{\oplus}$)\dotfill & $28.24^{+1.05}_{-1.06}$ & $26.1 \pm 2.2$ & $25.5\pm11.9$ & \dots & $31.3^{+2.7}_{-2.5} $ & \dots & \dots & \dots & \dots \\ [2pt] %
$M_{\rm p}\sin{i}$\,($M_{\rm Jup}$)\dotfill & \dots & \dots & \dots & \dots & \dots & \dots & $0.996^{+0.027}_{-0.026}$ & $1.05 \pm 0.05$ & $0.935\pm0.050$ \\ [2pt] %
$R_{\rm p}$\,($R_{\oplus}$)\dotfill & $5.27 \pm 0.11$ & $5.44\pm0.18$ & $5.46^{+0.47}_{-0.79}$ & $5.34_{-0.12}^{+0.15}$ & \dots & \dots & \dots & \dots & \dots \\ [2pt] %
$\rho_{\rm p}$\,(g\,cm$^{-3}$)\dotfill  & $1.062^{+0.080}_{-0.076}$ & $0.89 \pm 0.12$ & 0.87 & $1.13^{+0.21}_{-0.26}$ & \dots & \dots & \dots & \dots & \dots \\ [2pt] %
$g_{\rm p}$\,(m\,s$^{-2}$)\dotfill & $9.97^{+0.57}_{-0.55}$ & \dots & \dots & \dots & \dots & \dots & \dots & \dots & \dots \\ [2pt] %
$a$\,(au)\dotfill & $0.0445 \pm 0.0013$ & \dots & \dots & $0.045_{-0.0004}^{+0.0004}$ & $0.0450\pm0.0008$ & \dots &  $0.974^{+0.028}_{-0.029}$ & \dots & $0.10\pm0.02$ \\ 
$T_{\rm eq}^{(c)}$\,(K)\dotfill & $1037^{+17}_{-16}$ & \dots & 1136.57 & \dots & \dots & \dots & $222 \pm 4$ & \dots & \dots  \\  [2pt] %
TSM$^{(d)}$\dotfill & $95.45^{+6.42}_{-5.77}$ & \dots & 125.91 & \dots & \dots & \dots & \dots & \dots & \dots \\ [6pt]
%
%
\hline
\end{tabular}
}
\tablefoot{Notes are the same as Table~\ref{tab:planet_TOI-1272}. $\mu_{\rm{HIRES}}$ represents the offset between HARPS-N and HIRES RV measurements. The numbers in brackets represent the uncertainties in the preceding digits.}
\end{table*}

\section{Analysis and results}
\label{sec:analysis_characterisation}
In this section, we report our revision of the TOI-1272 and TOI-1694 planetary systems. Specifically, we redetermined the physical and orbital parameters of these two systems based on all available TESS photometry and HARPS-N RV measurements, described in Sect.~\ref{sec:observation}, and the HIRES data, which were taken from \citet{polanski2024}. We noted that \citet{vanzandt2025} published a TOI-1272 RV dataset with the same values as those of \citet{polanski2024}, with the difference that the first 21 spectra were left out and 5 additional, more recent spectra were added. Therefore, we used the dataset from \citet{polanski2024} as it contains a larger number of measurements.

\subsection{TOI-1272}
\label{sec:analysis_TOI-1272}
Our analysis of the TOI-1272 system is more detailed and in-depth compared with that of TOI-1694. This is because, while for TOI-1694 the results of our analysis agree with those of the literature (see Sect.~\ref{sec:analysis_TOI-1694}), in the case of TOI-1272 our analysis shows that the planet TOI-1272\,c is a false positive, contrary to what was identified in the discovery paper. 

\subsubsection{No significant signal at 8.7\,days in the HARPS-N data}
Our review of the properties of this system began with the analysis of the new data collected with the HARPS-N spectrograph. In Fig.~\ref{fig:RVtimeseries}, we show the time series of HARPS-N RVs and those of HIRES \citep{polanski2024}. It is noteworthy that the two data sets were collected in approximately the same time period. 

\begin{figure*}
\centering
\includegraphics[width=\hsize]{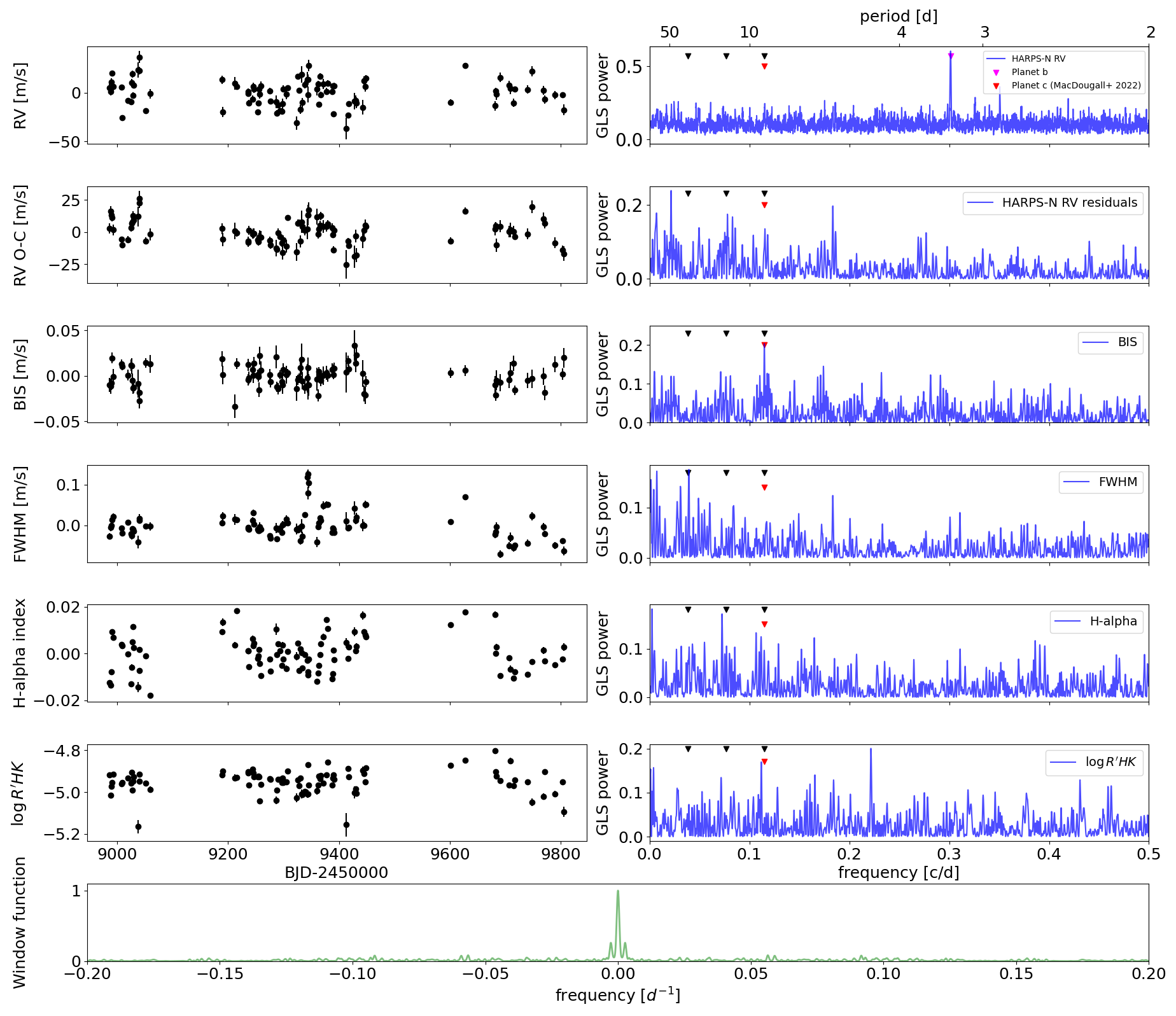}
\caption{\textit{Left panels}: Time series of the RVs and spectroscopic activity indicators derived from the HARPS-N spectra of TOI-1272, with the average values subtracted. The residual RVs, after subtracting the signal of the transiting planet b obtained with a Monte Carlo analysis, are shown in the second panel from the top. \textit{Right panels}: Corresponding GLS periodograms. The GLS of the original RVs is obtained by fitting the data with a Keplerian. The magenta and red triangles  refer to the orbital periods of the transiting planet TOI-1272\,b and the non-transiting planet TOI-1272\,c, which was identified by \cite{macDougall2022}, respectively. Black triangles identify the stellar rotation period and its first and second harmonic. \textit{Bottom panel}: Plot of the window function related to the HARPS-N observations.}
\label{fig:glsharpsactindrv}
\end{figure*}

We calculated the GLS periodogram using a Keplerian to model the HARPS-N time series, with the eccentricity sampled from a grid with a step size of 0.1. We found a significant peak at the orbital frequency of the transiting planet (first row of Fig.~\ref{fig:glsharpsactindrv}), with a least-square best-fit eccentricity of 0.4, close to the value measured by \cite{macDougall2022}. 

In the next step, we performed a more sophisticated analysis by jointly modelling HARPS-N RVs and TESS transits with a single Keplerian signal within a Bayesian framework, using the \texttt{juliet} package \citep{Espinoza2019} and sampling the posterior distribution with the nested sampler \texttt{dynesty} \citep{Speagle2020}. The GLS periodogram of the RV residuals is shown in the second row from the top of Fig. \ref{fig:glsharpsactindrv}. The main peak occurs at the period of 46.3 days with a FAP of $9.3\%$, which was estimated via a bootstrap simulation from 10\,000 mock time series, noting that there is no significant peak at the orbital period of the second planet reported by \cite{macDougall2022}, namely, at $P=8.69$\,days. We noticed that this period is likely compatible with being the second harmonic of the stellar rotation period as measured from the TESS light curve (see Sect.~\ref{sec:photometry1272}).

\subsubsection{Analysis of the activity indices}
The next step of our analysis was the characterisation of stellar activity by modelling the time series of the activity diagnostics extracted from the HARPS-N spectra, namely the full width at half maximum (FWHM) and the bisector span (BIS), measured from the CCF, and the activity indices calculated from the H-$\alpha$ and \ion{Ca}{ii}\,H\&K lines, the latter known as the $\log{ R^{\prime}_{HK}}$ index. Their time series and GLS periodograms are shown in Fig. \ref{fig:glsharpsactindrv}. The main peaks are located at 8.69\,d (BIS), 25.3\,d (FWHM), 389.9\,d (H-$\alpha$), and 4.5\,d ($\log R^{\prime}_{\rm HK}$), though none of them are statistically significant (bootstrap FAP $16\%$, $10\%$, $26\%$ and $23\%$, respectively). However, the peak value for the BIS is found at the same period as the signal attributed to TOI-1272\,c, and there is a weak-to-moderate anti-correlation between the BIS and the RV residual data ($\rho_{\rm Pearson}= -0.35$). 

The main peak of the FWHM periodogram is compatible with the stellar rotation period. These signals could have a quasi-periodic modulation; however, the GLS fits the data with a simple sinusoid. Thus, we modelled the four indices with a Gaussian Process (GP) regression using the quasi-periodic kernel, and a slightly different parametrisation than described above. In this case, a generic element of the quasi-periodic covariance matrix is defined as 
%
%
\begin{equation}
k_{QP}(t, t^{\prime}) = h^2\cdot e^{-\frac{\left(t-t^{\prime}\right)^2}{2\lambda^2} - \frac{\sin^{2}\left(\frac{\pi(t-t^{\prime})}{\theta}\right)}{2w^2}}+\, \left(\sigma^{2}(t)+\sigma^{2}_{\rm jit}\right)\cdot\delta_{t, t^{\prime}}\;. 
\end{equation}
Here, $t$ and $t^{\prime}$ represent two different epochs of observations, $\sigma$ is the uncertainty in the measurements (RVs or activity indices), and $\delta_{t, t^{\prime}}$ is the Kronecker delta. Any source of uncorrelated noise (either instrumental and astrophysical) can be modelled by adding a constant jitter term $\sigma_{\rm jit}$ in quadrature to $\sigma$. The GP hyper-parameters are: 
\begin{itemize}
\item[$\bullet$] $h$, which denotes the scale amplitude of the correlated signal; 
\item[$\bullet$] $\theta$, which represents the periodic time scale of the correlated signal and corresponds to the stellar rotation period; 
\item[$\bullet$] $w$, which describes the `weight' of the rotation period harmonic content within a complete stellar rotation (i.e. a low value of $w$ indicates that periodic variations contain a significant contribution from the harmonics of the rotation period); 
\item[$\bullet$] $\lambda$, which represents the decay timescale of the correlations. It is related to the temporal evolution of the magnetically active regions responsible for the correlated signal. 
\end{itemize}
We explored the full (hyper-)parameter space using the publicly available Monte Carlo nested sampler and the Bayesian inference tool {\tt MultiNest v3.10} (e.g. \citealt{Feroz2019}), through the {\tt pyMultiNest} wrapper \citep{Buchner2014}. To perform GP regression within the {\tt MultiNest} framework, we used the publicly available \texttt{python} module \texttt{george} v0.2.1 \citep{Ambikasaran2015}. The results are summarised in Table \ref{tab:gpactind1272}, which includes the values $\Delta\ln\mathcal{Z}$ that represent the difference between the Bayesian evidence of a model with and without the GP term. 

\begin{table}
\caption{Results of a GP regression analysis of spectroscopic activity indicators derived from HARPS-N spectra.} %
\label{tab:gpactind1272} %
\centering %
\resizebox{\hsize}{!}{
\begin{tabular}{ccccc}
\hline\hline
\noalign{\smallskip}
Parameter & \multicolumn{4}{c}{Activity diagnostic$^{a}$} \\
\hline \\ [-7pt]
& BIS & FWHM & H$\alpha$ & log$R^{\prime}_{\rm HK}$ \\[2pt]
\hline \\ [-8pt]
$h$ & 3.0$^{+3.3}_{-2.1}$ [m\,s$^{-1}$] & 34.2$^{+5.2}_{-4.4}$ [m\,s$^{-1}$] & 0.0080$^{+0.0019}_{-0.0014}$ & 0.029$^{+0.013}_{-0.017}$ \\ [2pt] %
$w$ & 0.47$^{+0.37}_{-0.32}$ & 0.55$^{+0.16}_{-0.11}$ & 0.40$^{+0.15}_{-0.11}$ & 0.41$^{+0.42}_{-0.28}$ \\ [2pt] %
$\lambda$ [d] & 523$^{+321}_{-392}$ & 22.5$^{+4.6}_{-5.2}$ & 67$^{+25}_{-20}$ & 206$^{+508}_{-138}$ \\ [2pt] %
$\theta$ [d] & 17.2$^{+10.4}_{-9.3}$ & 27.2$^{+2.1}_{-1.2}$ & 24.2$\pm$0.4 & 23.5$^{+3.9}_{-14.5}$ \\ [2pt] %
\hline \\ [-9pt] %
$\Delta \ln \mathcal{Z}^{b}$ & $-2.9$ & $+35.8$ & $+3.7$ & $-2.2$ \\ [2pt] %
\hline 
\end{tabular}  
}
\tablefoot{The hyper-parameters are those defined in Eq.\,(1). The Prior distributions are the following. $h$ (BIS and FWHM): $\mathcal{U}$(0,200) m\,s$^{-1}$; $h$  (H-$\alpha$): $\mathcal{U}$(0,0.5); $h$ ($\log R^{\prime}_{\rm HK}$): $\mathcal{U}$(0,0.5) dex; $w$ (all activity diagnostics): $\mathcal{U}$(0,1); $\lambda$ (all activity diagnostics): $\mathcal{U}$(0,1000) d; $\theta$ (all activity diagnostics): $\mathcal{U}$(0,35) d.
\tablefoottext{a}{The best-fit values are given as the median and 16$^{\rm th}$\,--\,84$^{\rm th}$ percentiles of the posterior distributions.}
\tablefoottext{b}{Difference between the Bayesian evidences of the models with and without GP regression, the latter being by a constant and representing the null hypothesis model.}}
\end{table}

We found that the FWHM is the activity indicator that shows a very significant quasi-periodic behaviour, with a well constrained stellar rotation period ($\theta$\,$=$\,$27.2^{+2.1}_{-1.2}$\,d), and a short evolutionary time-scale ($\lambda$\,$=$\,$22.5^{+4.6}_{-5.2}$\,d). The H-$\alpha$ index shows moderate quasi-periodic behaviour ($\theta$\,$=$\,$24.2\pm0.4$\,d; $\lambda$\,$=$\,$67^{+25}_{-20}$\,d). The BIS and $\log R^{\prime}_{\rm HK}$ do not show significant quasi-periodic variability, but their posteriors for $\theta$ are characterised by overdensity peaks related to the stellar rotation period or its harmonics. In particular, for $\log{R^{\prime}_{HK}}$ the posterior is bimodal (see Fig.~\ref{fig:posterior_prot}), with peaks in agreement with the stellar rotation period ($\sim 27$\,d) and its second harmonic ($\sim 9$\,d), which is very close to the orbital period of the putative planet c. These results show that the spectroscopic time series are affected by signals due to stellar activity and that the signal found in the HIRES RVs, which was attributed to a second planet, is actually due to activity. 

\begin{figure}
\centering
\includegraphics[width=1\linewidth]{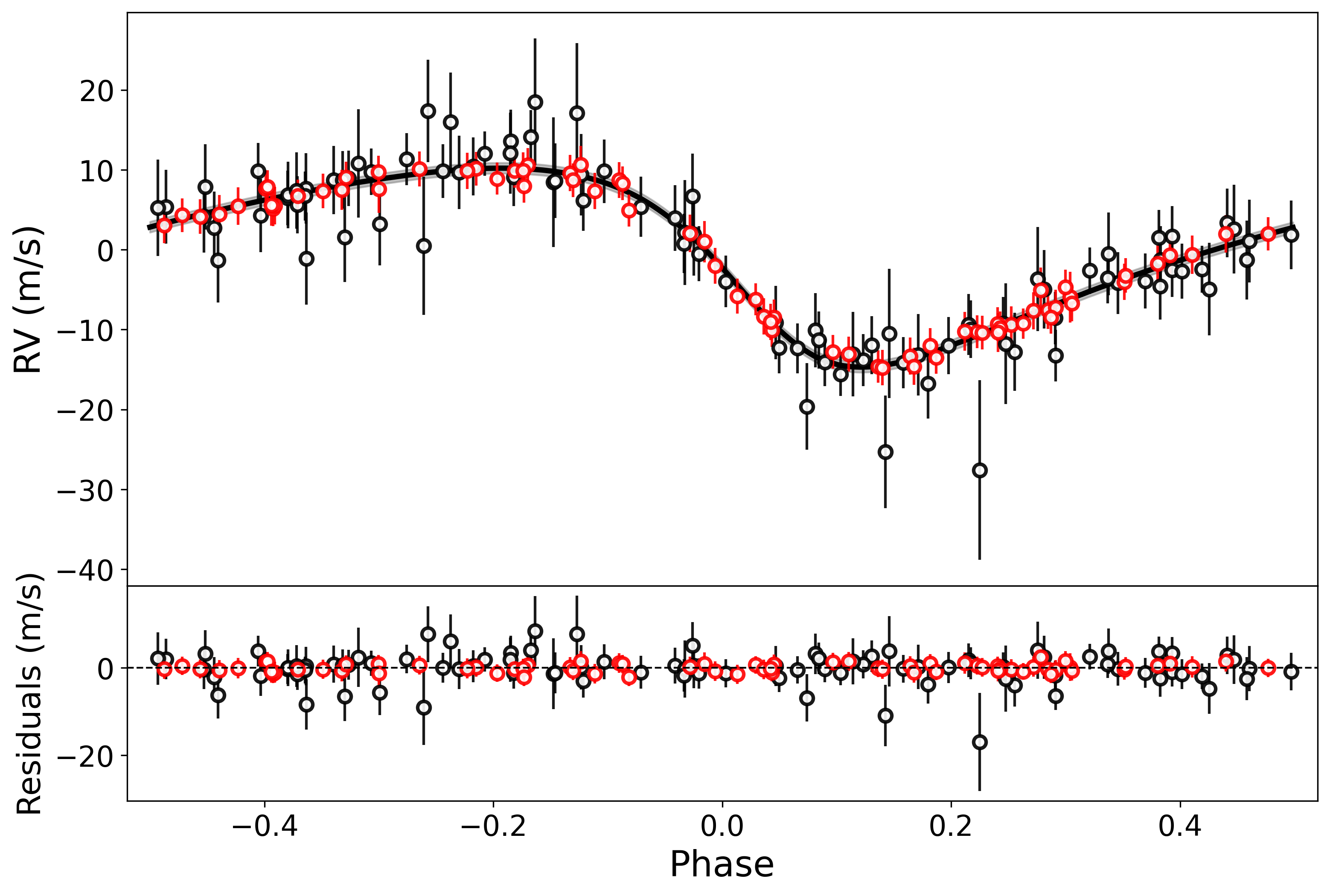}
\includegraphics[width=1\linewidth]{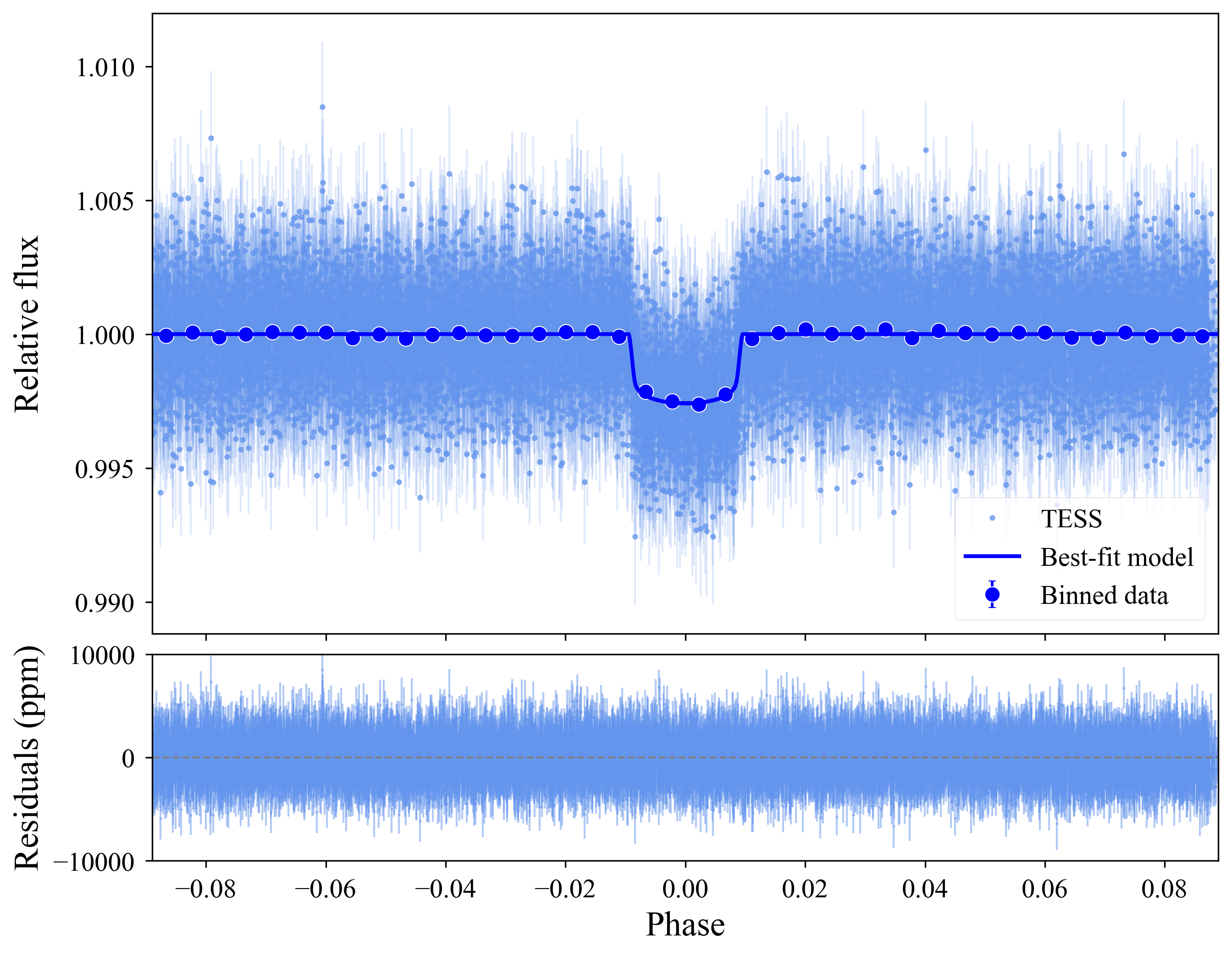}
\caption{{\it Top panel}: HARPS-N (HIRES) RV measurements of TOI-1272 in black (red) phase-folded to the period of planet b, together with our best-fit model. RV residuals from the best fit are also shown in parts per million. {\it Bottom panel}: Phase-folded unbinned TESS light curve for TOI-1272 (light-blue points). The solid blue line shows the model fit to the light curve. The residuals are also shown. The blue filled circles show the light curves binned in phase. In both the panels, the error bars include both the data uncertainty and the jitter derived from the analysis.} 
\label{fig:TOI-1272_phased}
\end{figure}

\begin{figure}
\centering
\includegraphics[width=1\linewidth]{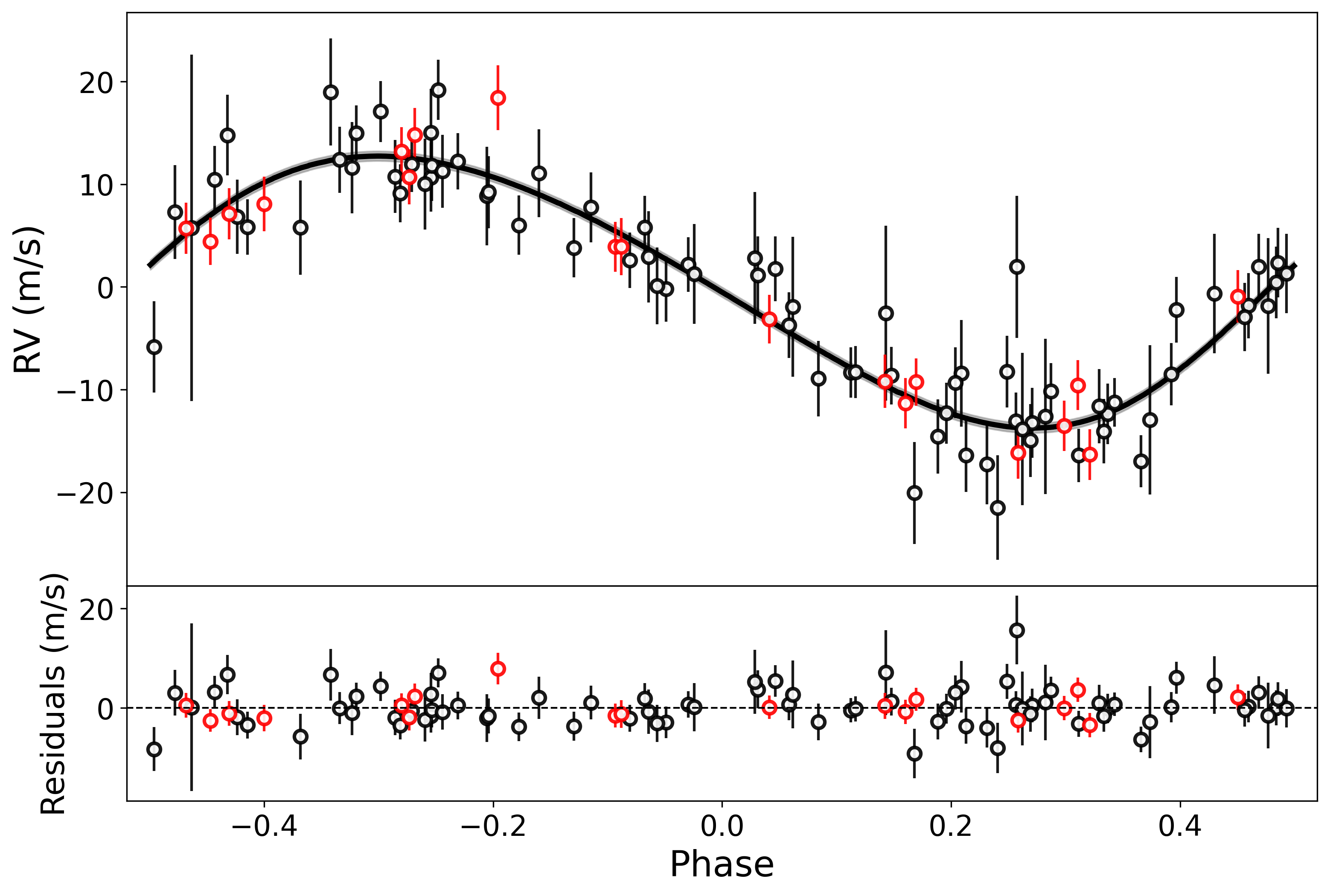}
\includegraphics[width=1\linewidth]{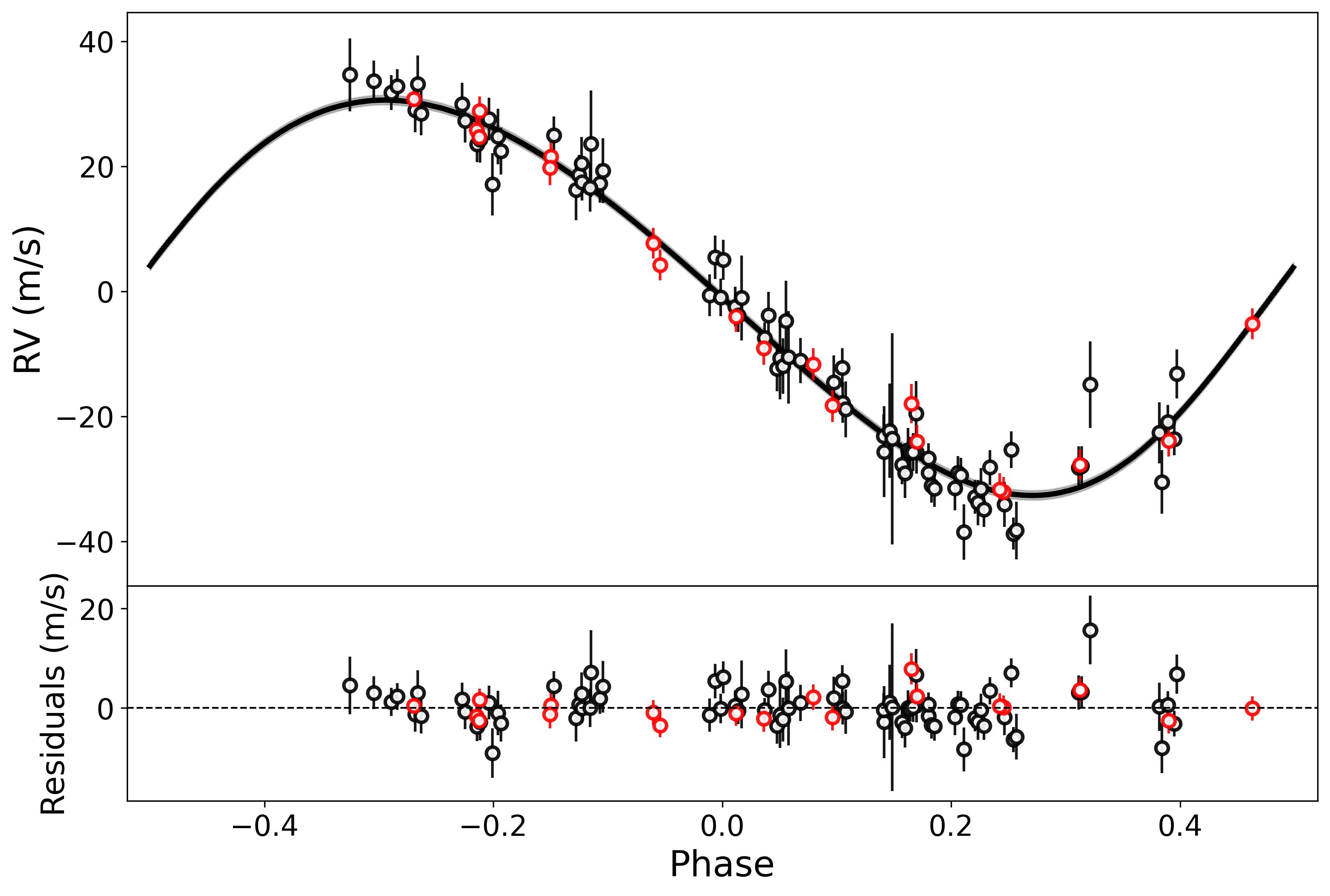}
\caption{Phase-folded RV measurements to the period of planet TOI-1694\,b ({\it top panel}) and TOI-1694\,c ({\it middle panel}), together with their residuals superimposed on the model. The black and red circles represent HARPS-N and HIRES data sets, respectively. 
The error bars include both the data uncertainty and the jitter derived from the analysis.}
\label{fig:TOI-1694_phased_rv}
\end{figure}

\subsubsection{Rejection of planet TOI-1272\,c}
\label{sec:rejection}
Motivated by the findings discussed above, we performed a new joint RV+transit photometry modelling as before, but this time including a GP quasi-periodic kernel applied to the RVs. We find that the model with the GP regression is statistically strongly favoured ($\Delta \ln \mathcal{Z}$=+13.7), and the posterior of the hyper-parameter $\theta$ shows a peak around $8-9$\,days, corresponding to a short time-scale signal with amplitude and periodicity compatible with that of the claimed planet c. Concerning the HARPS-N data, our conclusion is that the RV time series is affected by correlated noise due to activity, that would explain the result of \cite{macDougall2022} for their second signal in terms of activity instead of a second planet.

To verify this hypothesis, we analysed the Mount Wilson S-index and the RV time series calculated from the HIRES spectra (taken from \citealt{polanski2024}) within a GP framework as that we previously used in the case of HARPS-N RVs. We stress that this kind of analysis is lacking in \cite{macDougall2022}. For the S-index, we find that the GP model with a quasi-periodic kernel is largely favoured over the null hypothesis model represented by a constant offset ($\Delta \ln \mathcal{Z}$=+40), revealing the presence of a periodic modulation with a short timescale ($\theta=29.0^{+2.1}_{-1.5}$\,d; $\lambda=43\pm12$\,d) linked to the rotation period of the host star. For the RVs, we modelled the time series adopting three models: \textit{i)} only a Keplerian for the transiting planet; \textit{ii)} a Keplerian for the transiting planet and a GP quasi-periodic term; \textit{iii)} a Keplerian for the transiting planet and a sinusoid to model the signal of a possible second planetary. We found that model \textit{(ii)} is strongly favoured over the other two models ($\Delta \ln \mathcal{Z}>$10) and shows that the RVs are affected by a correlated signal with quasi-periodic properties ($\theta$\,$=$\,$26.1\pm0.2$\,d; $\lambda$\,$=$\,$64.6^{+25.9}_{-19.3}$\,d) and amplitude of $\sim$\,9\,m\,s$^{-1}$. We note that the signal with period $\sim$\,8.7\,d attributed to a second planet  \citep{macDougall2022} coincides with the second harmonic of $\theta$. The uncorrelated jitter is $\sim$\,2\,m\,s$^{-1}$, significantly lower than that found for the other two models ($\sim$\,8\,m\,s$^{-1}$).
The results for the HIRES-RV modelling (second scenario) are summarised in Table \ref{tab:rvgphires}. 

Our conclusion is that the second signal found by \cite{macDougall2022} in the HIRES RVs is real, but the interpretation is incorrect. This signal is better explained as due to quasi-periodic stellar activity, as was similarly found for the HARPS-N RVs, instead of invoking the existence of a second planet.

Simulations do show that a first- and second-harmonic signal, in addition to the star's rotation period, can be produced by starspots (see, e.g., \citealt{boisse2011}). Moreover, there are stars that exhibit first- and second-harmonic signals, in addition to the rotation period, in their observed RV series (see, e.g., \citealt{suarez2017,desidera2023,sozzetti2024}).

The amplitude of the TESS light curve of TOI-1272 for the sectors 15 and 16 is $\sim$\,2\%. It can be used to estimate the amplitude of the RV jitter induced by stellar activity by applying the results by \citet{Hojjatpanahetal20}. They show that an RV rms of $\ga$\,10\,m\,s$^{-1}$ is generally observed for such an amplitude of light modulation. The top-right panel of Fig.~4 of \citet{luhn2020} also indicates that a $S_{\rm HK}$ index equal to 0.331 \citep{macDougall2022} gives an RV rms of of about 7\,$-$\,10\,m\,s$^{-1}$. This conclusion is supported by the simulations by \citet{Desortetal07} for a K-type star with  $v\sin{i}$\,$=$\,1.1~km~s$^{-1}$ and a starspot filling factor of $f_{\rm r}$\,$\sim$\,1.2\% as derived from the TESS light curve and a starspot with an effective temperature of $\sim$\,3600~K, as they adopted in their Eq.\,(5). The simulations by \citet{Desortetal07} neglect the effect of the blueshift quenching on the RV variations, but it is remarkably smaller than in the solar case for a K2-type star such as TOI-1272 \citep{Meunieretal17}. Therefore, it is not expected to significantly modify our conclusion. The amplitude of the line bisector inverse slope variations, as obtained from Eq.\,(6) of \citet{Desortetal07}, is significantly smaller than in the third panel of Fig.~\ref{fig:glsharpsactindrv}; however, a peak-to-peak amplitude of $\sim$\,6.5\,m\,s$^{-1}$ is obtained when  $v\sin{i}$\,$=$\,2.1\,km\,s$^{-1}$ is adopted, which is still compatible with the uncertainty of our measurement of the rotational spectral line broadening.

\subsubsection{Apodised sine periodogram}
To further investigate the nature of the $P$\,$=$\,8.7\,d signal, we analysed the HARPS-N and HIRES datasets using the apodized-sine-periodogram formalism introduced by \citet{Gregory2016} and developed by \citet{Hara2022b}. This approach is designed to test whether a periodic signal is coherent over the entire time span of the observations, as expected for a Keplerian modulation, or is confined to a limited temporal window, as typically observed for activity-driven variability.
In practice, the model consists of a sinusoid multiplied by a temporal window function, $\mu(t) = w(\tau, t_0)\left[A \cos(\omega t) + B \sin(\omega t)\right]$,
%
%
where $w(\tau,t_0)$ is an apodisation function centred at $t_0$ with a characteristic timescale, $\tau$. By varying $\tau$ over a grid of values that span several times the value of $T_{\mathrm{obs}}$ down to shorter timescales, we can assess whether the signal is compatible with a constant-amplitude sinusoid ($\tau \gg T_{\mathrm{obs}}$) or rather favours a localised wavelet-like behaviour ($\tau \ll T_{\mathrm{obs}}$). For each pair $(\omega,\tau)$, the improvement with respect to the base model is quantified through the $\Delta\chi^2$ statistic, maximising over $t_0$.

The apodised-sine-periodogram analysis of the RV residuals, after subtracting the signal of planet TOI-1272\,b, shows that the 8.7\,d signal is prominent only in the first portion of the dataset (see Fig.\,\ref{fig:apodized}) and rapidly decreases in power when larger apodisation timescales are enforced. The best-fitting solutions correspond to low values of $\tau$, indicating that the signal is temporally localised rather than coherent over the full observing baseline. In contrast, solutions with $\tau \gtrsim T_{\mathrm{obs}}$, corresponding to a strictly periodic and phase-coherent signal as expected for a planetary companion, are not preferred by the data.
This behaviour further suggests that the 8.7-day modulation is not consistent with a stable Keplerian origin but is, instead, more likely associated with stellar activity evolving on relatively short timescales. If we remove the 8.7-day signal, which we associate with the second armonic of the rotation period of the parent star, a series of high-power peaks appear in the apodised periodogram around 13 days, i.e. the first armonic.

\subsubsection{Final parameters of the TOI-1272 system}
\label{sec:final_analysis_1272}
%
To derive final results for the TOI-1272 system, we performed a joint RV+transit photometry modelling with \texttt{juliet}, including a GP regression, using the RVs from both HIRES and HARPS-N and the complete TESS timeseries. This time, we employed the QP kernel as implemented in the \texttt{celerite} package \citep{Foreman2017}, i.e.
\begin{eqnarray}
k_{\rm QP}(t,t')=\frac{B}{2+C}\,e^{-\frac{t-t'}{L}} \left[\cos\left(\frac{2\pi(t-t')}{P_{\rm rot}}\right)+\left(1+C\right)\right] + \\
&& \hspace{-6.8cm} +\,\left(\sigma^{2}_{\rm RV}(t)+\sigma^{2}_{\rm jit}\right)\cdot\delta_{t, t^{\prime}}\;, \nonumber
\end{eqnarray}
where $B$, $C$ and $L$ are the amplitude, the constant scaling term, and the characteristic time-scale of the QP kernel, with $P_{\rm rot}$ being its characteristic period. The resulting parameters of the star and planet b are listed in Tables \ref{tab:star_TOI-1272} and \ref{tab:planet_TOI-1272}. Our estimations are all within the error bars of the literature determinations, but slightly more accurate.
HARPS-N and HIRES RVs are shown in the top panel of Fig.~\ref{fig:TOI-1272_phased} (phase-folded), together with the preferred global model (top panel) and its residual (bottom panel). 
The TESS light curve of TOI-1272, folded with the planet’s orbital period, is shown in the bottom panel of Fig.~\ref{fig:TOI-1272_phased} together with the best-fit transit model resulting from the global fit. 
%
%

\subsection{TOI-1694}
\label{sec:analysis_TOI-1694}
\begin{figure}
\centering
\includegraphics[width=1\linewidth]{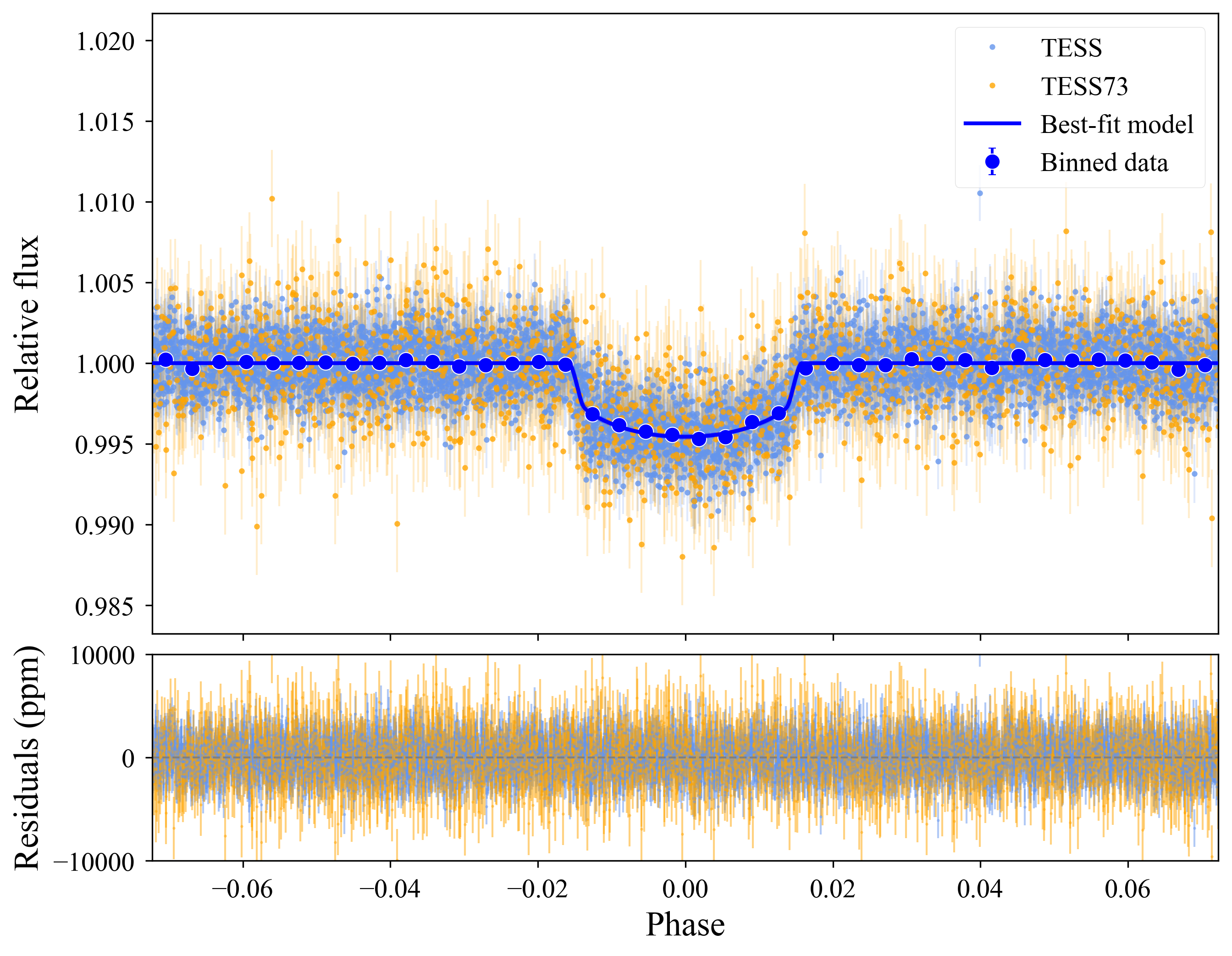}
\caption{Same as the bottom panel of Fig.~\ref{fig:TOI-1272_phased}, but for TOI-1694. The light-blue points refer to sectors 19 and 20, while green points refer to sector 73 (see Sect.~\ref{sec:photometry1694}).} 
\label{fig:TOI-1694_phased_lc}
\end{figure}

%
Unlike TOI-1272, both previously announced planetary signals for TOI-1694 \citep{VanZandt2023,polanski2024} are aptly retrieved within our HARPS-N dataset. Since no activity signal can be appreciated either in photometry or spectroscopy data, we performed a joint transit+RV fit of both HARPS-N and HIRES RVs to retrieve updated planet parameters, using the same approach described in Sect.\,\ref{sec:final_analysis_1272}, but without the use of any GP; see Figs. \ref{fig:TOI-1694_phased_rv}, \ref{fig:TOI-1694_phased_lc} and Table~\ref{tab:planet_TOI-1694}. We found that most of the parameters align well with those in the literature, although with a higher degree of accuracy (e.g. $M_{\rm b}=28.2\pm1.1\,M_{\oplus}$ vs $31.3^{+2.7}_{-2.5}\,M_{\oplus}$ or $P_{\rm orb,c}$\,=\,$386.2\pm2.1$\,d vs. $393.1\pm4.7$\,d from \citealt{polanski2024}). The correct value of the semi-major axis of TOI-1694\,c is reported here for the first time. Moreover, we find that the free eccentricity model for the hot Neptune TOI-1694\,b is statistically favoured over the circular one, resulting in a small but significant eccentricity $e_{\rm b}=0.122^{+0.025}_{-0.024}$. We note that the HIRES sampling for planet b was rather poor (Fig.\,\ref{fig:TOI-1694_phased_rv}), so this result largely stems from the larger and more complete HARPS-N dataset.

\subsection{Dynamical evolution}
The non-existence of the external planet, TOI-1272\,c, allows us to explain the presence of TOI-1272\,b in the Neptune desert as a result of planet-planet scattering followed by high-eccentricity migration. Specifically, in Sect.~6.3 of \citet{macDougall2022}, it is shown that the estimated age of TOI-1272 is compatible with its planet b having initially acquired an eccentricity of $\sim$\,0.8 followed by tidal dissipation inside the planet itself to reach the present value 0.3, assuming a constant planetary modified tidal quality factor $Q^{\prime}$\,$\sim$\,$10^{5}$, similar to those of Neptune and Uranus \citep{Ogilvie14}.

\begin{figure}
\centering
\includegraphics[width=1\linewidth]{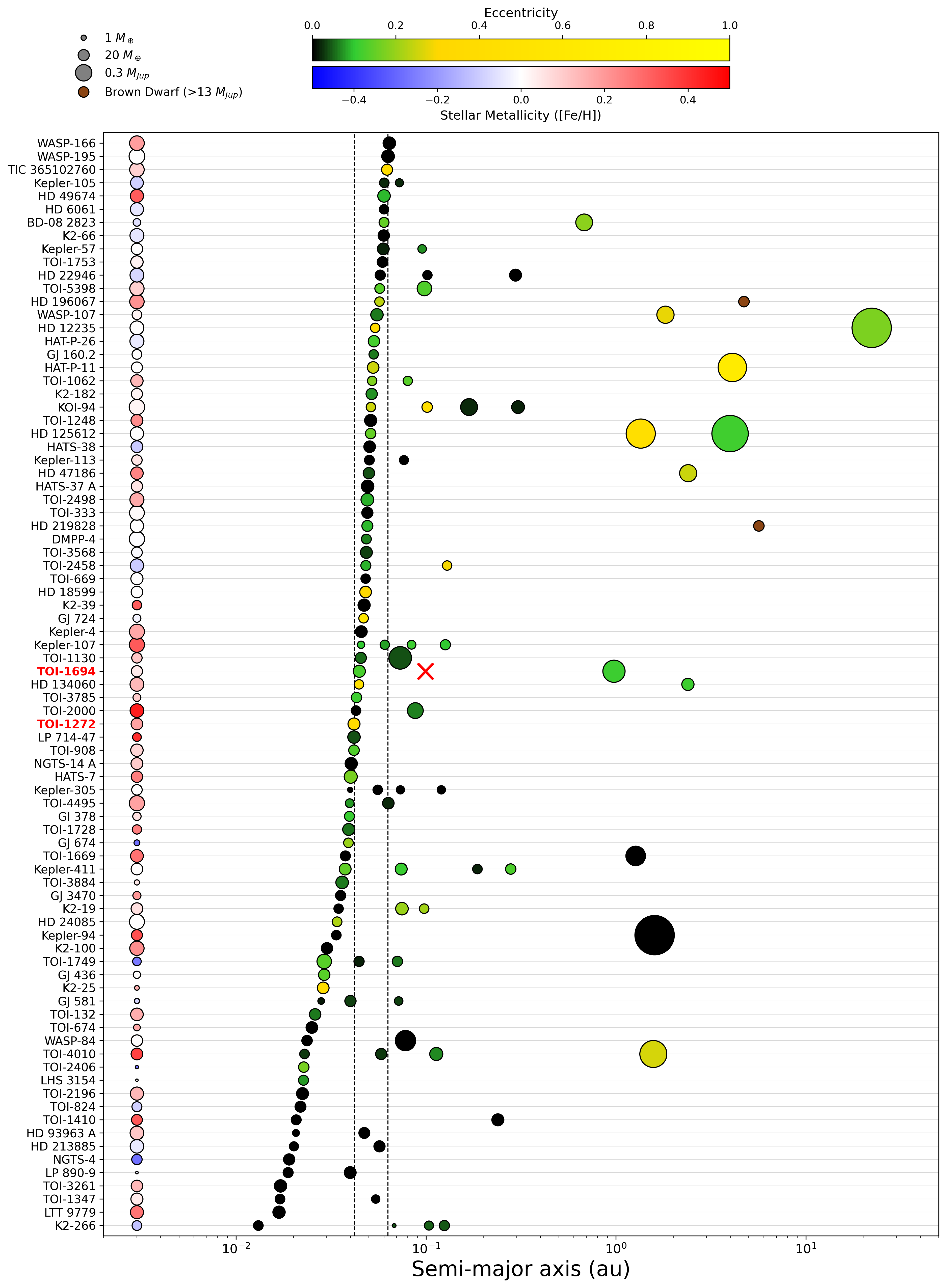}
\caption{Comparison of exoplanetary systems' architectures that have at least one Neptunian planet (defined as having a mass between 10 and 40 $M_{\oplus}$) with $P_{\rm orb}$\,$\lesssim$\,5.7\,d, therefore inside the ridge (delimited by the two dashed lines) and the desert. TOI-1272 and TOI-1694 are highlighted. Host stars' metallicity and size are illustrated by color and circle size, respectively. Planets' orbital eccentricity and mass are illustrated by color and circle size, respectively. Note that the TOI-1272 system hosts only one planet now. The red cross marks the wrong position of TOI-1694\,c reported in the exoplanet archives up to now (June 2026).}
\label{fig:architecture}
\end{figure}

The architecture of TOI-1694, instead, resembles that of WASP-107 \citep{piaulet2021P}, as illustrated in Fig~\ref{fig:architecture}. Both planets likely formed further out, beyond the snow line, and migrated inward, while the protoplanetary disc was still present, through a gas-driven disc migration. TOI-1694\,c likely underwent type II migration; as a massive body, it cleared a gap in the gas disc and moved inward as the disc material evolved viscously. Its current position suggests that it stopped migrating before the gas fully dissipated. Instead, being less massive, TOI-1694\,b likely migrated via Type I migration, which is much faster \citep[see, e.g.,][]{ward1997,kley2000,masset2001}, and would have been pushed toward the inner edge of the disc, near the magnetospheric cavity of the star \citep{kuchner2002}.
Planet-planet scattering could have affected planet TOI-1694\,c, if another giant planet was initially present in the system. In that case, the orbit of planet c could have been remarkably inclined to the initial plane of the disc, while the other massive planet could have been expelled from the system \citep[e.g.][]{DawsonJohnson18}.

However, the present non-zero eccentricities of planets TOI-1694\,b and \,c suggest that the system did not settle into a quiet circular configuration. The outer Jupiter-mass planet could have gravitationally perturbed the inner Neptune. If we again assume that $Q^{\prime}$\,$\sim$\,$10^{5}$, the tidal decay timescale for the present eccentricity of TOI-1694\,b is $\sim$\,325\,Myr \citep{Jackson08}, remarkably shorter than the estimated age of the system. This indicates that the eccentricity of planet b must be continuously excited. If the orbits of planets b and c were initially inclined, this could have triggered cycles of Kozai-Lidov (KL) oscillations that exchanged inclination for high eccentricity in the inner planet \citep{wu2003,fabrycky2007}. Given the relatively low eccentricity of the orbit of the outer planet c, we can apply Eq.~(27) of \citet{Naoz16} to estimate the timescale of the KL  oscillations and find a maximum period of $\sim$\,1.2\,$\times$\,$10^{4}$\,yr that corresponds to the minimum relative inclination of $\sim 40^{\circ}$ between the orbits of planet b and planet c to produce such oscillations. The precession of the line of the apsides of the inner planet is dominated by general relativity effects and its period is of $\sim$\,1.8\,$\times$\,$10^{4}$\,yr \citep{MardlingLin02}, comparable with but still longer than the period of Kozai oscillations, making it possible for the KL mechanism to account for the observed eccentricity of the inner planet. In other words, the system of TOI-1694 could be a rare planetary system caught in the final phase of the dynamical evolution produced by the KL mechanism.
In the past, when the semi-major axis of TOI-1694\,b was larger, the KL mechanism was capable of driving the eccentricity of the planet to very high values, so its periastron would come  extremely close to the star. Tidal friction would then drain orbital energy, shrinking the semi-major axis to the current value and gradually circularising the orbit. It remained slightly eccentric because the KL mechanism continued to operate with an efficiency that was gradually reduced as the period of the KL oscillations approached the period of the apsidal precession of the orbit of the planet.
In the future, once the KL mechanism and tidal dissipation have brought planet b still closer to the host star, the period of general relativity precession will become shorter than that of the KL oscillations, thereby preventing the KL mechanism from operating further \citep{Naoz16} and leaving planet b on its final orbit.

The possibility that the eccentricity of planet b is excited by the perturbations induced by planet c on a coplanar or low inclination ($\la 40^{\circ}$) orbit can be investigated by applying the model by \citet{mardling2007}. It gives an equilibrium eccentricity smaller than $10^{-3}$, that is too small to account for the observed eccentricity of planet b. Therefore, the interpretation based on the KL mechanism seems more likely.


Alternatively, if the TOI-1694 system originally contained a third planet that was subsequently ejected, the resulting planet-planet scattering followed by tidal dissipation could have affected also the less massive Neptune and brought it into the inner system, leaving the Jupiter-mass planet on its current slightly eccentric orbit \citep{rasio1996,chatterjee2008,DawsonJohnson18}. In this scenario, the present eccentricity of TOI-16941\,b would be a relic of that formation mechanism, implying that its modified tidal quality factor is $Q^{\prime}$\,$ \ga$\,$10^{6}$ to avoid a significant damping of the primordial eccentricity along the lifetime of the system.

\section{Conclusions}
\label{sec:conclusions}
Within the GAPS-Neptune and HONEI programmes, we used high-resolution spectrographs to observe a sample of TESS transiting-planet candidates, mostly hosting hot and warm Neptune-sized planets. This class of exoplanets exhibits significant diversity in physical properties, even extreme ones, indicating heterogeneous internal structures and atmospheric compositions. Our aim is to confirm the planetary nature of many of these candidates and accurately characterise their main orbital and physical properties.
In this way, we can enlarge the sample of well-studied Neptune-sized exoplanets. Thus, it becomes possible to deduce more robust information about their plausible formation and migration mechanisms.
In this work, we present new HARPS-N\,RV measurements of two already known planetary systems, TOI-1272 \citep{macDougall2022} and TOI-1694 \citep{VanZandt2023}, along with a comprehensive analysis of their properties. We started collecting spectra of these two targets with HARPS-N when they were still considered candidates, but other teams preceded us and published their results concerning these two planetary systems before we finished our analyses. However, neither of these two systems has been fully characterised by previous studies.
Several parameters were previously derived or have been incorrectly reported such as the semi-major axis of TOI-1694\,c. Moreover, a stellar activity signal from TOI-1272 was mistaken for a non-transiting planet. 

In our study, we joined our new RV measurements and those already published. It is worth emphasising that the new HARPS-N data cover the same time span as the data collected with HIRES for both TOI-1272 and TOI-1694. Finally, in our investigation, we considered new TESS photometric data for these two objects that were not available at the time of their discovery papers. Our main results are as follows.

We fully characterised the two stars, finding that both are K2\,V stars (Sect.~\ref{sec:host_stars}). Our estimations of their physical parameters are all within the error bars of previous determinations but are slightly more accurate. Their ages are not very-well constrained but indicate that the two stars are older than 2\,Gyr.
This conclusion is also supported by the low $v \sin{i_{\star}}$ ($<2$\,km\,s\,$^{-1}$) that we measured for both stars.

In the case of TOI-1272, our analysis of the HARPS-N data did not reveal any planetary signal other than that related to TOI-1272\,b, but revealed that the GLS periodogram of the BIS peaks exactly at the period of the planet c claimed by \citet{macDougall2022} and \citet{polanski2024}. We also reanalysed the HIRES data and found that the $\sim$\,8.7\,d signal coincides with the second harmonic of the stellar rotation period (see Sect.~\ref{sec:analysis_TOI-1272}). Our conclusion is that this signal should be considered due to stellar activity, the planetary interpretation should be discarded and the planet TOI-1272\,c should be rejected.

Although our estimated physical and orbital parameters for TOI-1272\,b, TOI-1694\,b and c are mostly consistent with previous determinations, they offer an enhanced level of precision. This is essentially due to the fact that our analysis is based on two and five times more RV measurements for TOI-1272 and TOI-1694 systems, respectively, as well as on photometric data from two and one more TESS sectors for TOI-1272 and TOI-1694 systems, respectively.

The increased number of RV measurements have allowed us to appreciate that the orbit of TOI-1694\,b is slightly eccentric ($e$\,=\,$0.122_{-0.024}^{+0.025}$). We also improved the measurements of the orbital period and eccentricity of TOI-1694\,c ($P_{\rm orb}$\,=\,$386\pm2$\,d; $e$\,=\,$0.111 \pm 0.025$). 

\begin{acknowledgements}
%
%
This work is based on observations made with the Italian Telescopio Nazionale Galileo (TNG) operated by the Fundaci\'{o}n Galileo Galilei (FGG) of the Istituto Nazionale di Astrofisica (INAF) at the Observatorio del Roque de los Muchachos (La Palma, Canary Islands, Spain). We acknowledge the INAF Italian centre for Astronomical Archives (IA2) for providing technical assistance and supporting activities of the GAPS collaboration. This work also includes data collected with the TESS mission, obtained from the Mikulski Archive for Space Telescopes data archive at the Space Telescope Science Institute (STScI). Funding for the TESS mission is provided by the NASA Explorer Program. STScI is operated by the Association of Universities for Research in Astronomy, Inc., under the NASA contract NAS 5-26555. 
The authors acknowledge the use of public TESS data from pipelines at the TESS Science Office and at the TESS Science Processing Operations Center. Resources supporting this work were provided by the NASA High-End Computing Program through the NASA Advanced Supercomputing Division at Ames Research Center for the production of the SPOC data products. 
This research has used the Exoplanet Follow-up Observation Program (ExoFOP; DOI: 10.26134/ExoFOP5) website, which is operated by the Caltech, under contract with the National Aeronautics and Space Administration under the Exoplanet Exploration Program.
This research used the NASA Exoplanet Archive, operated by the California Institute of Technology, under contract with the National Aeronautics and Space Administration under the Exoplanet Exploration Program. J.L.-B. is funded by the Spanish Ministry of Science and Universities (MICIU/AEI/10.13039/501100011033) and NextGenerationEU/PRTR grants PID2019-107061GB-C61, CNS2023-144309, and PID2023-150468NB-I00.
We acknowledge support from the European Union – NextGenerationEU (PRIN MUR 2022 project 20229R43BH) and the ``Programma di Ricerca Fondamentale INAF 2023''. We acknowledge financial contribution from the INAF Large Grant 2023 ``EXODEMO''.
The authors thank L.\,B.\,Handley for sharing their TESS detrended light curve of TOI-1694.
The authors thank the anonymous referee for a very useful report.
L.M. thanks the Turin Astrophysical Observatory for the kind hospitality and acknowledges support for this work by research funds from PRIN\,MUR\,2022 project 2022J4H55R.
\end{acknowledgements}

\bibliographystyle{aa} 
\bibliography{bib}

@ARTICLE{Naoz16,
       author = {{Naoz}, Smadar},
        title = "{The Eccentric Kozai-Lidov Effect and Its Applications}",
      journal = {\araa},
         year = 2016,
        month = sep,
       volume = {54},
        pages = {441-489},
          doi = {10.1146/annurev-astro-081915-023315},
archivePrefix = {arXiv},
       eprint = {1601.07175},
 primaryClass = {astro-ph.EP},
       adsurl = {https://ui.adsabs.harvard.edu/abs/2016ARA&A..54..441N}
}

@ARTICLE{MardlingLin02,
       author = {{Mardling}, Rosemary A. and {Lin}, D.~N.~C.},
        title = "{Calculating the Tidal, Spin, and Dynamical Evolution of Extrasolar Planetary Systems}",
      journal = {\apj},
         year = 2002,
        month = jul,
       volume = {573},
       number = {2},
        pages = {829-844},
          doi = {10.1086/340752},
       adsurl = {https://ui.adsabs.harvard.edu/abs/2002ApJ...573..829M}
}

@ARTICLE{Jackson08,
       author = {{Jackson}, Brian and {Greenberg}, Richard and {Barnes}, Rory},
        title = "{Tidal Evolution of Close-in Extrasolar Planets}",
      journal = {\apj},
         year = 2008,
        month = may,
       volume = {678},
       number = {2},
        pages = {1396-1406},
          doi = {10.1086/529187},
archivePrefix = {arXiv},
       eprint = {0802.1543},
 primaryClass = {astro-ph},
       adsurl = {https://ui.adsabs.harvard.edu/abs/2008ApJ...678.1396J}
}

@ARTICLE{Ogilvie14,
       author = {{Ogilvie}, Gordon I.},
        title = "{Tidal Dissipation in Stars and Giant Planets}",
      journal = {\araa},
         year = 2014,
        month = aug,
       volume = {52},
        pages = {171-210},
          doi = {10.1146/annurev-astro-081913-035941},
archivePrefix = {arXiv},
       eprint = {1406.2207},
 primaryClass = {astro-ph.SR},
       adsurl = {https://ui.adsabs.harvard.edu/abs/2014ARA&A..52..171O}
}

@ARTICLE{DawsonJohnson18,
       author = {{Dawson}, Rebekah I. and {Johnson}, John Asher},
        title = "{Origins of Hot Jupiters}",
      journal = {\araa},
         year = 2018,
        month = sep,
       volume = {56},
        pages = {175-221},
          doi = {10.1146/annurev-astro-081817-051853},
archivePrefix = {arXiv},
       eprint = {1801.06117},
 primaryClass = {astro-ph.EP},
       adsurl = {https://ui.adsabs.harvard.edu/abs/2018ARA&A..56..175D}
}

@ARTICLE{Paxton2015,
       author = {{Paxton}, Bill and {Marchant}, Pablo and {Schwab}, Josiah and {Bauer}, Evan B. and {Bildsten}, Lars and {Cantiello}, Matteo and {Dessart}, Luc and {Farmer}, R. and {Hu}, H. and {Langer}, N. and {Townsend}, R.~H.~D. and {Townsley}, Dean M. and {Timmes}, F.~X.},
        title = "{Modules for Experiments in Stellar Astrophysics (MESA): Binaries, Pulsations, and Explosions}",
      journal = {\apjs},
         year = 2015,
        month = sep,
       volume = {220},
       number = {1},
          eid = {15},
        pages = {15},
          doi = {10.1088/0067-0049/220/1/15},
archivePrefix = {arXiv},
       eprint = {1506.03146},
 primaryClass = {astro-ph.SR},
       adsurl = {https://ui.adsabs.harvard.edu/abs/2015ApJS..220...15P}
}

@MISC{2017ascl.soft10003E,
       author = {{Eastman}, Jason},
        title = "{EXOFASTv2: Generalized publication-quality exoplanet modeling code}",
 howpublished = {Astrophysics Source Code Library, record ascl:1710.003},
         year = 2017,
        month = oct,
          eid = {ascl:1710.003},
        pages = {ascl:1710.003},
archivePrefix = {ascl},
       eprint = {1710.003},
       adsurl = {https://ui.adsabs.harvard.edu/abs/2017ascl.soft10003E}
}

@ARTICLE{Eastman2019,
       author = {{Eastman}, Jason D. and {Rodriguez}, Joseph E. and {Agol}, Eric and {Stassun}, Keivan G. and {Beatty}, Thomas G. and {Vanderburg}, Andrew and {Gaudi}, B. Scott and {Collins}, Karen A. and {Luger}, Rodrigo},
        title = "{EXOFASTv2: A public, generalized, publication-quality exoplanet modeling code}",
      journal = {},
         year = 2019,
        month = jul,
          eid = {arXiv:1907.09480},
        pages = {arXiv:1907.09480},
          doi = {10.48550/arXiv.1907.09480},
archivePrefix = {arXiv},
       eprint = {1907.09480},
 primaryClass = {astro-ph.EP},
       adsurl = {https://ui.adsabs.harvard.edu/abs/2019arXiv190709480E}
}

@ARTICLE{Carvalho-Silvaetal25,
       author = {{Carvalho-Silva}, Gabriela and {Mel{\'e}ndez}, Jorge and {Rathsam}, Anne and {Shejeelammal}, J. and {Martos}, Giulia and {Lorenzo-Oliveira}, Diego and {Spina}, Lorenzo and {Ribeiro Alves}, D{\'e}bora},
        title = "{A New Age─Activity Relation For Solar Analogs that Accounts for Metallicity}",
      journal = {\apjl},
         year = 2025,
        month = apr,
       volume = {983},
       number = {2},
          eid = {L31},
        pages = {L31},
          doi = {10.3847/2041-8213/adc382},
archivePrefix = {arXiv},
       eprint = {2504.17482},
 primaryClass = {astro-ph.SR},
       adsurl = {https://ui.adsabs.harvard.edu/abs/2025ApJ...983L..31C}
}

@ARTICLE{Lorenzo-Oliveiraetal18,
       author = {{Lorenzo-Oliveira}, Diego and {Freitas}, Fabr{\'\i}cio C. and {Mel{\'e}ndez}, Jorge and {Bedell}, Megan and {Ram{\'\i}rez}, Iv{\'a}n and {Bean}, Jacob L. and {Asplund}, Martin and {Spina}, Lorenzo and {Dreizler}, Stefan and {Alves-Brito}, Alan and {Casagrande}, Luca},
        title = "{The Solar Twin Planet Search. The age-chromospheric activity relation}",
      journal = {\aap},
         year = 2018,
        month = nov,
       volume = {619},
          eid = {A73},
        pages = {A73},
          doi = {10.1051/0004-6361/201629294},
archivePrefix = {arXiv},
       eprint = {1806.08014},
 primaryClass = {astro-ph.SR},
       adsurl = {https://ui.adsabs.harvard.edu/abs/2018A&A...619A..73L}
}

@ARTICLE{Gruneretal23,
       author = {{Gruner}, D. and {Barnes}, S.~A. and {Janes}, K.~A.},
        title = "{Wide binaries demonstrate the consistency of rotational evolution between open cluster and field stars}",
      journal = {\aap},
         year = 2023,
        month = jul,
       volume = {675},
          eid = {A180},
        pages = {A180},
          doi = {10.1051/0004-6361/202346590},
archivePrefix = {arXiv},
       eprint = {2307.10836},
 primaryClass = {astro-ph.SR},
       adsurl = {https://ui.adsabs.harvard.edu/abs/2023A&A...675A.180G}
}

@ARTICLE{Desortetal07,
       author = {{Desort}, M. and {Lagrange}, A.-M. and {Galland}, F. and {Udry}, S. and {Mayor}, M.},
        title = "{Search for exoplanets with the radial-velocity technique: quantitative diagnostics of stellar activity}",
      journal = {\aap},
         year = 2007,
        month = oct,
       volume = {473},
       number = {3},
        pages = {983-993},
          doi = {10.1051/0004-6361:20078144},
archivePrefix = {arXiv},
       eprint = {0708.4338},
 primaryClass = {astro-ph},
       adsurl = {https://ui.adsabs.harvard.edu/abs/2007A&A...473..983D}
}

@ARTICLE{Hojjatpanahetal20,
       author = {{Hojjatpanah}, S. and {Oshagh}, M. and {Figueira}, P. and {Santos}, N.~C. and {Amazo-G{\'o}mez}, E.~M. and {Sousa}, S.~G. and {Adibekyan}, V. and {Akinsanmi}, B. and {Demangeon}, O. and {Faria}, J. and {Gomes da Silva}, J. and {Meunier}, N.},
        title = "{The correlation between photometric variability and radial velocity jitter. Based on TESS and HARPS observations}",
      journal = {\aap},
         year = 2020,
        month = jul,
       volume = {639},
          eid = {A35},
        pages = {A35},
          doi = {10.1051/0004-6361/202038035},
archivePrefix = {arXiv},
       eprint = {2005.10105},
 primaryClass = {astro-ph.EP},
       adsurl = {https://ui.adsabs.harvard.edu/abs/2020A&A...639A..35H}
}

@ARTICLE{Meunieretal17,
       author = {{Meunier}, N. and {Lagrange}, A.-M. and {Mbemba K.}, L. and {Alex}, M. and {Mignon}, L. and {Borgniet}, S.},
        title = "{Variability of stellar granulation and convective blueshift with spectral type and magnetic activity. I. K and G main sequence stars}",
      journal = {\aap},
         year = 2017,
        month = jan,
       volume = {597},
          eid = {A52},
        pages = {A52},
          doi = {10.1051/0004-6361/201629052},
archivePrefix = {arXiv},
       eprint = {1610.02168},
 primaryClass = {astro-ph.SR},
       adsurl = {https://ui.adsabs.harvard.edu/abs/2017A&A...597A..52M}
}

@ARTICLE{macDougall2022,
       author = {{MacDougall}, Mason G. and {Petigura}, Erik A. and {Fetherolf}, Tara and {Beard}, Corey and {Lubin}, Jack and {Angelo}, Isabel and {Batalha}, Natalie M. and {Behmard}, Aida and {Blunt}, Sarah and {Brinkman}, Casey and {Chontos}, Ashley and {Crossfield}, Ian J.~M. and {Dai}, Fei and {Dalba}, Paul A. and {Dressing}, Courtney and {Fulton}, Benjamin and {Giacalone}, Steven and {Hill}, Michelle L. and {Howard}, Andrew W. and {Huber}, Daniel and {Isaacson}, Howard and {Kane}, Stephen R. and {Kosiarek}, Molly and {Mayo}, Andrew and {Mo{\v{c}}nik}, Teo and {Akana Murphy}, Joseph M. and {Pidhorodetska}, Daria and {Polanski}, Alex and {Rice}, Malena and {Robertson}, Paul and {Rosenthal}, Lee J. and {Roy}, Arpita and {Rubenzahl}, Ryan A. and {Scarsdale}, Nicholas and {Turtelboom}, Emma V. and {Tyler}, Dakotah and {Van Zandt}, Judah and {Weiss}, Lauren M. and {Esparza-Borges}, Emma and {Fukui}, Akihiko and {Isogai}, Keisuke and {Kawauchi}, Kiyoe and {Mori}, Mayuko and {Murgas}, Felipe and {Narita}, Norio and {Nishiumi}, Taku and {Palle}, Enric and {Parviainen}, Hannu and {Watanabe}, Noriharu and {Jenkins}, Jon M. and {Latham}, David W. and {Ricker}, George R. and {Seager}, S. and {Vanderspek}, Roland K. and {Winn}, Joshua N. and {Bieryla}, Allyson and {Caldwell}, Douglas A. and {Dragomir}, Diana and {Fausnaugh}, M.~M. and {Mireles}, Ismael and {Rodriguez}, David R.},
        title = "{The TESS-Keck Survey. XIII. An Eccentric Hot Neptune with a Similar-mass Outer Companion around TOI-1272}",
      journal = {\aj},
         year = 2022,
        month = sep,
       volume = {164},
       number = {3},
          eid = {97},
        pages = {97},
          doi = {10.3847/1538-3881/ac7ce1},
archivePrefix = {arXiv},
       eprint = {2206.14327},
 primaryClass = {astro-ph.EP},
       adsurl = {https://ui.adsabs.harvard.edu/abs/2022AJ....164...97M}
}

@ARTICLE{VanZandt2023,
       author = {{Van Zandt}, Judah and {Petigura}, Erik A. and {MacDougall}, Mason and {Gilbert}, Gregory J. and {Lubin}, Jack and {Barclay}, Thomas and {Batalha}, Natalie M. and {Crossfield}, Ian J.~M. and {Dressing}, Courtney and {Fulton}, Benjamin and {Howard}, Andrew W. and {Huber}, Daniel and {Isaacson}, Howard and {Kane}, Stephen R. and {Robertson}, Paul and {Roy}, Arpita and {Weiss}, Lauren M. and {Behmard}, Aida and {Beard}, Corey and {Chontos}, Ashley and {Dai}, Fei and {Dalba}, Paul A. and {Fetherolf}, Tara and {Giacalone}, Steven and {Henze}, Christopher E. and {Hill}, Michelle L. and {Hirsch}, Lea A. and {Holcomb}, Rae and {Howell}, Steve B. and {Jenkins}, Jon M. and {Latham}, David W. and {Mayo}, Andrew and {Mireles}, Ismael and {Mo{\v{c}}nik}, Teo and {Murphy}, Joseph M. Akana and {Pidhorodetska}, Daria and {Polanski}, Alex S. and {Ricker}, George R. and {Rosenthal}, Lee J. and {Rubenzahl}, Ryan A. and {Seager}, S. and {Scarsdale}, Nicholas and {Turtelboom}, Emma V. and {Vanderspek}, Roland and {Winn}, Joshua N.},
        title = "{TESS-Keck Survey. XIV. Two Giant Exoplanets from the Distant Giants Survey}",
      journal = {\aj},
         year = 2023,
        month = feb,
       volume = {165},
       number = {2},
          eid = {60},
        pages = {60},
          doi = {10.3847/1538-3881/aca6ef},
archivePrefix = {arXiv},
       eprint = {2209.06958},
 primaryClass = {astro-ph.EP},
       adsurl = {https://ui.adsabs.harvard.edu/abs/2023AJ....165...60V}
}

@ARTICLE{Kipping2013,
       author = {{Kipping}, David M.},
        title = "{Efficient, uninformative sampling of limb darkening coefficients for two-parameter laws}",
      journal = {\mnras},
         year = 2013,
        month = nov,
       volume = {435},
       number = {3},
        pages = {2152-2160},
          doi = {10.1093/mnras/stt1435},
archivePrefix = {arXiv},
       eprint = {1308.0009},
 primaryClass = {astro-ph.SR},
       adsurl = {https://ui.adsabs.harvard.edu/abs/2013MNRAS.435.2152K}
}

@ARTICLE{Kempton2018,
       author = {{Kempton}, Eliza M. -R. and {Bean}, Jacob L. and {Louie}, Dana R. and {Deming}, Drake and {Koll}, Daniel D.~B. and {Mansfield}, Megan and {Christiansen}, Jessie L. and {L{\'o}pez-Morales}, Mercedes and {Swain}, Mark R. and {Zellem}, Robert T. and {Ballard}, Sarah and {Barclay}, Thomas and {Barstow}, Joanna K. and {Batalha}, Natasha E. and {Beatty}, Thomas G. and {Berta-Thompson}, Zach and {Birkby}, Jayne and {Buchhave}, Lars A. and {Charbonneau}, David and {Cowan}, Nicolas B. and {Crossfield}, Ian and {de Val-Borro}, Miguel and {Doyon}, Ren{\'e} and {Dragomir}, Diana and {Gaidos}, Eric and {Heng}, Kevin and {Hu}, Renyu and {Kane}, Stephen R. and {Kreidberg}, Laura and {Mallonn}, Matthias and {Morley}, Caroline V. and {Narita}, Norio and {Nascimbeni}, Valerio and {Pall{\'e}}, Enric and {Quintana}, Elisa V. and {Rauscher}, Emily and {Seager}, Sara and {Shkolnik}, Evgenya L. and {Sing}, David K. and {Sozzetti}, Alessandro and {Stassun}, Keivan G. and {Valenti}, Jeff A. and {von Essen}, Carolina},
        title = "{A Framework for Prioritizing the TESS Planetary Candidates Most Amenable to Atmospheric Characterization}",
      journal = {\pasp},
         year = 2018,
        month = nov,
       volume = {130},
       number = {993},
        pages = {114401},
          doi = {10.1088/1538-3873/aadf6f},
archivePrefix = {arXiv},
       eprint = {1805.03671},
 primaryClass = {astro-ph.EP},
       adsurl = {https://ui.adsabs.harvard.edu/abs/2018PASP..130k4401K}
}

@ARTICLE{zechmeister2009,
       author = {{Zechmeister}, M. and {K{\"u}rster}, M.},
        title = "{The generalised Lomb-Scargle periodogram. A new formalism for the floating-mean and Keplerian periodograms}",
      journal = {\aap},
         year = 2009,
        month = mar,
       volume = {496},
       number = {2},
        pages = {577-584},
          doi = {10.1051/0004-6361:200811296},
archivePrefix = {arXiv},
       eprint = {0901.2573},
 primaryClass = {astro-ph.IM},
       adsurl = {https://ui.adsabs.harvard.edu/abs/2009A&A...496..577Z}
}

@ARTICLE{handley2025,
       author = {{Handley}, Luke B. and {Howard}, Andrew W. and {Rubenzahl}, Ryan A. and {Dai}, Fei and {Tyler}, Dakotah and {Lee}, Rena A. and {Giacalone}, Steven and {Isaacson}, Howard and {Fulton}, Benjamin and {Householder}, Aaron and {Halverson}, Samuel and {Roy}, Arpita and {Walawender}, Josh},
        title = "{An Obliquity Measurement of the Hot Neptune TOI-1694b}",
      journal = {\aj},
         year = 2025,
        month = apr,
       volume = {169},
       number = {4},
          eid = {212},
        pages = {212},
          doi = {10.3847/1538-3881/adb71b},
archivePrefix = {arXiv},
       eprint = {2412.07950},
 primaryClass = {astro-ph.EP},
       adsurl = {https://ui.adsabs.harvard.edu/abs/2025AJ....169..212H}
}

@ARTICLE{mistry2023,
       author = {{Mistry}, Priyashkumar and {Pathak}, Kamlesh and {Prasad}, Aniket and {Lekkas}, Georgios and {Bhattarai}, Surendra and {Gharat}, Sarvesh and {Maity}, Mousam and {Kumar}, Dhruv and {Collins}, Karen A. and {Schwarz}, Richard P. and {Mann}, Christopher R. and {Furlan}, Elise and {Howell}, Steve B. and {Ciardi}, David and {Bieryla}, Allyson and {Matthews}, Elisabeth C. and {Gonzales}, Erica and {Ziegler}, Carl and {Crossfield}, Ian and {Giacalone}, Steven and {Tan}, Thiam-Guan and {Evans}, Phil and {He{\l}miniak}, Krzysztof G. and {Collins}, Kevin I. and {Narita}, Norio and {Fukui}, Akihiko and {Pozuelos}, Francisco J. and {Dressing}, Courtney and {Soubkiou}, Abderahmane and {Benkhaldoun}, Zouhair and {Schlieder}, Joshua E. and {Suarez}, Olga and {Barkaoui}, Khalid and {Palle}, Enric and {Murgas}, Felipe and {Srdoc}, Gregor and {Goliguzova}, Maria V. and {Strakhov}, Ivan A. and {Gnilka}, Crystal and {Lester}, Kathryn and {Littlefield}, Colin and {Scott}, Nic and {Matson}, Rachel and {Gillon}, Micha{\"e}l and {Jehin}, Emmanuel and {Timmermans}, Mathilde and {Ghachoui}, Mourad and {Abe}, Lyu and {Bendjoya}, Philippe and {Guillot}, Tristan and {Triaud}, Amaury H.~M.~J.},
        title = "{VaTEST. II. Statistical Validation of 11 TESS-detected Exoplanets Orbiting K-type Stars}",
      journal = {\aj},
         year = 2023,
        month = jul,
       volume = {166},
       number = {1},
          eid = {9},
        pages = {9},
          doi = {10.3847/1538-3881/acd548},
archivePrefix = {arXiv},
       eprint = {2301.09865},
 primaryClass = {astro-ph.EP},
       adsurl = {https://ui.adsabs.harvard.edu/abs/2023AJ....166....9M}
}

@ARTICLE{polanski2024,
       author = {{Polanski}, Alex S. and {Lubin}, Jack and {Beard}, Corey and {Akana Murphy}, Joseph M. and {Rubenzahl}, Ryan and {Hill}, Michelle L. and {Crossfield}, Ian J.~M. and {Chontos}, Ashley and {Robertson}, Paul and {Isaacson}, Howard and {Kane}, Stephen R. and {Ciardi}, David R. and {Batalha}, Natalie M. and {Dressing}, Courtney and {Fulton}, Benjamin and {Howard}, Andrew W. and {Huber}, Daniel and {Petigura}, Erik A. and {Weiss}, Lauren M. and {Angelo}, Isabel and {Behmard}, Aida and {Blunt}, Sarah and {Brinkman}, Casey L. and {Dai}, Fei and {Dalba}, Paul A. and {Fetherolf}, Tara and {Giacalone}, Steven and {Hirsch}, Lea A. and {Holcomb}, Rae and {Kosiarek}, Molly R. and {Mayo}, Andrew W. and {MacDougall}, Mason G. and {Mo{\v{c}}nik}, Teo and {Pidhorodetska}, Daria and {Rice}, Malena and {Rosenthal}, Lee J. and {Scarsdale}, Nicholas and {Turtelboom}, Emma V. and {Tyler}, Dakotah and {Van Zandt}, Judah and {Yee}, Samuel W. and {Coria}, David R. and {Dulz}, Shannon D. and {Hartman}, Joel D. and {Householder}, Aaron and {Lange}, Sarah and {Langford}, Andrew and {Louden}, Emma M. and {Siegel}, Jared C. and {Gilbert}, Emily A. and {Gonzales}, Erica J. and {Schlieder}, Joshua E. and {Boyle}, Andrew W. and {Christiansen}, Jessie L. and {Clark}, Catherine A. and {Fernandes}, Rachel B. and {Lund}, Michael B. and {Savel}, Arjun B. and {Gill}, Holden and {Beichman}, Charles and {Matson}, Rachel and {Matthews}, Elisabeth C. and {Furlan}, E. and {Howell}, Steve B. and {Scott}, Nicholas J. and {Everett}, Mark E. and {Livingston}, John H. and {Ershova}, Irina O. and {Cheryasov}, Dmitry V. and {Safonov}, Boris and {Lillo-Box}, Jorge and {Barrado}, David and {Morales-Calder{\'o}n}, Mar{\'\i}a},
        title = "{The TESS-Keck Survey. XX. 15 New TESS Planets and a Uniform RV Analysis of All Survey Targets}",
      journal = {\apjs},
         year = 2024,
        month = jun,
       volume = {272},
       number = {2},
          eid = {32},
        pages = {32},
          doi = {10.3847/1538-4365/ad4484},
archivePrefix = {arXiv},
       eprint = {2405.14786},
 primaryClass = {astro-ph.EP},
       adsurl = {https://ui.adsabs.harvard.edu/abs/2024ApJS..272...32P}
}

@ARTICLE{macDougall2023,
       author = {{MacDougall}, Mason G. and {Petigura}, Erik A. and {Gilbert}, Gregory J. and {Angelo}, Isabel and {Batalha}, Natalie M. and {Beard}, Corey and {Behmard}, Aida and {Blunt}, Sarah and {Brinkman}, Casey and {Chontos}, Ashley and {Crossfield}, Ian J.~M. and {Dai}, Fei and {Dalba}, Paul A. and {Dressing}, Courtney and {Fetherolf}, Tara and {Fulton}, Benjamin and {Giacalone}, Steven and {Hill}, Michelle L. and {Holcomb}, Rae and {Howard}, Andrew W. and {Huber}, Daniel and {Isaacson}, Howard and {Kane}, Stephen R. and {Kosiarek}, Molly and {Lubin}, Jack and {Mayo}, Andrew and {Mo{\v{c}}nik}, Teo and {Akana Murphy}, Joseph M. and {Pidhorodetska}, Daria and {Polanski}, Alex S. and {Rice}, Malena and {Robertson}, Paul and {Rosenthal}, Lee J. and {Roy}, Arpita and {Rubenzahl}, Ryan A. and {Scarsdale}, Nicholas and {Turtelboom}, Emma V. and {Tyler}, Dakotah and {Van Zandt}, Judah and {Weiss}, Lauren M. and {Yee}, Samuel W.},
        title = "{The TESS-Keck Survey. XV. Precise Properties of 108 TESS Planets and Their Host Stars}",
      journal = {\aj},
         year = 2023,
        month = jul,
       volume = {166},
       number = {1},
          eid = {33},
        pages = {33},
          doi = {10.3847/1538-3881/acd557},
archivePrefix = {arXiv},
       eprint = {2306.00251},
 primaryClass = {astro-ph.EP},
       adsurl = {https://ui.adsabs.harvard.edu/abs/2023AJ....166...33M}
}

@ARTICLE{Feroz2019,
       author = {{Feroz}, Farhan and {Hobson}, Michael P. and {Cameron}, Ewan and {Pettitt}, Anthony N.},
        title = "{Importance Nested Sampling and the MultiNest Algorithm}",
      journal = {OJAp},
         year = 2019,
        month = nov,
       volume = {2},
       number = {1},
          eid = {10},
        pages = {10},
          doi = {10.21105/astro.1306.2144},
archivePrefix = {arXiv},
       eprint = {1306.2144},
 primaryClass = {astro-ph.IM},
       adsurl = {https://ui.adsabs.harvard.edu/abs/2019OJAp....2E..10F}
}

@ARTICLE{Buchner2014,
       author = {{Buchner}, J. and {Georgakakis}, A. and {Nandra}, K. and {Hsu}, L. and
         {Rangel}, C. and {Brightman}, M. and {Merloni}, A. and {Salvato}, M. and
         {Donley}, J. and {Kocevski}, D.},
        title = "{X-ray spectral modelling of the AGN obscuring region in the CDFS: Bayesian model selection and catalogue}",
      journal = {\aap},
         year = 2014,
        month = apr,
       volume = {564},
          eid = {A125},
        pages = {A125},
          doi = {10.1051/0004-6361/201322971},
archivePrefix = {arXiv},
       eprint = {1402.0004},
 primaryClass = {astro-ph.HE},
       adsurl = {https://ui.adsabs.harvard.edu/abs/2014A&A...564A.125B}
}

@ARTICLE{Ambikasaran2015,
       author = {{Ambikasaran}, Sivaram and {Foreman-Mackey}, Daniel and
         {Greengard}, Leslie and {Hogg}, David W. and {O'Neil}, Michael},
        title = "{Fast Direct Methods for Gaussian Processes}",
      journal = {IEEE Transactions on Pattern Analysis and Machine Intel.},
         year = "2015",
        month = "Jun",
       volume = {38},
        pages = {252},
          doi = {10.1109/TPAMI.2015.2448083},
archivePrefix = {arXiv},
       eprint = {1403.6015},
 primaryClass = {math.NA},
       adsurl = {https://ui.adsabs.harvard.edu/\#abs/2015ITPAM..38..252A}
}

@INPROCEEDINGS{vogt1994,
       author = {{Vogt}, S.~S. and {Allen}, S.~L. and {Bigelow}, B.~C. and {Bresee}, L. and {Brown}, B. and {Cantrall}, T. and {Conrad}, A. and {Couture}, M. and {Delaney}, C. and {Epps}, H.~W. and {Hilyard}, D. and {Hilyard}, D.~F. and {Horn}, E. and {Jern}, N. and {Kanto}, D. and {Keane}, M.~J. and {Kibrick}, R.~I. and {Lewis}, J.~W. and {Osborne}, J. and {Pardeilhan}, G.~H. and {Pfister}, T. and {Ricketts}, T. and {Robinson}, L.~B. and {Stover}, R.~J. and {Tucker}, D. and {Ward}, J. and {Wei}, M.~Z.},
        title = "{HIRES: the high-resolution echelle spectrometer on the Keck 10-m Telescope}",
    booktitle = {Instrum. in Astron. VIII},
         year = 1994,
       editor = {{Crawford}, David L. and {Craine}, Eric R.},
       series = {SPIE Conference Series},
       volume = {2198},
        month = jun,
        pages = {362},
          doi = {10.1117/12.176725},
       adsurl = {https://ui.adsabs.harvard.edu/abs/1994SPIE.2198..362V}
}

@ARTICLE{chontos2022,
       author = {{Chontos}, Ashley and {Murphy}, Joseph M. Akana and {MacDougall}, Mason G. and {Fetherolf}, Tara and {Van Zandt}, Judah and {Rubenzahl}, Ryan A. and {Beard}, Corey and {Huber}, Daniel and {Batalha}, Natalie M. and {Crossfield}, Ian J.~M. and {Dressing}, Courtney D. and {Fulton}, Benjamin and {Howard}, Andrew W. and {Isaacson}, Howard and {Kane}, Stephen R. and {Petigura}, Erik A. and {Robertson}, Paul and {Roy}, Arpita and {Weiss}, Lauren M. and {Behmard}, Aida and {Dai}, Fei and {Dalba}, Paul A. and {Giacalone}, Steven and {Hill}, Michelle L. and {Lubin}, Jack and {Mayo}, Andrew and {Mo{\v{c}}nik}, Teo and {Polanski}, Alex S. and {Rosenthal}, Lee J. and {Scarsdale}, Nicholas and {Turtelboom}, Emma V. and {Ricker}, George R. and {Vanderspek}, Roland and {Latham}, David W. and {Seager}, Sara and {Winn}, Joshua N. and {Jenkins}, Jon M. and {Quinn}, Samuel N. and {Guerrero}, Natalia M. and {Collins}, Karen A. and {Ciardi}, David R. and {Shporer}, Avi and {Goeke}, Robert F. and {Levine}, Alan M. and {Ting}, Eric B. and {Bieryla}, Allyson and {Collins}, Kevin I. and {Kielkopf}, John F. and {Barkaoui}, Khalid and {Benni}, Paul and {Esparza-Borges}, Emma and {Conti}, Dennis M. and {Hooton}, Matthew J. and {Kagetani}, Taiki and {Laloum}, Didier and {Marino}, Giuseppe and {Massey}, Bob and {Murgas}, Felipe and {Papini}, Riccardo and {Schwarz}, Richard P. and {Srdoc}, Gregor and {Stockdale}, Chris and {Wang}, Gavin and {Wittrock}, Justin M. and {Zou}, Yujie},
        title = "{The TESS-Keck Survey: Science Goals and Target Selection}",
      journal = {\aj},
         year = 2022,
        month = jun,
       volume = {163},
       number = {6},
          eid = {297},
        pages = {297},
          doi = {10.3847/1538-3881/ac6266},
archivePrefix = {arXiv},
       eprint = {2106.06156},
 primaryClass = {astro-ph.EP},
       adsurl = {https://ui.adsabs.harvard.edu/abs/2022AJ....163..297C}
}

@INPROCEEDINGS{gibson2024,
       author = {{Gibson}, Steven R. and {Howard}, Andrew W. and {Rider}, Kodi and {Halverson}, Samuel and {Roy}, Arpita and {Baker}, Ashley D. and {Edelstein}, Jerry and {Smith}, Christopher and {Fulton}, Benjamin J. and {Walawender}, Josh and {Brodheim}, Max and {Brown}, Matthew and {Chan}, Dwight and {Dai}, Fei and {Deich}, William and {Gottschalk}, Colby and {Grillo}, Jason and {Hale}, David and {Hill}, Grant and {Holden}, Bradford and {Householder}, Aaron and {Isaacson}, Howard and {Ishikawa}, Yuzo and {Jelinsky}, Sharon and {Kassis}, Marc and {Kaye}, Stephen and {Laher}, Russ and {Lanclos}, Kyle and {Lee}, Chien-Hsiu and {Lilley}, Scott and {McCarney}, Benjamin and {Miller}, Timothy N. and {Payne}, Joel and {Petigura}, Erik and {Poppett}, Claire and {Raffanti}, Michael P. and {Rubenzahl}, Ryan and {Sandford}, Dale and {Schwab}, Christian and {Shaum}, Abby P. and {Sirk}, Martin M. and {Smith}, Roger and {Thorne}, Jim and {Valliant}, John and {Vandenberg}, Adam and {Wang}, Shin-Ywan and {Wishnow}, Edward H. and {Wold}, Truman and {Yeh}, Sherry and {Baca}, Steve and {Beichman}, Charles and {Berriman}, Bruce and {Brown}, Thomas and {Casey}, Kelleen and {Chin}, Jason and {Chong}, James and {Cowley}, David and {Devenot}, Mark and {Elwir}, Hamza and {Finstad}, Daniel and {Fraysse}, Matthew and {James}, Ean and {Jhoti}, Elisha and {Killian}, Joe and {Levine}, Obie and {Li}, Adela Chenyang and {Marin}, Eduardo and {Milner}, Steven and {Nance}, Craig and {O'Hanlon}, Timothy J. and {Orr}, Daniel and {Ortiz-Soto}, Roberto and {Payne}, Tom and {Pember}, Jacob and {Raskin}, Gert and {Savage}, Maureen and {Seifahrt}, Andreas and {Smith}, Brett and {Storesund}, Rob and {St{\"u}rmer}, Julian and {Suominen}, Nick and {Tehero}, Jerez and {Von Boeckmann}, Tod and {Wages}, Keith and {Weisfeiler}, Marie and {Wilcox}, Mavourneen and {Wizinowich}, Peter and {Wolfenberger}, Anna},
        title = "{System design of the Keck Planet Finder}",
    booktitle = {Ground-based and Airborne Instrumentation for Astronomy X},
         year = 2024,
       editor = {{Bryant}, Julia J. and {Motohara}, Kentaro and {Vernet}, Jo{\"e}l. R.~D.},
       series = {SPIE Conference Series},
       volume = {13096},
        month = jul,
          eid = {1309609},
        pages = {1309609},
          doi = {10.1117/12.3017841},
       adsurl = {https://ui.adsabs.harvard.edu/abs/2024SPIE13096E..09G}
}

@INPROCEEDINGS{cosentino2012,
       author = {{Cosentino}, Rosario and {Lovis}, Christophe and {Pepe}, Francesco and {Collier Cameron}, Andrew and {Latham}, David W. and {Molinari}, Emilio and {Udry}, Stephane and {Bezawada}, Naidu and {Black}, Martin and {Born}, Andy and {Buchschacher}, Nicolas and {Charbonneau}, Dave and {Figueira}, Pedro and {Fleury}, Michel and {Galli}, Alberto and {Gallie}, Angus and {Gao}, Xiaofeng and {Ghedina}, Adriano and {Gonzalez}, Carlos and {Gonzalez}, Manuel and {Guerra}, Jose and {Henry}, David and {Horne}, Keith and {Hughes}, Ian and {Kelly}, Dennis and {Lodi}, Marcello and {Lunney}, David and {Maire}, Charles and {Mayor}, Michel and {Micela}, Giusi and {Ordway}, Mark P. and {Peacock}, John and {Phillips}, David and {Piotto}, Giampaolo and {Pollacco}, Don and {Queloz}, Didier and {Rice}, Ken and {Riverol}, Carlos and {Riverol}, Luis and {San Juan}, Jose and {Sasselov}, Dimitar and {Segransan}, Damien and {Sozzetti}, Alessandro and {Sosnowska}, Danuta and {Stobie}, Brian and {Szentgyorgyi}, Andrew and {Vick}, Andy and {Weber}, Luc},
        title = "{Harps-N: the new planet hunter at TNG}",
    booktitle = {Ground-based and Airborne Instrumentation for Astronomy IV},
         year = 2012,
       editor = {{McLean}, Ian S. and {Ramsay}, Suzanne K. and {Takami}, Hideki},
       series = {SPIE Conference Series},
       volume = {8446},
        month = sep,
          eid = {84461V},
        pages = {84461V},
          doi = {10.1117/12.925738},
       adsurl = {https://ui.adsabs.harvard.edu/abs/2012SPIE.8446E..1VC}
}

@ARTICLE{dumusque2021,
       author = {{Dumusque}, X. and {Cretignier}, M. and {Sosnowska}, D. and {Buchschacher}, N. and {Lovis}, C. and {Phillips}, D.~F. and {Pepe}, F. and {Alesina}, F. and {Buchhave}, L.~A. and {Burnier}, J. and {Cecconi}, M. and {Cegla}, H.~M. and {Cloutier}, R. and {Collier Cameron}, A. and {Cosentino}, R. and {Ghedina}, A. and {Gonz{\'a}lez}, M. and {Haywood}, R.~D. and {Latham}, D.~W. and {Lodi}, M. and {L{\'o}pez-Morales}, M. and {Maldonado}, J. and {Malavolta}, L. and {Micela}, G. and {Molinari}, E. and {Mortier}, A. and {P{\'e}rez Ventura}, H. and {Pinamonti}, M. and {Poretti}, E. and {Rice}, K. and {Riverol}, L. and {Riverol}, C. and {San Juan}, J. and {S{\'e}gransan}, D. and {Sozzetti}, A. and {Thompson}, S.~J. and {Udry}, S. and {Wilson}, T.~G.},
        title = "{Three years of HARPS-N high-resolution spectroscopy and precise radial velocity data for the Sun}",
      journal = {\aap},
         year = 2021,
        month = apr,
       volume = {648},
          eid = {A103},
        pages = {A103},
          doi = {10.1051/0004-6361/202039350},
archivePrefix = {arXiv},
       eprint = {2009.01945},
 primaryClass = {astro-ph.SR},
       adsurl = {https://ui.adsabs.harvard.edu/abs/2021A&A...648A.103D}
}

@ARTICLE{Biazzoetal2022,
       author = {{Biazzo}, K. and {D'Orazi}, V. and {Desidera}, S. and {Turrini}, D. and {Benatti}, S. and {Gratton}, R. and {Magrini}, L. and {Sozzetti}, A. and {Baratella}, M. and {Bonomo}, A.~S. and {Borsa}, F. and {Claudi}, R. and {Covino}, E. and {Damasso}, M. and {Di Mauro}, M.~P. and {Lanza}, A.~F. and {Maggio}, A. and {Malavolta}, L. and {Maldonado}, J. and {Marzari}, F. and {Micela}, G. and {Poretti}, E. and {Vitello}, F. and {Affer}, L. and {Bignamini}, A. and {Carleo}, I. and {Cosentino}, R. and {Fiorenzano}, A.~F.~M. and {Giacobbe}, P. and {Harutyunyan}, A. and {Leto}, G. and {Mancini}, L. and {Molinari}, E. and {Molinaro}, M. and {Nardiello}, D. and {Nascimbeni}, V. and {Pagano}, I. and {Pedani}, M. and {Piotto}, G. and {Rainer}, M. and {Scandariato}, G.},
        title = "{The GAPS Programme at TNG. XXXV. Fundamental properties of transiting exoplanet host stars}",
      journal = {\aap},
         year = 2022,
        month = aug,
       volume = {664},
          eid = {A161},
        pages = {A161},
          doi = {10.1051/0004-6361/202243467},
archivePrefix = {arXiv},
       eprint = {2205.15796},
 primaryClass = {astro-ph.SR},
       adsurl = {https://ui.adsabs.harvard.edu/abs/2022A&A...664A.161B}
}

@INPROCEEDINGS{CastelliKurucz2003,
       author = {{Castelli}, F. and {Kurucz}, R.~L.},
        title = "{New Grids of ATLAS9 Model Atmospheres}",
    booktitle = {Modelling of Stellar Atmospheres},
         year = 2003,
       editor = {{Piskunov}, N. and {Weiss}, W.~W. and {Gray}, D.~F.},
       series = {IAU Symposium},
       volume = {210},
        month = jan,
        pages = {A20},
          doi = {10.48550/arXiv.astro-ph/0405087},
archivePrefix = {arXiv},
       eprint = {astro-ph/0405087},
 primaryClass = {astro-ph},
       adsurl = {https://ui.adsabs.harvard.edu/abs/2003IAUS..210P.A20C}
}

@ARTICLE{Sneden1973,
       author = {{Sneden}, C.},
        title = "{The nitrogen abundance of the very metal-poor star HD 122563.}",
      journal = {\apj},
         year = 1973,
        month = sep,
       volume = {184},
        pages = {839},
          doi = {10.1086/152374},
       adsurl = {https://ui.adsabs.harvard.edu/abs/1973ApJ...184..839S}
}

@ARTICLE{Sousaetal2015,
       author = {{Sousa}, S.~G. and {Santos}, N.~C. and {Adibekyan}, V. and {Delgado-Mena}, E. and {Israelian}, G. and {Israelian}, G. and {Israelian}, G.},
        title = "{ARES v2: new features and improved performance}",
      journal = {\aap},
         year = 2015,
        month = may,
       volume = {577},
          eid = {A67},
        pages = {A67},
          doi = {10.1051/0004-6361/201425463},
archivePrefix = {arXiv},
       eprint = {1504.02725},
 primaryClass = {astro-ph.IM},
       adsurl = {https://ui.adsabs.harvard.edu/abs/2015A&A...577A..67S}
}

@ARTICLE{Breweretal2016,
       author = {{Brewer}, John M. and {Fischer}, Debra A. and {Valenti}, Jeff A. and {Piskunov}, Nikolai},
        title = "{Spectral Properties of Cool Stars: Extended Abundance Analysis of 1,617 Planet-search Stars}",
      journal = {\apjs},
         year = 2016,
        month = aug,
       volume = {225},
       number = {2},
          eid = {32},
        pages = {32},
          doi = {10.3847/0067-0049/225/2/32},
archivePrefix = {arXiv},
       eprint = {1606.07929},
 primaryClass = {astro-ph.SR},
       adsurl = {https://ui.adsabs.harvard.edu/abs/2016ApJS..225...32B}
}

@ARTICLE{bonomo2014,
       author = {{Bonomo}, A.~S. and {Sozzetti}, A. and {Lovis}, C. and {Malavolta}, L. and {Rice}, K. and {Buchhave}, L.~A. and {Sasselov}, D. and {Cameron}, A.~C. and {Latham}, D.~W. and {Molinari}, E. and {Pepe}, F. and {Udry}, S. and {Affer}, L. and {Charbonneau}, D. and {Cosentino}, R. and {Dressing}, C.~D. and {Dumusque}, X. and {Figueira}, P. and {Fiorenzano}, A.~F.~M. and {Gettel}, S. and {Harutyunyan}, A. and {Haywood}, R.~D. and {Horne}, K. and {Lopez-Morales}, M. and {Mayor}, M. and {Micela}, G. and {Motalebi}, F. and {Nascimbeni}, V. and {Phillips}, D.~F. and {Piotto}, G. and {Pollacco}, D. and {Queloz}, D. and {S{\'e}gransan}, D. and {Szentgyorgyi}, A. and {Watson}, C.},
        title = "{Characterization of the planetary system Kepler-101 with HARPS-N. A hot super-Neptune with an Earth-sized low-mass companion}",
      journal = {\aap},
         year = 2014,
        month = dec,
       volume = {572},
          eid = {A2},
        pages = {A2},
          doi = {10.1051/0004-6361/201424617},
archivePrefix = {arXiv},
       eprint = {1409.4592},
 primaryClass = {astro-ph.EP},
       adsurl = {https://ui.adsabs.harvard.edu/abs/2014A&A...572A...2B}
}

@ARTICLE{bakos2015,
       author = {{Bakos}, G. {\'A}. and {Penev}, K. and {Bayliss}, D. and {Hartman}, J.~D. and {Zhou}, G. and {Brahm}, R. and {Mancini}, L. and {de Val-Borro}, M. and {Bhatti}, W. and {Jord{\'a}n}, A. and {Rabus}, M. and {Espinoza}, N. and {Csubry}, Z. and {Howard}, A.~W. and {Fulton}, B.~J. and {Buchhave}, L.~A. and {Ciceri}, S. and {Henning}, T. and {Schmidt}, B. and {Isaacson}, H. and {Noyes}, R.~W. and {Marcy}, G.~W. and {Suc}, V. and {Howe}, A.~R. and {Burrows}, A.~S. and {L{\'a}z{\'a}r}, J. and {Papp}, I. and {S{\'a}ri}, P.},
        title = "{HATS-7b: A Hot Super Neptune Transiting a Quiet K Dwarf Star}",
      journal = {\apj},
         year = 2015,
        month = nov,
       volume = {813},
       number = {2},
          eid = {111},
        pages = {111},
          doi = {10.1088/0004-637X/813/2/111},
archivePrefix = {arXiv},
       eprint = {1507.01024},
 primaryClass = {astro-ph.EP},
       adsurl = {https://ui.adsabs.harvard.edu/abs/2015ApJ...813..111B}
}

@ARTICLE{bayliss2015,
       author = {{Bayliss}, D. and {Hartman}, J.~D. and {Bakos}, G. {\'A}. and {Penev}, K. and {Zhou}, G. and {Brahm}, R. and {Rabus}, M. and {Jord{\'a}n}, A. and {Mancini}, L. and {de Val-Borro}, M. and {Bhatti}, W. and {Espinoza}, N. and {Csubry}, Z. and {Howard}, A.~W. and {Fulton}, B.~J. and {Buchhave}, L.~A. and {Henning}, T. and {Schmidt}, B. and {Ciceri}, S. and {Noyes}, R.~W. and {Isaacson}, H. and {Marcy}, G.~W. and {Suc}, V. and {L{\'a}z{\'a}r}, J. and {Papp}, I. and {S{\'a}ri}, P.},
        title = "{HATS-8b: A Low-density Transiting Super-Neptune}",
      journal = {\aj},
         year = 2015,
        month = aug,
       volume = {150},
       number = {2},
          eid = {49},
        pages = {49},
          doi = {10.1088/0004-6256/150/2/49},
archivePrefix = {arXiv},
       eprint = {1506.01334},
 primaryClass = {astro-ph.EP},
       adsurl = {https://ui.adsabs.harvard.edu/abs/2015AJ....150...49B}
}

@ARTICLE{knudstrup2023,
       author = {{Knudstrup}, E. and {Gandolfi}, D. and {Nowak}, G. and {Persson}, C.~M. and {Furlan}, E. and {Livingston}, J. and {Matthews}, E. and {Lundkvist}, M.~S. and {Winther}, M.~L. and {R{\o}rsted}, J.~L. and {Albrecht}, S.~H. and {Goffo}, E. and {Carleo}, I. and {Deeg}, H.~J. and {Collins}, K.~A. and {Narita}, N. and {Isaacson}, H. and {Redfield}, S. and {Dai}, F. and {Hirano}, T. and {Akana Murphy}, J.~M. and {Beard}, C. and {Buchhave}, L.~A. and {Cary}, S. and {Chontos}, A. and {Crossfield}, I. and {Cochran}, W.~D. and {Conti}, D. and {Dalba}, P.~A. and {Esposito}, M. and {Fajardo-Acosta}, S. and {Giacalone}, S. and {Grunblatt}, S.~K. and {Guerra}, P. and {Hatzes}, A.~P. and {Holcomb}, R. and {Horta}, F.~G. and {Howard}, A.~W. and {Huber}, D. and {Jenkins}, J.~M. and {Kab{\'a}th}, P. and {Kane}, S. and {Korth}, J. and {Lam}, K.~W.~F. and {Lester}, K.~V. and {Matson}, R. and {McLeod}, K.~K. and {Orell-Miquel}, J. and {Murgas}, F. and {Palle}, E. and {Polanski}, A.~S. and {Ricker}, G. and {Robertson}, P. and {Rubenzahl}, R. and {Schlieder}, J.~E. and {Seager}, S. and {Smith}, A.~M.~S. and {Tenenbaum}, P. and {Turtelboom}, E. and {Vanderspek}, R. and {Weiss}, L. and {Winn}, J.},
        title = "{Radial velocity confirmation of a hot super-Neptune discovered by TESS with a warm Saturn-mass companion}",
      journal = {\mnras},
         year = 2023,
        month = mar,
       volume = {519},
       number = {4},
        pages = {5637-5655},
          doi = {10.1093/mnras/stac3684},
archivePrefix = {arXiv},
       eprint = {2211.17035},
 primaryClass = {astro-ph.EP},
       adsurl = {https://ui.adsabs.harvard.edu/abs/2023MNRAS.519.5637K}
}

@ARTICLE{castro2024b,
       author = {{Castro-Gonz{\'a}lez}, A. and {Lillo-Box}, J. and {Armstrong}, D.~J. and {Acu{\~n}a}, L. and {Aguichine}, A. and {Bourrier}, V. and {Gandhi}, S. and {Sousa}, S.~G. and {Delgado-Mena}, E. and {Moya}, A. and {Adibekyan}, V. and {Correia}, A.~C.~M. and {Barrado}, D. and {Damasso}, M. and {Winn}, J.~N. and {Santos}, N.~C. and {Barkaoui}, K. and {Barros}, S.~C.~C. and {Benkhaldoun}, Z. and {Bouchy}, F. and {Brice{\~n}o}, C. and {Caldwell}, D.~A. and {Collins}, K.~A. and {Essack}, Z. and {Ghachoui}, M. and {Gillon}, M. and {Hounsell}, R. and {Jehin}, E. and {Jenkins}, J.~M. and {Keniger}, M.~A.~F. and {Law}, N. and {Mann}, A.~W. and {Nielsen}, L.~D. and {Pozuelos}, F.~J. and {Schanche}, N. and {Seager}, S. and {Tan}, T.-G. and {Timmermans}, M. and {Villase{\~n}or}, J. and {Watkins}, C.~N. and {Ziegler}, C.},
        title = "{TOI-5005 b: A super-Neptune in the savanna near the ridge}",
      journal = {\aap},
         year = 2024,
        month = nov,
       volume = {691},
          eid = {A233},
        pages = {A233},
          doi = {10.1051/0004-6361/202451656},
archivePrefix = {arXiv},
       eprint = {2409.18129},
 primaryClass = {astro-ph.EP},
       adsurl = {https://ui.adsabs.harvard.edu/abs/2024A&A...691A.233C}
}

@ARTICLE{latham2011,
       author = {{Latham}, David W. and {Rowe}, Jason F. and {Quinn}, Samuel N. and {Batalha}, Natalie M. and {Borucki}, William J. and {Brown}, Timothy M. and {Bryson}, Stephen T. and {Buchhave}, Lars A. and {Caldwell}, Douglas A. and {Carter}, Joshua A. and {Christiansen}, Jessie L. and {Ciardi}, David R. and {Cochran}, William D. and {Dunham}, Edward W. and {Fabrycky}, Daniel C. and {Ford}, Eric B. and {Gautier}, III, Thomas N. and {Gilliland}, Ronald L. and {Holman}, Matthew J. and {Howell}, Steve B. and {Ibrahim}, Khadeejah A. and {Isaacson}, Howard and {Jenkins}, Jon M. and {Koch}, David G. and {Lissauer}, Jack J. and {Marcy}, Geoffrey W. and {Quintana}, Elisa V. and {Ragozzine}, Darin and {Sasselov}, Dimitar and {Shporer}, Avi and {Steffen}, Jason H. and {Welsh}, William F. and {Wohler}, Bill},
        title = "{A First Comparison of Kepler Planet Candidates in Single and Multiple Systems}",
      journal = {\apjl},
         year = 2011,
        month = may,
       volume = {732},
       number = {2},
          eid = {L24},
        pages = {L24},
          doi = {10.1088/2041-8205/732/2/L24},
archivePrefix = {arXiv},
       eprint = {1103.3896},
 primaryClass = {astro-ph.EP},
       adsurl = {https://ui.adsabs.harvard.edu/abs/2011ApJ...732L..24L}
}

@ARTICLE{szabo2011,
       author = {{Szab{\'o}}, Gy. M. and {Kiss}, L.~L.},
        title = "{A Short-period Censor of Sub-Jupiter Mass Exoplanets with Low Density}",
      journal = {\apjl},
         year = 2011,
        month = feb,
       volume = {727},
       number = {2},
          eid = {L44},
        pages = {L44},
          doi = {10.1088/2041-8205/727/2/L44},
archivePrefix = {arXiv},
       eprint = {1012.4791},
 primaryClass = {astro-ph.EP},
       adsurl = {https://ui.adsabs.harvard.edu/abs/2011ApJ...727L..44S}
}

@ARTICLE{matsakos2016,
       author = {{Matsakos}, Titos and {K{\"o}nigl}, Arieh},
        title = "{On the Origin of the Sub-Jovian Desert in the Orbital-period-Planetary-mass Plane}",
      journal = {\apjl},
         year = 2016,
        month = mar,
       volume = {820},
       number = {1},
          eid = {L8},
        pages = {L8},
          doi = {10.3847/2041-8205/820/1/L8},
archivePrefix = {arXiv},
       eprint = {1603.00414},
 primaryClass = {astro-ph.EP},
       adsurl = {https://ui.adsabs.harvard.edu/abs/2016ApJ...820L...8M}
}

@ARTICLE{owen2018,
       author = {{Owen}, James E. and {Lai}, Dong},
        title = "{Photoevaporation and high-eccentricity migration created the sub-Jovian desert}",
      journal = {\mnras},
         year = 2018,
        month = oct,
       volume = {479},
       number = {4},
        pages = {5012-5021},
          doi = {10.1093/mnras/sty1760},
archivePrefix = {arXiv},
       eprint = {1807.00012},
 primaryClass = {astro-ph.EP},
       adsurl = {https://ui.adsabs.harvard.edu/abs/2018MNRAS.479.5012O}
}

@ARTICLE{castro2024a,
       author = {{Castro-Gonz{\'a}lez}, A. and {Bourrier}, V. and {Lillo-Box}, J. and {Delisle}, J.-B. and {Armstrong}, D.~J. and {Barrado}, D. and {Correia}, A.~C.~M.},
        title = "{Mapping the exo-Neptunian landscape: A ridge between the desert and savanna}",
      journal = {\aap},
         year = 2024,
        month = sep,
       volume = {689},
          eid = {A250},
        pages = {A250},
          doi = {10.1051/0004-6361/202450957},
archivePrefix = {arXiv},
       eprint = {2409.10517},
 primaryClass = {astro-ph.EP},
       adsurl = {https://ui.adsabs.harvard.edu/abs/2024A&A...689A.250C}
}

@ARTICLE{Kochanek2017,
       author = {{Kochanek}, C.~S. and {Shappee}, B.~J. and {Stanek}, K.~Z. and {Holoien}, T.~W. -S. and {Thompson}, Todd A. and {Prieto}, J.~L. and {Dong}, Subo and {Shields}, J.~V. and {Will}, D. and {Britt}, C. and {Perzanowski}, D. and {Pojma{\'n}ski}, G.},
        title = "{The All-Sky Automated Survey for Supernovae (ASAS-SN) Light Curve Server v1.0}",
      journal = {\pasp},
         year = 2017,
        month = oct,
       volume = {129},
       number = {980},
        pages = {104502},
          doi = {10.1088/1538-3873/aa80d9},
archivePrefix = {arXiv},
       eprint = {1706.07060},
 primaryClass = {astro-ph.SR},
       adsurl = {https://ui.adsabs.harvard.edu/abs/2017PASP..129j4502K}
}

@ARTICLE{Speagle2020,
       author = {{Speagle}, Joshua S.},
        title = "{DYNESTY: a dynamic nested sampling package for estimating Bayesian posteriors and evidences}",
      journal = {\mnras},
         year = 2020,
        month = apr,
       volume = {493},
       number = {3},
        pages = {3132-3158},
          doi = {10.1093/mnras/staa278},
archivePrefix = {arXiv},
       eprint = {1904.02180},
 primaryClass = {astro-ph.IM},
       adsurl = {https://ui.adsabs.harvard.edu/abs/2020MNRAS.493.3132S}
}

@ARTICLE{Espinoza2019,
       author = {{Espinoza}, N{\'e}stor and {Kossakowski}, Diana and {Brahm}, Rafael},
        title = "{juliet: a versatile modelling tool for transiting and non-transiting exoplanetary systems}",
      journal = {\mnras},
         year = 2019,
        month = dec,
       volume = {490},
       number = {2},
        pages = {2262-2283},
          doi = {10.1093/mnras/stz2688},
archivePrefix = {arXiv},
       eprint = {1812.08549},
 primaryClass = {astro-ph.EP},
       adsurl = {https://ui.adsabs.harvard.edu/abs/2019MNRAS.490.2262E}
}

@ARTICLE{Foreman2017,
       author = {{Foreman-Mackey}, Daniel and {Agol}, Eric and {Ambikasaran}, Sivaram and {Angus}, Ruth and {Angus}, Ruth and {Angus}, Ruth},
        title = "{Fast and Scalable Gaussian Process Modeling with Applications to Astronomical Time Series}",
      journal = {\aj},
         year = 2017,
        month = dec,
       volume = {154},
       number = {6},
          eid = {220},
        pages = {220},
          doi = {10.3847/1538-3881/aa9332},
archivePrefix = {arXiv},
       eprint = {1703.09710},
 primaryClass = {astro-ph.IM},
       adsurl = {https://ui.adsabs.harvard.edu/abs/2017AJ....154..220F}
}

@ARTICLE{Gregory2016,
   author = {{Gregory}, P.~C.},
    title = "{An Apodized Kepler Periodogram for Separating Planetary and Stellar Activity Signals}",
  journal = {ArXiv e-prints},
archivePrefix = "arXiv",
   eprint = {1601.08105},
 primaryClass = "astro-ph.SR",
     year = 2016,
    month = jan,
   adsurl = {http://adsabs.harvard.edu/abs/2016arXiv160108105G}
}

@ARTICLE{hara2022b,
       author = {{Hara}, Nathan C. and {Delisle}, Jean-Baptiste and {Unger}, Nicolas and {Dumusque}, Xavier},
        title = "{Testing whether a signal is strictly periodic. Application to disentangling planets and stellar activity in radial velocities}",
      journal = {\aap},
         year = 2022,
        month = feb,
       volume = {658},
          eid = {A177},
        pages = {A177},
          doi = {10.1051/0004-6361/202141197},
archivePrefix = {arXiv},
       eprint = {2106.01365},
 primaryClass = {astro-ph.EP},
       adsurl = {https://ui.adsabs.harvard.edu/abs/2022A&A...658A.177H}
}

@ARTICLE{vissapragada2025,
       author = {{Vissapragada}, Shreyas and {Behmard}, Aida},
        title = "{The Hottest Neptunes Orbit Metal-rich Stars}",
      journal = {\aj},
         year = 2025,
        month = feb,
       volume = {169},
       number = {2},
          eid = {117},
        pages = {117},
          doi = {10.3847/1538-3881/ada143},
archivePrefix = {arXiv},
       eprint = {2412.13245},
 primaryClass = {astro-ph.EP},
       adsurl = {https://ui.adsabs.harvard.edu/abs/2025AJ....169..117V}
}

@ARTICLE{doyle2025,
       author = {{Doyle}, Lauren and {Armstrong}, David J. and {Acu{\~n}a}, Lorena and {Osborn}, Ares and {Sousa}, S{\'e}rgio A.~G. and {Castro-Gonz{\'a}lez}, Amadeo and {Bourrier}, Vincent and {Alves}, Douglas and {Barrado}, David and {Barros}, Susana C.~C. and {Bayliss}, Daniel and {Cui}, Kaiming and {Demangeon}, Olivier and {D{\'\i}az}, Rodrigo F. and {Dumusque}, Xavier and {Eeles-Nolle}, Fintan and {Gill}, Samuel and {Hacker}, Alejandro and {Jenkins}, James S. and {Keniger}, Marcelo Aron Fetzner and {Lafarga}, Marina and {Lillo-Box}, Jorge and {Lockley}, Isobel and {Nielsen}, Louise D. and {Parc}, L{\'e}na and {Rodrigues}, Jos{\'e} and {Santerne}, Alexandre and {Santos}, Nuno C. and {Wheatley}, Peter J.},
        title = "{Exploring the Neptunian desert: insights from a homogeneous planetary sample}",
      journal = {\mnras},
         year = 2025,
        month = jun,
       volume = {539},
       number = {4},
        pages = {3138-3156},
          doi = {10.1093/mnras/staf670},
archivePrefix = {arXiv},
       eprint = {2504.16164},
 primaryClass = {astro-ph.EP},
       adsurl = {https://ui.adsabs.harvard.edu/abs/2025MNRAS.539.3138D}
}

@ARTICLE{dong2018,
       author = {{Dong}, Subo and {Xie}, Ji-Wei and {Zhou}, Ji-Lin and {Zheng}, Zheng and {Luo}, Ali},
        title = "{LAMOST telescope reveals that Neptunian cousins of hot Jupiters are mostly single offspring of stars that are rich in heavy elements}",
      journal = {Proceedings of the National Academy of Science},
         year = 2018,
        month = jan,
       volume = {115},
       number = {2},
        pages = {266-271},
          doi = {10.1073/pnas.1711406115},
archivePrefix = {arXiv},
       eprint = {1706.07807},
 primaryClass = {astro-ph.EP},
       adsurl = {https://ui.adsabs.harvard.edu/abs/2018PNAS..115..266D}
}

@ARTICLE{southworth2011,
       author = {{Southworth}, John},
        title = "{Homogeneous studies of transiting extrasolar planets - IV. Thirty systems with space-based light curves}",
      journal = {\mnras},
         year = 2011,
        month = nov,
       volume = {417},
       number = {3},
        pages = {2166-2196},
          doi = {10.1111/j.1365-2966.2011.19399.x},
archivePrefix = {arXiv},
       eprint = {1107.1235},
 primaryClass = {astro-ph.EP},
       adsurl = {https://ui.adsabs.harvard.edu/abs/2011MNRAS.417.2166S}
}

@ARTICLE{Stumpe2012,
       author = {{Stumpe}, Martin C. and {Smith}, Jeffrey C. and {Van Cleve}, Jeffrey E. and {Twicken}, Joseph D. and {Barclay}, Thomas S. and {Fanelli}, Michael N. and {Girouard}, Forrest R. and {Jenkins}, Jon M. and {Kolodziejczak}, Jeffery J. and {McCauliff}, Sean D. and {Morris}, Robert L.},
        title = "{Kepler Presearch Data Conditioning I{\textemdash}Architecture and Algorithms for Error Correction in Kepler Light Curves}",
      journal = {\pasp},
         year = 2012,
        month = sep,
       volume = {124},
       number = {919},
        pages = {985},
          doi = {10.1086/667698},
archivePrefix = {arXiv},
       eprint = {1203.1382},
 primaryClass = {astro-ph.IM},
       adsurl = {https://ui.adsabs.harvard.edu/abs/2012PASP..124..985S}
}

@ARTICLE{Stumpe2014,
       author = {{Stumpe}, Martin C. and {Smith}, Jeffrey C. and {Catanzarite}, Joseph H. and {Van Cleve}, Jeffrey E. and {Jenkins}, Jon M. and {Twicken}, Joseph D. and {Girouard}, Forrest R.},
        title = "{Multiscale Systematic Error Correction via Wavelet-Based Bandsplitting in Kepler Data}",
      journal = {\pasp},
         year = 2014,
        month = jan,
       volume = {126},
       number = {935},
        pages = {100},
          doi = {10.1086/674989},
       adsurl = {https://ui.adsabs.harvard.edu/abs/2014PASP..126..100S}
}

@ARTICLE{Smith2012,
       author = {{Smith}, Jeffrey C. and {Stumpe}, Martin C. and {Van Cleve}, Jeffrey E. and {Jenkins}, Jon M. and {Barclay}, Thomas S. and {Fanelli}, Michael N. and {Girouard}, Forrest R. and {Kolodziejczak}, Jeffery J. and {McCauliff}, Sean D. and {Morris}, Robert L. and {Twicken}, Joseph D.},
        title = "{Kepler Presearch Data Conditioning II - A Bayesian Approach to Systematic Error Correction}",
      journal = {\pasp},
         year = 2012,
        month = sep,
       volume = {124},
       number = {919},
        pages = {1000},
          doi = {10.1086/667697},
archivePrefix = {arXiv},
       eprint = {1203.1383},
 primaryClass = {astro-ph.IM},
       adsurl = {https://ui.adsabs.harvard.edu/abs/2012PASP..124.1000S}
}

@misc{lightkurve,
       author = {{Lightkurve Collaboration} and {Cardoso}, Jos{\'e} and {Hedges}, Christina and {Gully-Santiago}, Michael and {Saunders}, Nicholas and {Cody}, Ann Marie and {Barclay}, Thomas and {Hall}, Oliver and {Sagear}, Sheila and {Turtelboom}, Emma and {Zhang}, Johnny and {Tzanidakis}, Andy and {Mighell}, Ken and {Coughlin}, Jeff and {Bell}, Keaton and {Berta-Thompson}, Zach and {Williams}, Peter and {Dotson}, Jessie and {Barentsen}, Geert},
        title = "{Lightkurve: Kepler and TESS time series analysis in Python}",
 howpublished = {ASCL, record ascl:1812.013},
         year = 2018,
        month = dec,
          eid = {ascl:1812.013},
       adsurl = {https://ui.adsabs.harvard.edu/abs/2018ascl.soft12013L}
}

@ARTICLE{Naponiello2026a,
       author = {{Naponiello}, Luca},
        title = "{A homogeneous transit-timing-variation investigation of all TESS systems with a confirmed single-transiting planet}",
      journal = {\aap},
         year = 2026,
        month = jan,
       volume = {705},
          eid = {A5},
        pages = {A5},
          doi = {10.1051/0004-6361/202557512},
archivePrefix = {arXiv},
       eprint = {2511.16504},
 primaryClass = {astro-ph.EP},
       adsurl = {https://ui.adsabs.harvard.edu/abs/2026A&A...705A...5N}
}

@ARTICLE{armstrong2020,
       author = {{Armstrong}, David J. and {Lopez}, Th{\'e}o A. and {Adibekyan}, Vardan and {Booth}, Richard A. and {Bryant}, Edward M. and {Collins}, Karen A. and {Deleuil}, Magali and {Emsenhuber}, Alexandre and {Huang}, Chelsea X. and {King}, George W. and {Lillo-Box}, Jorge and {Lissauer}, Jack J. and {Matthews}, Elisabeth and {Mousis}, Olivier and {Nielsen}, Louise D. and {Osborn}, Hugh and {Otegi}, Jon and {Santos}, Nuno C. and {Sousa}, S{\'e}rgio G. and {Stassun}, Keivan G. and {Veras}, Dimitri and {Ziegler}, Carl and {Acton}, Jack S. and {Almenara}, Jose M. and {Anderson}, David R. and {Barrado}, David and {Barros}, Susana C.~C. and {Bayliss}, Daniel and {Belardi}, Claudia and {Bouchy}, Francois and {Brice{\~n}o}, C{\'e}sar and {Brogi}, Matteo and {Brown}, David J.~A. and {Burleigh}, Matthew R. and {Casewell}, Sarah L. and {Chaushev}, Alexander and {Ciardi}, David R. and {Collins}, Kevin I. and {Col{\'o}n}, Knicole D. and {Cooke}, Benjamin F. and {Crossfield}, Ian J.~M. and {D{\'\i}az}, Rodrigo F. and {Delgado Mena}, Elisa and {Demangeon}, Olivier D.~S. and {Dorn}, Caroline and {Dumusque}, Xavier and {Eigm{\"u}ller}, Philipp and {Fausnaugh}, Michael and {Figueira}, Pedro and {Gan}, Tianjun and {Gandhi}, Siddharth and {Gill}, Samuel and {Gonzales}, Erica J. and {Goad}, Michael R. and {G{\"u}nther}, Maximilian N. and {Helled}, Ravit and {Hojjatpanah}, Saeed and {Howell}, Steve B. and {Jackman}, James and {Jenkins}, James S. and {Jenkins}, Jon M. and {Jensen}, Eric L.~N. and {Kennedy}, Grant M. and {Latham}, David W. and {Law}, Nicholas and {Lendl}, Monika and {Lozovsky}, Michael and {Mann}, Andrew W. and {Moyano}, Maximiliano and {McCormac}, James and {Meru}, Farzana and {Mordasini}, Christoph and {Osborn}, Ares and {Pollacco}, Don and {Queloz}, Didier and {Raynard}, Liam and {Ricker}, George R. and {Rowden}, Pamela and {Santerne}, Alexandre and {Schlieder}, Joshua E. and {Seager}, Sara and {Sha}, Lizhou and {Tan}, Thiam-Guan and {Tilbrook}, Rosanna H. and {Ting}, Eric and {Udry}, St{\'e}phane and {Vanderspek}, Roland and {Watson}, Christopher A. and {West}, Richard G. and {Wilson}, Paul A. and {Winn}, Joshua N. and {Wheatley}, Peter and {Villasenor}, Jesus Noel and {Vines}, Jose I. and {Zhan}, Zhuchang},
        title = "{A remnant planetary core in the hot-Neptune desert}",
      journal = {\nat},
         year = 2020,
        month = jul,
       volume = {583},
       number = {7814},
        pages = {39-42},
          doi = {10.1038/s41586-020-2421-7},
archivePrefix = {arXiv},
       eprint = {2003.10314},
 primaryClass = {astro-ph.EP},
       adsurl = {https://ui.adsabs.harvard.edu/abs/2020Natur.583...39A}
}

@ARTICLE{osborn2023,
       author = {{Osborn}, Ares and {Armstrong}, David J. and {Fern{\'a}ndez}, Jorge and {Knierim}, Henrik and {Adibekyan}, Vardan and {Collins}, Karen A. and {Delgado-Mena}, Elisa and {Fridlund}, Malcolm and {Gomes da Silva}, Jo{\~a}o and {Hellier}, Coel and {Jackson}, David G. and {King}, George W. and {Lillo-Box}, Jorge and {Matson}, Rachel A. and {Matthews}, Elisabeth C. and {Santos}, Nuno C. and {Sousa}, S{\'e}rgio G. and {Stassun}, Keivan G. and {Tan}, Thiam-Guan and {Ricker}, George R. and {Vanderspek}, Roland and {Latham}, David W. and {Seager}, Sara and {Winn}, Joshua N. and {Jenkins}, Jon M. and {Bayliss}, Daniel and {Bouma}, Luke G. and {Ciardi}, David R. and {Collins}, Kevin I. and {Col{\'o}n}, Knicole D. and {Crossfield}, Ian J.~M. and {Demangeon}, Olivier D.~S. and {D{\'\i}az}, Rodrigo F. and {Dorn}, Caroline and {Dumusque}, Xavier and {Keniger}, Marcelo Aron Fetzner and {Figueira}, Pedro and {Gan}, Tianjun and {Goeke}, Robert F. and {Hadjigeorghiou}, Andreas and {Hawthorn}, Faith and {Helled}, Ravit and {Howell}, Steve B. and {Nielsen}, Louise D. and {Osborn}, Hugh P. and {Quinn}, Samuel N. and {Sefako}, Ramotholo and {Shporer}, Avi and {Str{\o}m}, Paul A. and {Twicken}, Joseph D. and {Vanderburg}, Andrew and {Wheatley}, Peter J.},
        title = "{TOI-332 b: a super dense Neptune found deep within the Neptunian desert}",
      journal = {\mnras},
         year = 2023,
        month = nov,
       volume = {526},
       number = {1},
        pages = {548-566},
          doi = {10.1093/mnras/stad2575},
archivePrefix = {arXiv},
       eprint = {2308.12137},
 primaryClass = {astro-ph.EP},
       adsurl = {https://ui.adsabs.harvard.edu/abs/2023MNRAS.526..548O}
}

@ARTICLE{naponiello2023,
       author = {{Naponiello}, Luca and {Mancini}, Luigi and {Sozzetti}, Alessandro and {Bonomo}, Aldo S. and {Morbidelli}, Alessandro and {Dou}, Jingyao and {Zeng}, Li and {Leinhardt}, Zoe M. and {Biazzo}, Katia and {Cubillos}, Patricio E. and {Pinamonti}, Matteo and {Locci}, Daniele and {Maggio}, Antonio and {Damasso}, Mario and {Lanza}, Antonino F. and {Lissauer}, Jack J. and {Collins}, Karen A. and {Carter}, Philip J. and {Jensen}, Eric L.~N. and {Bignamini}, Andrea and {Boschin}, Walter and {Bouma}, Luke G. and {Ciardi}, David R. and {Cosentino}, Rosario and {Crossfield}, Ian and {Desidera}, Silvano and {Dumusque}, Xavier and {Fiorenzano}, Aldo F.~M. and {Fukui}, Akihiko and {Giacobbe}, Paolo and {Gnilka}, Crystal L. and {Ghedina}, Adriano and {Guilluy}, Gloria and {Harutyunyan}, Avet and {Howell}, Steve B. and {Jenkins}, Jon M. and {Lund}, Michael B. and {Kielkopf}, John F. and {Lester}, Katie V. and {Malavolta}, Luca and {Mann}, Andrew W. and {Matson}, Rachel A. and {Matthews}, Elisabeth C. and {Nardiello}, Domenico and {Narita}, Norio and {Pace}, Emanuele and {Pagano}, Isabella and {Palle}, Enric and {Pedani}, Marco and {Seager}, Sara and {Schlieder}, Joshua E. and {Schwarz}, Richard P. and {Shporer}, Avi and {Twicken}, Joseph D. and {Winn}, Joshua N. and {Ziegler}, Carl and {Zingales}, Tiziano},
        title = "{A super-massive Neptune-sized planet}",
      journal = {\nat},
         year = 2023,
        month = oct,
       volume = {622},
       number = {7982},
        pages = {255-260},
          doi = {10.1038/s41586-023-06499-2},
archivePrefix = {arXiv},
       eprint = {2309.01464},
 primaryClass = {astro-ph.EP},
       adsurl = {https://ui.adsabs.harvard.edu/abs/2023Natur.622..255N}
}

@ARTICLE{hallatt2026,
       author = {{Hallatt}, Tim and {Millholland}, Sarah},
        title = "{Shedding Light on Desert Dwellers}",
      journal = {\apj},
         year = 2026,
        month = feb,
       volume = {997},
       number = {2},
          eid = {139},
        pages = {139},
          doi = {10.3847/1538-4357/adfb75},
archivePrefix = {arXiv},
       eprint = {2509.22893},
 primaryClass = {astro-ph.EP},
       adsurl = {https://ui.adsabs.harvard.edu/abs/2026ApJ...997..139H}
}

@ARTICLE{ionov2018,
       author = {{Ionov}, Dmitry and {Pavlyuchenkov}, Yaroslav and {Shematovich}, Valery},
        title = "{Survival of a planet in short-period Neptunian desert under effect of photoevaporation}",
      journal = {\mnras},
         year = 2018,
        month = jun,
       volume = {476},
       number = {4},
        pages = {5639-5644},
          doi = {10.1093/mnras/sty626},
archivePrefix = {arXiv},
       eprint = {1803.04278},
 primaryClass = {astro-ph.EP},
       adsurl = {https://ui.adsabs.harvard.edu/abs/2018MNRAS.476.5639I}
}

@ARTICLE{koskinen2022,
       author = {{Koskinen}, Tommi T. and {Lavvas}, Panayotis and {Huang}, Chenliang and {Bergsten}, Galen and {Fernandes}, Rachel B. and {Young}, Mitchell E.},
        title = "{Mass Loss by Atmospheric Escape from Extremely Close-in Planets}",
      journal = {\apj},
         year = 2022,
        month = apr,
       volume = {929},
       number = {1},
          eid = {52},
        pages = {52},
          doi = {10.3847/1538-4357/ac4f45},
archivePrefix = {arXiv},
       eprint = {2203.06302},
 primaryClass = {astro-ph.EP},
       adsurl = {https://ui.adsabs.harvard.edu/abs/2022ApJ...929...52K}
}

@ARTICLE{thorngren2023,
       author = {{Thorngren}, Daniel P. and {Lee}, Eve J. and {Lopez}, Eric D.},
        title = "{Removal of Hot Saturns in Mass-Radius Plane by Runaway Mass Loss}",
      journal = {\apjl},
         year = 2023,
        month = mar,
       volume = {945},
       number = {2},
          eid = {L36},
        pages = {L36},
          doi = {10.3847/2041-8213/acbd35},
archivePrefix = {arXiv},
       eprint = {2211.11770},
 primaryClass = {astro-ph.EP},
       adsurl = {https://ui.adsabs.harvard.edu/abs/2023ApJ...945L..36T}
}

@ARTICLE{ricker2015,
       author = {{Ricker}, George R. and {Winn}, Joshua N. and {Vanderspek}, Roland and {Latham}, David W. and {Bakos}, G{\'a}sp{\'a}r {\'A}. and {Bean}, Jacob L. and {Berta-Thompson}, Zachory K. and {Brown}, Timothy M. and {Buchhave}, Lars and {Butler}, Nathaniel R. and {Butler}, R. Paul and {Chaplin}, William J. and {Charbonneau}, David and {Christensen-Dalsgaard}, J{\o}rgen and {Clampin}, Mark and {Deming}, Drake and {Doty}, John and {De Lee}, Nathan and {Dressing}, Courtney and {Dunham}, Edward W. and {Endl}, Michael and {Fressin}, Francois and {Ge}, Jian and {Henning}, Thomas and {Holman}, Matthew J. and {Howard}, Andrew W. and {Ida}, Shigeru and {Jenkins}, Jon M. and {Jernigan}, Garrett and {Johnson}, John Asher and {Kaltenegger}, Lisa and {Kawai}, Nobuyuki and {Kjeldsen}, Hans and {Laughlin}, Gregory and {Levine}, Alan M. and {Lin}, Douglas and {Lissauer}, Jack J. and {MacQueen}, Phillip and {Marcy}, Geoffrey and {McCullough}, Peter R. and {Morton}, Timothy D. and {Narita}, Norio and {Paegert}, Martin and {Palle}, Enric and {Pepe}, Francesco and {Pepper}, Joshua and {Quirrenbach}, Andreas and {Rinehart}, Stephen A. and {Sasselov}, Dimitar and {Sato}, Bun'ei and {Seager}, Sara and {Sozzetti}, Alessandro and {Stassun}, Keivan G. and {Sullivan}, Peter and {Szentgyorgyi}, Andrew and {Torres}, Guillermo and {Udry}, Stephane and {Villasenor}, Joel},
        title = "{Transiting Exoplanet Survey Satellite (TESS)}",
      journal = {JATIS},
         year = 2015,
        month = jan,
       volume = {1},
          eid = {014003},
        pages = {014003},
          doi = {10.1117/1.JATIS.1.1.014003},
       adsurl = {https://ui.adsabs.harvard.edu/abs/2015JATIS...1a4003R}
}

@ARTICLE{jenkins2020,
       author = {{Jenkins}, James S. and {D{\'\i}az}, Mat{\'\i}as R. and {Kurtovic}, Nicol{\'a}s T. and {Espinoza}, N{\'e}stor and {Vines}, Jose I. and {Rojas}, Pablo A. Pe{\~n}a and {Brahm}, Rafael and {Torres}, Pascal and {Cort{\'e}s-Zuleta}, P{\'\i}a and {Soto}, Maritza G. and {Lopez}, Eric D. and {King}, George W. and {Wheatley}, Peter J. and {Winn}, Joshua N. and {Ciardi}, David R. and {Ricker}, George and {Vanderspek}, Roland and {Latham}, David W. and {Seager}, Sara and {Jenkins}, Jon M. and {Beichman}, Charles A. and {Bieryla}, Allyson and {Burke}, Christopher J. and {Christiansen}, Jessie L. and {Henze}, Christopher E. and {Klaus}, Todd C. and {McCauliff}, Sean and {Mori}, Mayuko and {Narita}, Norio and {Nishiumi}, Taku and {Tamura}, Motohide and {de Leon}, Jerome Pitogo and {Quinn}, Samuel N. and {Villase{\~n}or}, Jesus Noel and {Vezie}, Michael and {Lissauer}, Jack J. and {Collins}, Karen A. and {Collins}, Kevin I. and {Isopi}, Giovanni and {Mallia}, Franco and {Ercolino}, Andrea and {Petrovich}, Cristobal and {Jord{\'a}n}, Andr{\'e}s and {Acton}, Jack S. and {Armstrong}, David J. and {Bayliss}, Daniel and {Bouchy}, Fran{\c{c}}ois and {Belardi}, Claudia and {Bryant}, Edward M. and {Burleigh}, Matthew R. and {Cabrera}, Juan and {Casewell}, Sarah L. and {Chaushev}, Alexander and {Cooke}, Benjamin F. and {Eigm{\"u}ller}, Philipp and {Erikson}, Anders and {Foxell}, Emma and {G{\"a}nsicke}, Boris T. and {Gill}, Samuel and {Gillen}, Edward and {G{\"u}nther}, Maximilian N. and {Goad}, Michael R. and {Hooton}, Matthew J. and {Jackman}, James A.~G. and {Louden}, Tom and {McCormac}, James and {Moyano}, Maximiliano and {Nielsen}, Louise D. and {Pollacco}, Don and {Queloz}, Didier and {Rauer}, Heike and {Raynard}, Liam and {Smith}, Alexis M.~S. and {Tilbrook}, Rosanna H. and {Titz-Weider}, Ruth and {Turner}, Oliver and {Udry}, St{\'e}phane and {Walker}, Simon. R. and {Watson}, Christopher A. and {West}, Richard G. and {Palle}, Enric and {Ziegler}, Carl and {Law}, Nicholas and {Mann}, Andrew W.},
        title = "{An ultrahot Neptune in the Neptune desert}",
      journal = {Nat. Astron.},
         year = 2020,
        month = jan,
       volume = {4},
        pages = {1148-1157},
          doi = {10.1038/s41550-020-1142-z},
archivePrefix = {arXiv},
       eprint = {2009.12832},
 primaryClass = {astro-ph.EP},
       adsurl = {https://ui.adsabs.harvard.edu/abs/2020NatAs...4.1148J}
}

@ARTICLE{nabbie2024,
       author = {{Nabbie}, Emma and {Huang}, Chelsea X. and {Burt}, Jennifer A. and {Armstrong}, David J. and {Mamajek}, Eric E. and {Adibekyan}, Vardan and {Sousa}, S{\'e}rgio G. and {Lopez}, Eric D. and {Thorngren}, Daniel and {Fern{\'a}ndez Fern{\'a}ndez}, Jorge and {Li}, Gongjie and {Jenkins}, James S. and {Vines}, Jose I. and {Gomes da Silva}, Jo{\~a}o and {Wittenmyer}, Robert A. and {Bayliss}, Daniel and {Brice{\~n}o}, C{\'e}sar and {Collins}, Karen A. and {Dumusque}, Xavier and {Horne}, Keith and {Keniger}, Marcelo Aron F. and {Law}, Nicholas and {Lillo-Box}, Jorge and {Liu}, Shang-Fei and {Mann}, Andrew W. and {Nielsen}, Louise D. and {Osborn}, Ares and {Relles}, Howard M. and {Rodrigues}, Jos{\'e} J. and {Serrano Bell}, Juan and {Srdoc}, Gregor and {Stockdale}, Chris and {Str{\o}m}, Paul A. and {Watkins}, Cristilyn N. and {Wheatley}, Peter J. and {Wright}, Duncan J. and {Zhou}, George and {Ziegler}, Carl and {Ricker}, George and {Seager}, Sara and {Vanderspek}, Roland and {Winn}, Joshua N. and {Jenkins}, Jon M. and {Fausnaugh}, Michael and {Kunimoto}, Michelle and {Osborn}, Hugh P. and {Quinn}, Samuel N. and {Wohler}, Bill},
        title = "{Surviving in the Hot-Neptune Desert: The Discovery of the Ultrahot Neptune TOI-3261b}",
      journal = {\aj},
         year = 2024,
        month = sep,
       volume = {168},
       number = {3},
          eid = {132},
        pages = {132},
          doi = {10.3847/1538-3881/ad60be},
archivePrefix = {arXiv},
       eprint = {2407.04225},
 primaryClass = {astro-ph.EP},
       adsurl = {https://ui.adsabs.harvard.edu/abs/2024AJ....168..132N}
}

@ARTICLE{damasso2015,
       author = {{Damasso}, M. and {Biazzo}, K. and {Bonomo}, A.~S. and {Desidera}, S. and {Lanza}, A.~F. and {Nascimbeni}, V. and {Esposito}, M. and {Scandariato}, G. and {Sozzetti}, A. and {Cosentino}, R. and {Gratton}, R. and {Malavolta}, L. and {Rainer}, M. and {Gandolfi}, D. and {Poretti}, E. and {Zanmar Sanchez}, R. and {Ribas}, I. and {Santos}, N. and {Affer}, L. and {Andreuzzi}, G. and {Barbieri}, M. and {Bedin}, L.~R. and {Benatti}, S. and {Bernagozzi}, A. and {Bertolini}, E. and {Bonavita}, M. and {Borsa}, F. and {Borsato}, L. and {Boschin}, W. and {Calcidese}, P. and {Carbognani}, A. and {Cenadelli}, D. and {Christille}, J.~M. and {Claudi}, R.~U. and {Covino}, E. and {Cunial}, A. and {Giacobbe}, P. and {Granata}, V. and {Harutyunyan}, A. and {Lattanzi}, M.~G. and {Leto}, G. and {Libralato}, M. and {Lodato}, G. and {Lorenzi}, V. and {Mancini}, L. and {Martinez Fiorenzano}, A.~F. and {Marzari}, F. and {Masiero}, S. and {Micela}, G. and {Molinari}, E. and {Molinaro}, M. and {Munari}, U. and {Murabito}, S. and {Pagano}, I. and {Pedani}, M. and {Piotto}, G. and {Rosenberg}, A. and {Silvotti}, R. and {Southworth}, J.},
        title = "{The GAPS programme with HARPS-N at TNG. V. A comprehensive analysis of the XO-2 stellar and planetary systems}",
      journal = {\aap},
         year = 2015,
        month = mar,
       volume = {575},
          eid = {A111},
        pages = {A111},
          doi = {10.1051/0004-6361/201425332},
archivePrefix = {arXiv},
       eprint = {1501.01424},
 primaryClass = {astro-ph.SR},
       adsurl = {https://ui.adsabs.harvard.edu/abs/2015A&A...575A.111D}
}

@ARTICLE{esposito2017,
       author = {{Esposito}, M. and {Covino}, E. and {Desidera}, S. and {Mancini}, L. and {Nascimbeni}, V. and {Zanmar Sanchez}, R. and {Biazzo}, K. and {Lanza}, A.~F. and {Leto}, G. and {Southworth}, J. and {Bonomo}, A.~S. and {Su{\'a}rez Mascare{\~n}o}, A. and {Boccato}, C. and {Cosentino}, R. and {Claudi}, R.~U. and {Gratton}, R. and {Maggio}, A. and {Micela}, G. and {Molinari}, E. and {Pagano}, I. and {Piotto}, G. and {Poretti}, E. and {Smareglia}, R. and {Sozzetti}, A. and {Affer}, L. and {Anderson}, D.~R. and {Andreuzzi}, G. and {Benatti}, S. and {Bignamini}, A. and {Borsa}, F. and {Borsato}, L. and {Ciceri}, S. and {Damasso}, M. and {di Fabrizio}, L. and {Giacobbe}, P. and {Granata}, V. and {Harutyunyan}, A. and {Henning}, T. and {Malavolta}, L. and {Maldonado}, J. and {Martinez Fiorenzano}, A. and {Masiero}, S. and {Molaro}, P. and {Molinaro}, M. and {Pedani}, M. and {Rainer}, M. and {Scandariato}, G. and {Turner}, O.~D.},
        title = "{The GAPS Programme with HARPS-N at TNG. XIII. The orbital obliquity of three close-in massive planets hosted by dwarf K-type stars: WASP-43, HAT-P-20 and Qatar-2}",
      journal = {\aap},
         year = 2017,
        month = may,
       volume = {601},
          eid = {A53},
        pages = {A53},
          doi = {10.1051/0004-6361/201629720},
archivePrefix = {arXiv},
       eprint = {1702.03136},
 primaryClass = {astro-ph.EP},
       adsurl = {https://ui.adsabs.harvard.edu/abs/2017A&A...601A..53E}
}

@ARTICLE{naponiello2022,
       author = {{Naponiello}, L. and {Mancini}, L. and {Damasso}, M. and {Bonomo}, A.~S. and {Sozzetti}, A. and {Nardiello}, D. and {Biazzo}, K. and {Stognone}, R.~G. and {Lillo-Box}, J. and {Lanza}, A.~F. and {Poretti}, E. and {Lissauer}, J.~J. and {Zeng}, L. and {Bieryla}, A. and {H{\'e}brard}, G. and {Basilicata}, M. and {Benatti}, S. and {Bignamini}, A. and {Borsa}, F. and {Claudi}, R. and {Cosentino}, R. and {Covino}, E. and {de Gurtubai}, A. and {Delfosse}, X. and {Desidera}, S. and {Dragomir}, D. and {Eastman}, J.~D. and {Essack}, Z. and {Fiorenzano}, A.~F.~M. and {Giacobbe}, P. and {Harutyunyan}, A. and {Heidari}, N. and {Hellier}, C. and {Jenkins}, J.~M. and {Knapic}, C. and {K{\"o}nig}, P.-C. and {Latham}, D.~W. and {Magazz{\`u}}, A. and {Maggio}, A. and {Maldonado}, J. and {Micela}, G. and {Molinari}, E. and {Molinaro}, M. and {Morgan}, E.~H. and {Moutou}, C. and {Nascimbeni}, V. and {Pace}, E. and {Pagano}, I. and {Pedani}, M. and {Piotto}, G. and {Pinamonti}, M. and {Quintana}, E.~V. and {Rainer}, M. and {Ricker}, G.~R. and {Seager}, S. and {Twicken}, J.~D. and {Vanderspek}, R. and {Winn}, J.~N.},
        title = "{The GAPS programme at TNG. XL. A puffy and warm Neptune-sized planet and an outer Neptune-mass candidate orbiting the solar-type star TOI-1422}",
      journal = {\aap},
         year = 2022,
        month = nov,
       volume = {667},
          eid = {A8},
        pages = {A8},
          doi = {10.1051/0004-6361/202244079},
archivePrefix = {arXiv},
       eprint = {2207.03293},
 primaryClass = {astro-ph.EP},
       adsurl = {https://ui.adsabs.harvard.edu/abs/2022A&A...667A...8N}
}

@ARTICLE{naponiello2025a,
       author = {{Naponiello}, L. and {Bonomo}, A.~S. and {Mancini}, L. and {Steinmeyer}, M.-L. and {Biazzo}, K. and {Polychroni}, D. and {Dorn}, C. and {Turrini}, D. and {Lanza}, A.~F. and {Sozzetti}, A. and {Desidera}, S. and {Damasso}, M. and {Collins}, K.~A. and {Carleo}, I. and {Collins}, K.~I. and {Colombo}, S. and {D'Arpa}, M.~C. and {Dumusque}, X. and {Gonz{\'a}lez}, M. and {Guilluy}, G. and {Lorenzi}, V. and {Mantovan}, G. and {Nardiello}, D. and {Pinamonti}, M. and {Schwarz}, R.~P. and {Singh}, V. and {Watkins}, C.~N. and {Zingales}, T.},
        title = "{The GAPS programme at TNG: LXIV. An inner eccentric sub-Neptune and an outer sub-Neptune-mass candidate around BD+00 444 (TOI-2443)}",
      journal = {\aap},
         year = 2025,
        month = jan,
       volume = {693},
          eid = {A7},
        pages = {A7},
          doi = {10.1051/0004-6361/202451859},
archivePrefix = {arXiv},
       eprint = {2411.09417},
 primaryClass = {astro-ph.EP},
       adsurl = {https://ui.adsabs.harvard.edu/abs/2025A&A...693A...7N}
}

@ARTICLE{naponiello2026b,
       author = {{Naponiello}, L. and {Leonardi}, P. and {Damasso}, M. and {Steinmeyer}, M.-L. and {Stalport}, M. and {Dorn}, C. and {Bonomo}, A.~S. and {Mancini}, L. and {Sozzetti}, A. and {Benatti}, S. and {Colombo}, S. and {Cosentino}, R.},
        title = "{A 34.6-d transiting sub-Neptune in the TOI-1422 planetary system}",
      journal = {\mnras},
         year = 2026,
        month = jan,
       volume = {545},
       number = {2},
          eid = {staf2030},
        pages = {staf2030},
          doi = {10.1093/mnras/staf2030},
archivePrefix = {arXiv},
       eprint = {2511.11492},
 primaryClass = {astro-ph.EP},
       adsurl = {https://ui.adsabs.harvard.edu/abs/2026MNRAS.545f2030N}
}

@ARTICLE{naponiello2025b,
       author = {{Naponiello}, L. and {Vissapragada}, S. and {Bonomo}, A.~S. and {Steinmeyer}, M.-L. and {Filomeno}, S. and {D'Orazi}, V. and {Dorn}, C. and {Sozzetti}, A. and {Mancini}, L. and {Lanza}, A.~F. and {Biazzo}, K. and {Watkins}, C.~N. and {H{\'e}brard}, G. and {Lissauer}, J.~J. and {Howell}, S.~B. and {Ciardi}, D.~R. and {Mantovan}, G. and {Baker}, D. and {Bourrier}, V. and {Buchhave}, L.~A. and {Clark}, C.~A. and {Collins}, K.~A. and {Cosentino}, R. and {Damasso}, M. and {Dumusque}, X. and {Fiorenzano}, A. and {Forveille}, T. and {Heidari}, N. and {Latham}, D.~W. and {Littlefield}, C. and {L{\'o}pez-Morales}, M. and {Lund}, M.~B. and {Malavolta}, L. and {Manni}, F. and {Nardiello}, D. and {Pinamonti}, M. and {Yee}, S.~W. and {Zambelli}, R. and {Ziegler}, C. and {Zingales}, T.},
        title = "{The Hot-Neptune Initiative (HONEI): I. Two hot sub-Neptunes on a close-in eccentric orbit (TOI-5800 b) and a farther-out circular orbit (TOI-5817 b)}",
      journal = {\aap},
         year = 2025,
        month = sep,
       volume = {701},
          eid = {A79},
        pages = {A79},
          doi = {10.1051/0004-6361/202555523},
archivePrefix = {arXiv},
       eprint = {2505.10123},
 primaryClass = {astro-ph.EP},
       adsurl = {https://ui.adsabs.harvard.edu/abs/2025A&A...701A..79N}
}

@ARTICLE{manni2025,
       author = {{Manni}, F. and {Naponiello}, L. and {Mancini}, L. and {Vissapragada}, S. and {Biazzo}, K. and {Bonomo}, A.~S. and {Polychroni}, D. and {Turrini}, D. and {Locci}, D. and {Maggio}, A. and {D'Orazi}, V. and {Damasso}, M. and {Brice{\~n}o}, C. and {Ciardi}, D.~R. and {Clark}, C.~A. and {Collins}, K.~A. and {Latham}, D.~W. and {Law}, N. and {L{\'o}pez-Morales}, M. and {Lund}, M.~B. and {Malavolta}, L. and {Mann}, A.~W. and {Mantovan}, G. and {Nardiello}, D. and {Pinamonti}, M. and {Radford}, D.~J. and {Schwarz}, R.~P. and {Shporer}, A. and {Sozzetti}, A. and {Watkins}, C.~N. and {Yee}, S.~W. and {Ziegler}, C. and {Zingales}, T.},
        title = "{The Hot Neptune Initiative (HONEI): II. TOI-5795 b: A hot super-Neptune orbiting a metal-poor star}",
      journal = {\aap},
         year = 2025,
        month = sep,
       volume = {701},
          eid = {A230},
        pages = {A230},
          doi = {10.1051/0004-6361/202556081},
archivePrefix = {arXiv},
       eprint = {2507.23413},
 primaryClass = {astro-ph.EP},
       adsurl = {https://ui.adsabs.harvard.edu/abs/2025A&A...701A.230M}
}

@ARTICLE{ward1997,
       author = {{Ward}, William R.},
        title = "{Protoplanet Migration by Nebula Tides}",
      journal = {\icarus},
         year = 1997,
        month = apr,
       volume = {126},
       number = {2},
        pages = {261-281},
          doi = {10.1006/icar.1996.5647},
       adsurl = {https://ui.adsabs.harvard.edu/abs/1997Icar..126..261W}
}

@ARTICLE{kley2000,
       author = {{Kley}, W.},
        title = "{On the migration of a system of protoplanets}",
      journal = {\mnras},
         year = 2000,
        month = apr,
       volume = {313},
       number = {4},
        pages = {L47-L51},
          doi = {10.1046/j.1365-8711.2000.03495.x},
archivePrefix = {arXiv},
       eprint = {astro-ph/9910155},
 primaryClass = {astro-ph},
       adsurl = {https://ui.adsabs.harvard.edu/abs/2000MNRAS.313L..47K}
}

@ARTICLE{masset2001,
       author = {{Masset}, F. and {Snellgrove}, M.},
        title = "{Reversing type II migration: resonance trapping of a lighter giant protoplanet}",
      journal = {\mnras},
         year = 2001,
        month = feb,
       volume = {320},
       number = {4},
        pages = {L55-L59},
          doi = {10.1046/j.1365-8711.2001.04159.x},
archivePrefix = {arXiv},
       eprint = {astro-ph/0003421},
 primaryClass = {astro-ph},
       adsurl = {https://ui.adsabs.harvard.edu/abs/2001MNRAS.320L..55M}
}

@ARTICLE{kuchner2002,
       author = {{Kuchner}, Marc J. and {Lecar}, Myron},
        title = "{Halting Planet Migration in the Evacuated Centers of Protoplanetary Disks}",
      journal = {\apjl},
         year = 2002,
        month = jul,
       volume = {574},
       number = {1},
        pages = {L87-L89},
          doi = {10.1086/342370},
archivePrefix = {arXiv},
       eprint = {astro-ph/0206232},
 primaryClass = {astro-ph},
       adsurl = {https://ui.adsabs.harvard.edu/abs/2002ApJ...574L..87K}
}

@ARTICLE{wu2003,
       author = {{Wu}, Y. and {Murray}, N.},
        title = "{Planet Migration and Binary Companions: The Case of HD 80606b}",
      journal = {\apj},
         year = 2003,
        month = may,
       volume = {589},
       number = {1},
        pages = {605-614},
          doi = {10.1086/374598},
archivePrefix = {arXiv},
       eprint = {astro-ph/0303010},
 primaryClass = {astro-ph},
       adsurl = {https://ui.adsabs.harvard.edu/abs/2003ApJ...589..605W}
}

@ARTICLE{fabrycky2007,
       author = {{Fabrycky}, Daniel and {Tremaine}, Scott},
        title = "{Shrinking Binary and Planetary Orbits by Kozai Cycles with Tidal Friction}",
      journal = {\apj},
         year = 2007,
        month = nov,
       volume = {669},
       number = {2},
        pages = {1298-1315},
          doi = {10.1086/521702},
archivePrefix = {arXiv},
       eprint = {0705.4285},
 primaryClass = {astro-ph},
       adsurl = {https://ui.adsabs.harvard.edu/abs/2007ApJ...669.1298F}
}

@ARTICLE{mardling2007,
       author = {{Mardling}, Rosemary A.},
        title = "{Long-term tidal evolution of short-period planets with companions}",
      journal = {\mnras},
         year = 2007,
        month = dec,
       volume = {382},
       number = {4},
        pages = {1768-1790},
          doi = {10.1111/j.1365-2966.2007.12500.x},
archivePrefix = {arXiv},
       eprint = {0706.0224},
 primaryClass = {astro-ph},
       adsurl = {https://ui.adsabs.harvard.edu/abs/2007MNRAS.382.1768M}
}

@ARTICLE{rasio1996,
       author = {{Rasio}, Frederic A. and {Ford}, Eric B.},
        title = "{Dynamical instabilities and the formation of extrasolar planetary systems}",
      journal = {Science},
         year = 1996,
        month = nov,
       volume = {274},
        pages = {954-956},
          doi = {10.1126/science.274.5289.954},
       adsurl = {https://ui.adsabs.harvard.edu/abs/1996Sci...274..954R}
}

@ARTICLE{chatterjee2008,
       author = {{Chatterjee}, Sourav and {Ford}, Eric B. and {Matsumura}, Soko and {Rasio}, Frederic A.},
        title = "{Dynamical Outcomes of Planet-Planet Scattering}",
      journal = {\apj},
         year = 2008,
        month = oct,
       volume = {686},
       number = {1},
        pages = {580-602},
          doi = {10.1086/590227},
archivePrefix = {arXiv},
       eprint = {astro-ph/0703166},
 primaryClass = {astro-ph},
       adsurl = {https://ui.adsabs.harvard.edu/abs/2008ApJ...686..580C}
}

@ARTICLE{piaulet2021P,
       author = {{Piaulet}, Caroline and {Benneke}, Bj{\"o}rn and {Rubenzahl}, Ryan A. and {Howard}, Andrew W. and {Lee}, Eve J. and {Thorngren}, Daniel and {Angus}, Ruth and {Peterson}, Merrin and {Schlieder}, Joshua E. and {Werner}, Michael and {Kreidberg}, Laura and {Jaouni}, Tareq and {Crossfield}, Ian J.~M. and {Ciardi}, David R. and {Petigura}, Erik A. and {Livingston}, John and {Dressing}, Courtney D. and {Fulton}, Benjamin J. and {Beichman}, Charles and {Christiansen}, Jessie L. and {Gorjian}, Varoujan and {Hardegree-Ullman}, Kevin K. and {Krick}, Jessica and {Sinukoff}, Evan},
        title = "{WASP-107b's Density Is Even Lower: A Case Study for the Physics of Planetary Gas Envelope Accretion and Orbital Migration}",
      journal = {\aj},
         year = 2021,
        month = feb,
       volume = {161},
       number = {2},
          eid = {70},
        pages = {70},
          doi = {10.3847/1538-3881/abcd3c},
archivePrefix = {arXiv},
       eprint = {2011.13444},
 primaryClass = {astro-ph.EP},
       adsurl = {https://ui.adsabs.harvard.edu/abs/2021AJ....161...70P}
}

@article{hormuth08,
    adsurl = {http://adsabs.harvard.edu/abs/2008JPhCS.131a2051H},
    archiveprefix = {arXiv},
    author = {{Hormuth}, F. and {Brandner}, W. and {Hippler}, S. and {Henning}, T.},
    doi = {10.1088/1742-6596/131/1/012051},
    eprint = {0807.0504},
    journal = {JPCS},
    month = oct,
    number = 1,
    pages = {012051},
    title = {{AstraLux - the Calar Alto 2.2-m telescope Lucky Imaging Camera}},
    volume = 131,
    year = 2008}

@article{Lillo-Box12,
    adsurl = {http://ads.nao.ac.jp/abs/2012A%26A...546A..10L},
    archiveprefix = {arXiv},
    author = {{Lillo-Box}, J. and {Barrado}, D. and {Bouy}, H.},
    doi = {10.1051/0004-6361/201219631},
    eid = {A10},
    eprint = {1208.0242},
    journal = {\aap},
    month = oct,
    pages = {A10},
    primaryclass = {astro-ph.EP},
    title = {{Multiplicity in transiting planet-host stars. A lucky imaging study of Kepler candidates}},
    volume = 546,
    year = 2012}

@article{lillo-box14,
    adsurl = {http://adsabs.harvard.edu/abs/2014A%26A...566A.103L},
    archiveprefix = {arXiv},
    author = {{Lillo-Box}, J. and {Barrado}, D. and {Bouy}, H.},
    doi = {10.1051/0004-6361/201423497},
    eid = {A103},
    eprint = {1405.3120},
    journal = {\aap},
    month = jun,
    pages = {A103},
    primaryclass = {astro-ph.EP},
    title = {{High-resolution imaging of Kepler planet host candidates. A comprehensive comparison of different techniques}},
    volume = 566,
    year = 2014}

@article{lillo-box24,
    adsurl = {https://ui.adsabs.harvard.edu/abs/2024A&A...686A.232L},
    archiveprefix = {arXiv},
    author = {{Lillo-Box}, J. and {Morales-Calder{\'o}n}, M. and {Barrado}, D. and {Balsalobre-Ruza}, O. and {Castro-Gonz{\'a}lez}, A. and {Mendigut{\'\i}a}, I. and {Hu{\'e}lamo}, N. and {Montesinos}, B. and {Vioque}, M.},
    doi = {10.1051/0004-6361/202449687},
    eid = {A232},
    eprint = {2404.06316},
    journal = {\aap},
    month = jun,
    pages = {A232},
    primaryclass = {astro-ph.EP},
    title = {{The AstraLux-TESS high spatial resolution imaging survey. Search for stellar companions of 215 planet candidates from TESS}},
    volume = {686},
    year = 2024}

@ARTICLE{butters2010,
       author = {{Butters}, O.~W. and {West}, R.~G. and {Anderson}, D.~R. and {Collier Cameron}, A. and {Clarkson}, W.~I. and {Enoch}, B. and {Haswell}, C.~A. and {Hellier}, C. and {Horne}, K. and {Joshi}, Y. and {Kane}, S.~R. and {Lister}, T.~A. and {Maxted}, P.~F.~L. and {Parley}, N. and {Pollacco}, D. and {Smalley}, B. and {Street}, R.~A. and {Todd}, I. and {Wheatley}, P.~J. and {Wilson}, D.~M.},
        title = "{The first WASP public data release}",
      journal = {\aap},
         year = 2010,
        month = sep,
       volume = {520},
          eid = {L10},
        pages = {L10},
          doi = {10.1051/0004-6361/201015655},
archivePrefix = {arXiv},
       eprint = {1009.5306},
 primaryClass = {astro-ph.EP},
       adsurl = {https://ui.adsabs.harvard.edu/abs/2010A&A...520L..10B}
}

@ARTICLE{wizinowich2000,
       author = {{Wizinowich}, P. and {Acton}, D.~S. and {Shelton}, C. and {Stomski}, P. and {Gathright}, J. and {Ho}, K. and {Lupton}, W. and {Tsubota}, K. and {Lai}, O. and {Max}, C. and {Brase}, J. and {An}, J. and {Avicola}, K. and {Olivier}, S. and {Gavel}, D. and {Macintosh}, B. and {Ghez}, A. and {Larkin}, J.},
        title = "{First Light Adaptive Optics Images from the Keck II Telescope: A New Era of High Angular Resolution Imagery}",
      journal = {\pasp},
         year = 2000,
        month = mar,
       volume = {112},
       number = {769},
        pages = {315-319},
          doi = {10.1086/316543},
       adsurl = {https://ui.adsabs.harvard.edu/abs/2000PASP..112..315W}
}

@ARTICLE{scott2021,
       author = {{Scott}, Nicholas J. and {Howell}, Steve B. and {Gnilka}, Crystal L. and {Stephens}, Andrew W. and {Salinas}, Ricardo and {Matson}, Rachel A. and {Furlan}, Elise and {Horch}, Elliott P. and {Everett}, Mark E. and {Ciardi}, David R. and {Mills}, Dave and {Quigley}, Emmett A.},
        title = "{Twin High-resolution, High-speed Imagers for the Gemini Telescopes: Instrument description and science verification results}",
      journal = {FSPAS},
         year = 2021,
        month = sep,
       volume = {8},
          eid = {138},
        pages = {138},
          doi = {10.3389/fspas.2021.716560},
       adsurl = {https://ui.adsabs.harvard.edu/abs/2021FrASS...8..138S}
}

@ARTICLE{pecaut2013,
       author = {{Pecaut}, Mark J. and {Mamajek}, Eric E.},
        title = "{Intrinsic Colors, Temperatures, and Bolometric Corrections of Pre-main-sequence Stars}",
      journal = {\apjs},
         year = 2013,
        month = sep,
       volume = {208},
       number = {1},
          eid = {9},
        pages = {9},
          doi = {10.1088/0067-0049/208/1/9},
archivePrefix = {arXiv},
       eprint = {1307.2657},
 primaryClass = {astro-ph.SR},
       adsurl = {https://ui.adsabs.harvard.edu/abs/2013ApJS..208....9P}
}

@ARTICLE{evans2016,
       author = {{Evans}, D.~F. and {Southworth}, J. and {Maxted}, P.~F.~L. and {Skottfelt}, J. and {Hundertmark}, M. and {J{\o}rgensen}, U.~G. and {Dominik}, M. and {Alsubai}, K.~A. and {Andersen}, M.~I. and {Bozza}, V. and {Bramich}, D.~M. and {Burgdorf}, M.~J. and {Ciceri}, S. and {D'Ago}, G. and {Figuera Jaimes}, R. and {Gu}, S.-H. and {Haugb{\o}lle}, T. and {Hinse}, T.~C. and {Juncher}, D. and {Kains}, N. and {Kerins}, E. and {Korhonen}, H. and {Kuffmeier}, M. and {Mancini}, L. and {Peixinho}, N. and {Popovas}, A. and {Rabus}, M. and {Rahvar}, S. and {Schmidt}, R.~W. and {Snodgrass}, C. and {Starkey}, D. and {Surdej}, J. and {Tronsgaard}, R. and {von Essen}, C. and {Wang}, Yi-Bo and {Wertz}, O.},
        title = "{High-resolution Imaging of Transiting Extrasolar Planetary systems (HITEP). I. Lucky imaging observations of 101 systems in the southern hemisphere}",
      journal = {\aap},
         year = 2016,
        month = may,
       volume = {589},
          eid = {A58},
        pages = {A58},
          doi = {10.1051/0004-6361/201527970},
archivePrefix = {arXiv},
       eprint = {1603.03274},
 primaryClass = {astro-ph.EP},
       adsurl = {https://ui.adsabs.harvard.edu/abs/2016A&A...589A..58E}
}

@dataset{henden2016,
       author = {{Henden}, A.~A. and {Templeton}, M. and {Terrell}, D. and {Smith}, T.~C. and {Levine}, S. and {Welch}, D.},
        title = "{VizieR Online Data Catalog: AAVSO Photometric All Sky Survey (APASS) DR9 (Henden+, 2016)}",
 howpublished = {VizieR On-line Data Catalog: II/336.  Originally published in: 2015AAS...22533616H},
         year = 2016,
        month = jan,
          eid = {II/336},
       adsurl = {https://ui.adsabs.harvard.edu/abs/2016yCat.2336....0H}
}

@dataset{cutri2021,
       author = {{Cutri}, R.~M. and {Wright}, E.~L. and {Conrow}, T. and {Fowler}, J.~W. and {Eisenhardt}, P.~R.~M. and {Grillmair}, C. and {Kirkpatrick}, J.~D. and {Masci}, F. and {McCallon}, H.~L. and {Wheelock}, S.~L. and {Fajardo-Acosta}, S. and {Yan}, L. and {Benford}, D. and {Harbut}, M. and {Jarrett}, T. and {Lake}, S. and {Leisawitz}, D. and {Ressler}, M.~E. and {Stanford}, S.~A. and {Tsai}, C.-W. and {Liu}, F. and {Helou}, G. and {Mainzer}, A. and {Gettngs}, D. and {Gonzalez}, A. and {Hoffman}, D. and {Marsh}, K.~A. and {Padgett}, D. and {Skrutskie}, M.~F. and {Beck}, R. and {Papin}, M. and {Wittman}, M.},
        title = "{VizieR Online Data Catalog: AllWISE Data Release (Cutri+ 2013)}",
 howpublished = {VizieR On-line Data Catalog: II/328.  Originally published in: IPAC/Caltech (2013)},
         year = 2021,
        month = feb,
          eid = {II/328},
       adsurl = {https://ui.adsabs.harvard.edu/abs/2014yCat.2328....0C}
}

@ARTICLE{luhn2020,
       author = {{Luhn}, Jacob K. and {Wright}, Jason T. and {Howard}, Andrew W. and {Isaacson}, Howard},
        title = "{Astrophysical Insights into Radial Velocity Jitter from an Analysis of 600 Planet-search Stars}",
      journal = {\aj},
         year = 2020,
        month = may,
       volume = {159},
       number = {5},
          eid = {235},
        pages = {235},
          doi = {10.3847/1538-3881/ab855a},
archivePrefix = {arXiv},
       eprint = {2004.13734},
 primaryClass = {astro-ph.EP},
       adsurl = {https://ui.adsabs.harvard.edu/abs/2020AJ....159..235L}
}

@ARTICLE{boisse2011,
       author = {{Boisse}, I. and {Bouchy}, F. and {H{\'e}brard}, G. and {Bonfils}, X. and {Santos}, N. and {Vauclair}, S.},
        title = "{Disentangling between stellar activity and planetary signals}",
      journal = {\aap},
         year = 2011,
        month = apr,
       volume = {528},
          eid = {A4},
        pages = {A4},
          doi = {10.1051/0004-6361/201014354},
       adsurl = {https://ui.adsabs.harvard.edu/abs/2011A&A...528A...4B}
}

@ARTICLE{suarez2017,
       author = {{Su{\'a}rez Mascare{\~n}o}, A. and {Rebolo}, R. and {Gonz{\'a}lez Hern{\'a}ndez}, J.~I. and {Esposito}, M.},
        title = "{Characterization of the radial velocity signal induced by rotation in late-type dwarfs}",
      journal = {\mnras},
         year = 2017,
        month = jul,
       volume = {468},
       number = {4},
        pages = {4772-4781},
          doi = {10.1093/mnras/stx771},
archivePrefix = {arXiv},
       eprint = {1703.08884},
 primaryClass = {astro-ph.EP},
       adsurl = {https://ui.adsabs.harvard.edu/abs/2017MNRAS.468.4772S}
}

@ARTICLE{desidera2023,
       author = {{Desidera}, S. and {Damasso}, M. and {Gratton}, R. and {Benatti}, S. and {Nardiello}, D. and {D'Orazi}, V. and {Lanza}, A.~F. and {Locci}, D. and {Marzari}, F. and {Mesa}, D. and {Messina}, S. and {Pillitteri}, I. and {Sozzetti}, A. and {Girard}, J. and {Maggio}, A. and {Micela}, G. and {Malavolta}, L. and {Nascimbeni}, V. and {Pinamonti}, M. and {Squicciarini}, V. and {Alcal{\'a}}, J. and {Biazzo}, K. and {Bohn}, A. and {Bonavita}, M. and {Brooks}, K. and {Chauvin}, G. and {Covino}, E. and {Delorme}, P. and {Hagelberg}, J. and {Janson}, M. and {Lagrange}, A.-M. and {Lazzoni}, C.},
        title = "{TOI-179: A young system with a transiting compact Neptune-mass planet and a low-mass companion in outer orbit}",
      journal = {\aap},
         year = 2023,
        month = jul,
       volume = {675},
          eid = {A158},
        pages = {A158},
          doi = {10.1051/0004-6361/202244611},
archivePrefix = {arXiv},
       eprint = {2210.07933},
 primaryClass = {astro-ph.EP},
       adsurl = {https://ui.adsabs.harvard.edu/abs/2023A&A...675A.158D}
}

@ARTICLE{sozzetti2024,
       author = {{Sozzetti}, A. and {Damasso}, M. and {Fern{\'a}ndez Fern{\'a}ndez}, J. and {Mortier}, A. and {John}, A. Anna and {Cubillos}, P.~E. and {Wilson}, T.~G. and {Pinamonti}, M. and {Nielsen}, L. and {Bonomo}, A.~S. and {Freckelton}, A.~V. and {Cameron}, A. Collier and {Armstrong}, D. and {Vanderburg}, A. and {Bayliss}, D. and {Dumusque}, X. and {Ghedina}, A. and {Keniger}, M.~A.~F. and {Latham}, D.~W. and {L{\'o}pez Morales}, M. and {Malavolta}, L. and {Osborn}, A. and {Pepe}, F. and {Rabino}, R. and {Str{\o}m}, P.~A. and {Udry}, S. and {Wheatley}, P.},
        title = "{K2-370 b: a strongly irradiated sub-Neptune transiting a very active solar-type star}",
      journal = {\mnras},
         year = 2024,
        month = nov,
       volume = {535},
       number = {1},
        pages = {531-550},
          doi = {10.1093/mnras/stae2323},
       adsurl = {https://ui.adsabs.harvard.edu/abs/2024MNRAS.535..531S}
}

@ARTICLE{skrutskie2006,
       author = {{Skrutskie}, M.~F. and {Cutri}, R.~M. and {Stiening}, R. and {Weinberg}, M.~D. and {Schneider}, S. and {Carpenter}, J.~M. and {Beichman}, C. and {Capps}, R. and {Chester}, T. and {Elias}, J. and {Huchra}, J. and {Liebert}, J. and {Lonsdale}, C. and {Monet}, D.~G. and {Price}, S. and {Seitzer}, P. and {Jarrett}, T. and {Kirkpatrick}, J.~D. and {Gizis}, J.~E. and {Howard}, E. and {Evans}, T. and {Fowler}, J. and {Fullmer}, L. and {Hurt}, R. and {Light}, R. and {Kopan}, E.~L. and {Marsh}, K.~A. and {McCallon}, H.~L. and {Tam}, R. and {Van Dyk}, S. and {Wheelock}, S.},
        title = "{The Two Micron All Sky Survey (2MASS)}",
      journal = {\aj},
         year = 2006,
        month = feb,
       volume = {131},
       number = {2},
        pages = {1163-1183},
          doi = {10.1086/498708},
       adsurl = {https://ui.adsabs.harvard.edu/abs/2006AJ....131.1163S}
}

@ARTICLE{vanzandt2025,
       author = {{Van Zandt}, Judah and {Petigura}, Erik A. and {Lubin}, Jack and {Weiss}, Lauren M. and {Turtelboom}, Emma V. and {Fetherolf}, Tara and {Murphy}, Joseph M. Akana and {Crossfield}, Ian J.~M. and {Gilbert}, Gregory J. and {Mo{\v{c}}nik}, Teo and {Batalha}, Natalie M. and {Dressing}, Courtney and {Fulton}, Benjamin and {Howard}, Andrew W. and {Huber}, Daniel and {Isaacson}, Howard and {Kane}, Stephen R. and {Robertson}, Paul and {Roy}, Arpita and {Angelo}, Isabel and {Behmard}, Aida and {Beard}, Corey and {Chontos}, Ashley and {Dai}, Fei and {Giacalone}, Steven and {Hill}, Michelle L. and {Holcomb}, Rae and {Howell}, Steve B. and {Mayo}, Andrew W. and {Pidhorodetska}, Daria and {Polanski}, Alex S. and {Rogers}, James and {Rosenthal}, Lee J. and {Rubenzahl}, Ryan A. and {Scarsdale}, Nicholas and {Tyler}, Dakotah and {Yee}, Samuel W. and {Zink}, Jon},
        title = "{The TESS─Keck Survey. XXIV. Outer Giants May Be More Prevalent in the Presence of Inner Small Planets}",
      journal = {\aj},
         year = 2025,
        month = may,
       volume = {169},
       number = {5},
          eid = {235},
        pages = {235},
          doi = {10.3847/1538-3881/adbbed},
archivePrefix = {arXiv},
       eprint = {2501.06342},
 primaryClass = {astro-ph.EP},
       adsurl = {https://ui.adsabs.harvard.edu/abs/2025AJ....169..235V}
}

\begin{appendix}

\onecolumn

\section{Additional tables}
Tables~\ref{tab:RV_TOI-1272} and \ref{tab:RV_TOI-1694} contain our RV measurements of TOI-1272 and TOI-1694, respectively, obtained with HARPS-N (this work) and the corresponding activity indices, including H$\alpha$ and $\log{R^{\prime}HK}$ indices. FWHM, `Contrast', and BIS are the full width at half maximum, contrast, and bisector span of the cross-correlation function, respectively.
Table~\ref{tab:rvgphires} contains priors and best-fitting values for the GP regression analysis of the HIRES-RV dataset.

\tiny{
\begin{longtable}{crccccccc}
\caption{TOI-1272 HARPS-N RV data points and activity indices.}
\label{tab:RV_TOI-1272} \\
\hline\hline \\ [-8pt] %
BJD$_{\rm TDB}$ & $T_{\rm exp}$ & RV & S/N & FWHM & Contrast & BIS & H$\alpha$ & $\log{R^{\prime}_{\rm HK}}$  \\
$-2458000$ & (s) & (m\,s$^{-1}$) & & (m\,s$^{-1}$) & (m\,s$^{-1}$) &  (m\,s$^{-1}$)\\ [2 pt]
\hline \\ [-8pt] %
\endfirsthead
\caption{continued.}\\
\hline \\ [-8pt] %
BJD$_{\rm TDB}$ & $T_{\rm exp}$ & RV & S/N & FWHM & Contrast & BIS & H$\alpha$ & $\log{R^{\prime}_{\rm HK}}$  \\
$-2458000$ & (s) & (m\,s$^{-1}$) & & (m\,s$^{-1}$) & (m\,s$^{-1}$) &  (m\,s$^{-1}$)\\ [2 pt]
\hline \\ [-8pt] %
\endhead
\hline
\endfoot
 986.58852 & 900  & $1907.55 \pm 3.92$ & 25.13 & $6.7845 \pm 0.0078$ & $65.956 \pm 0.076$ & $-0.0405 \pm 0.0078$ & $0.2755 \pm 0.0007$ & $-4.9166 \pm 0.0087$ \\
 988.52652 & 900  & $1903.66 \pm 4.32$ & 22.81 & $6.8066 \pm 0.0086$ & $65.754 \pm 0.083$ & $-0.0419 \pm 0.0086$ & $0.2745 \pm 0.0008$ & $-5.0165 \pm 0.0120$ \\
 989.52656 & 900  & $1913.12 \pm 3.86$ & 25.64 & $6.8117 \pm 0.0077$ & $65.593 \pm 0.074$ & $-0.0383 \pm 0.0077$ & $0.2802 \pm 0.0009$ & $-4.9712 \pm 0.0088$ \\
 990.49060 & 900  & $1922.10 \pm 3.00$ & 30.46 & $6.8253 \pm 0.0060$ & $65.618 \pm 0.058$ & $-0.0110 \pm 0.0060$ & $0.2972 \pm 0.0010$ & $-4.9535 \pm 0.0057$ \\
 993.38802 & 900  & $1908.39 \pm 4.20$ & 23.79 & $6.8332 \pm 0.0084$ & $65.656 \pm 0.081$ & $-0.0315 \pm 0.0084$ & $0.2947 \pm 0.0010$ & $-4.9138 \pm 0.0083$ \\
1007.46485 & 900  & $1907.81 \pm 2.18$ & 39.43 & $6.8046 \pm 0.0044$ & $65.609 \pm 0.042$ & $-0.0171 \pm 0.0044$ & $0.2919 \pm 0.0006$ & $-4.9616 \pm 0.0035$ \\
1008.49555 & 900  & $1876.69 \pm 2.15$ & 39.36 & $6.7923 \pm 0.0043$ & $65.747 \pm 0.042$ & $-0.0204 \pm 0.0043$ & $0.2912 \pm 0.0007$ & $-4.9528 \pm 0.0035$ \\
1019.40396 & 900  & $1894.33 \pm 2.84$ & 32.03 & $6.8196 \pm 0.0057$ & $65.765 \pm 0.055$ & $-0.0298 \pm 0.0057$ & $0.2877 \pm 0.0008$ & $-4.9336 \pm 0.0050$ \\
1025.38779 & 900  & $1892.93 \pm 3.14$ & 29.64 & $6.7916 \pm 0.0063$ & $65.692 \pm 0.061$ & $-0.0187 \pm 0.0063$ & $0.2749 \pm 0.0009$ & $-4.9563 \pm 0.0061$ \\
1026.43727 & 900  & $1912.43 \pm 4.29$ & 23.43 & $6.7863 \pm 0.0086$ & $65.837 \pm 0.083$ & $-0.0198 \pm 0.0086$ & $0.2820 \pm 0.0010$ & $-4.9071 \pm 0.0087$ \\
1027.43564 & 900  & $1921.43 \pm 3.54$ & 27.05 & $6.8035 \pm 0.0071$ & $65.822 \pm 0.068$ & $-0.0353 \pm 0.0071$ & $0.2929 \pm 0.0010$ & $-4.9916 \pm 0.0080$ \\
1028.48198 & 900  & $1899.66 \pm 3.19$ & 29.89 & $6.8016 \pm 0.0064$ & $65.812 \pm 0.062$ & $-0.0441 \pm 0.0064$ & $0.2993 \pm 0.0010$ & $-4.9469 \pm 0.0065$ \\
1029.37983 & 900  & $1909.88 \pm 2.98$ & 31.15 & $6.7973 \pm 0.0060$ & $65.042 \pm 0.057$ & $-0.0435 \pm 0.0060$ & $0.2904 \pm 0.0008$ & $-4.9270 \pm 0.0052$ \\
1037.45355 & 900  & $1925.74 \pm 7.74$ & 15.42 & $6.7708 \pm 0.0150$ & $65.385 \pm 0.150$ & $-0.0390 \pm 0.0150$ & $0.2736 \pm 0.0020$ & $-5.1651 \pm 0.0320$ \\
1039.46048 & 900  & $1924.88 \pm 3.92$ & 25.49 & $6.8238 \pm 0.0078$ & $65.750 \pm 0.076$ & $-0.0582 \pm 0.0078$ & $0.2897 \pm 0.0009$ & $-4.9484 \pm 0.0084$ \\
1040.45973 & 900  & $1938.71 \pm 6.18$ & 17.88 & $6.8280 \pm 0.0120$ & $65.631 \pm 0.120$ & $-0.0491 \pm 0.0120$ & $0.2805 \pm 0.0010$ & $-4.9160 \pm 0.0140$ \\
1051.42514 & 900  & $1884.17 \pm 2.70$ & 33.84 & $6.8095 \pm 0.0054$ & $65.811 \pm 0.052$ & $-0.0163 \pm 0.0054$ & $0.2868 \pm 0.0006$ & $-4.9584 \pm 0.0052$ \\
1059.41986 & 900  & $1901.35 \pm 4.90$ & 21.26 & $6.8094 \pm 0.0098$ & $65.360 \pm 0.094$ & $-0.0176 \pm 0.0098$ & $0.2701 \pm 0.0009$ & $-4.9867 \pm 0.0130$ \\
1189.77035 & 900  & $1915.84 \pm 4.25$ & 24.39 & $6.8178 \pm 0.0085$ & $65.691 \pm 0.082$ & $-0.0121 \pm 0.0085$ & $0.2973 \pm 0.0010$ & $-4.9187 \pm 0.0089$ \\
1190.77743 & 900  & $1882.57 \pm 5.14$ & 20.58 & $6.8346 \pm 0.0100$ & $65.412 \pm 0.098$ & $-0.0295 \pm 0.0100$ & $0.3012 \pm 0.0020$ & $-4.8998 \pm 0.0100$ \\
1212.68852 & 900  & $1912.04 \pm 6.57$ & 17.37 & $6.8266 \pm 0.0130$ & $64.908 \pm 0.120$ & $-0.0640 \pm 0.0130$ & $0.2916 \pm 0.0010$ & $-4.9318 \pm 0.0160$ \\
1215.79989 & 900  & $1908.74 \pm 2.94$ & 31.81 & $6.8255 \pm 0.0059$ & $65.179 \pm 0.056$ & $-0.0173 \pm 0.0059$ & $0.3061 \pm 0.0008$ & $-4.9297 \pm 0.0050$ \\
1235.72475 & 900  & $1901.37 \pm 3.14$ & 30.42 & $6.8070 \pm 0.0063$ & $65.717 \pm 0.061$ & $-0.0352 \pm 0.0063$ & $0.2890 \pm 0.0008$ & $-4.9012 \pm 0.0054$ \\
1236.71988 & 900  & $1903.29 \pm 3.35$ & 29.03 & $6.8044 \pm 0.0067$ & $65.726 \pm 0.065$ & $-0.0184 \pm 0.0067$ & $0.2891 \pm 0.0009$ & $-4.9077 \pm 0.0061$ \\
1237.66880 & 1200 & $1891.82 \pm 3.42$ & 28.56 & $6.8025 \pm 0.0068$ & $65.706 \pm 0.066$ & $-0.0327 \pm 0.0068$ & $0.2823 \pm 0.0008$ & $-4.9673 \pm 0.0070$ \\
1244.73262 & 900  & $1895.97 \pm 3.46$ & 28.02 & $6.8241 \pm 0.0069$ & $65.220 \pm 0.066$ & $-0.0235 \pm 0.0069$ & $0.2943 \pm 0.0010$ & $-4.8912 \pm 0.0058$ \\
1245.69744 & 1200 & $1908.76 \pm 4.06$ & 25.04 & $6.8423 \pm 0.0081$ & $65.027 \pm 0.077$ & $-0.0303 \pm 0.0081$ & $0.2912 \pm 0.0010$ & $-4.9083 \pm 0.0077$ \\
1246.76659 & 900  & $1905.89 \pm 3.73$ & 26.77 & $6.8140 \pm 0.0075$ & $65.435 \pm 0.072$ & $-0.0171 \pm 0.0075$ & $0.2923 \pm 0.0009$ & $-4.9267 \pm 0.0073$ \\
1253.75308 & 900  & $1882.98 \pm 2.61$ & 35.06 & $6.8020 \pm 0.0052$ & $65.750 \pm 0.050$ & $-0.0293 \pm 0.0052$ & $0.2858 \pm 0.0007$ & $-4.9256 \pm 0.0044$ \\
1254.76150 & 900  & $1891.54 \pm 2.96$ & 31.94 & $6.7983 \pm 0.0059$ & $65.682 \pm 0.057$ & $-0.0313 \pm 0.0059$ & $0.2870 \pm 0.0007$ & $-4.9232 \pm 0.0052$ \\
1255.72537 & 900  & $1906.28 \pm 3.88$ & 25.58 & $6.7988 \pm 0.0078$ & $65.799 \pm 0.075$ & $-0.0458 \pm 0.0078$ & $0.2897 \pm 0.0008$ & $-4.9276 \pm 0.0079$ \\
1256.76902 & 900  & $1901.08 \pm 4.90$ & 21.90 & $6.8087 \pm 0.0098$ & $65.747 \pm 0.095$ & $-0.0084 \pm 0.0098$ & $0.2837 \pm 0.0010$ & $-5.0440 \pm 0.0140$ \\
1259.75827 & 900  & $1909.13 \pm 4.78$ & 22.09 & $6.8015 \pm 0.0096$ & $65.783 \pm 0.093$ & $-0.0242 \pm 0.0096$ & $0.2785 \pm 0.0010$ & $-4.9620 \pm 0.0110$ \\
1275.64817 & 900  & $1904.04 \pm 3.15$ & 30.09 & $6.7870 \pm 0.0063$ & $65.691 \pm 0.061$ & $-0.0296 \pm 0.0063$ & $0.2856 \pm 0.0010$ & $-4.9354 \pm 0.0058$ \\
1276.68013 & 900  & $1893.49 \pm 2.97$ & 31.56 & $6.7807 \pm 0.0059$ & $65.743 \pm 0.058$ & $-0.0371 \pm 0.0059$ & $0.2805 \pm 0.0009$ & $-4.9405 \pm 0.0055$ \\
1286.58701 & 900  & $1893.24 \pm 6.34$ & 17.18 & $6.8049 \pm 0.0130$ & $65.756 \pm 0.120$ & $-0.0099 \pm 0.0130$ & $0.2983 \pm 0.0030$ & $-5.0381 \pm 0.0180$ \\
1287.66008 & 900  & $1881.52 \pm 2.80$ & 32.92 & $6.7785 \pm 0.0056$ & $65.857 \pm 0.054$ & $-0.0425 \pm 0.0056$ & $0.2874 \pm 0.0007$ & $-4.9896 \pm 0.0057$ \\
1290.73131 & 900  & $1889.97 \pm 3.18$ & 30.56 & $6.8027 \pm 0.0064$ & $65.675 \pm 0.061$ & $-0.0446 \pm 0.0064$ & $0.2920 \pm 0.0008$ & $-4.9384 \pm 0.0061$ \\
1293.62401 & 900  & $1883.77 \pm 2.40$ & 37.37 & $6.8046 \pm 0.0048$ & $65.611 \pm 0.046$ & $-0.0293 \pm 0.0048$ & $0.2891 \pm 0.0007$ & $-4.9397 \pm 0.0039$ \\
1297.45698 & 900  & $1883.58 \pm 2.83$ & 33.22 & $6.7945 \pm 0.0057$ & $65.587 \pm 0.055$ & $-0.0371 \pm 0.0057$ & $0.2829 \pm 0.0007$ & $-4.9537 \pm 0.0054$ \\
1298.49833 & 900  & $1891.02 \pm 5.00$ & 21.61 & $6.8126 \pm 0.0100$ & $65.377 \pm 0.096$ & $-0.0405 \pm 0.0100$ & $0.2915 \pm 0.0010$ & $-4.9331 \pm 0.0110$ \\
1299.61699 & 900  & $1906.12 \pm 3.58$ & 27.92 & $6.8143 \pm 0.0072$ & $65.269 \pm 0.069$ & $-0.0239 \pm 0.0072$ & $0.2855 \pm 0.0007$ & $-4.9544 \pm 0.0071$ \\
1305.59864 & 900  & $1900750 \pm 4.89$ & 21.96 & $6.8274 \pm 0.0098$ & $65.338 \pm 0.094$ & $-0.0292 \pm 0.0098$ & $0.2816 \pm 0.0010$ & $-4.9525 \pm 0.0110$ \\
1307.65647 & 900  & $1907.59 \pm 2.32$ & 37.69 & $6.8181 \pm 0.0046$ & $65.577 \pm 0.045$ & $-0.0275 \pm 0.0046$ & $0.2888 \pm 0.0004$ & $-4.9007 \pm 0.0036$ \\
1323.64325 & 900  & $1871.50 \pm 6.83$ & 17.02 & $6.8025 \pm 0.0140$ & $65.959 \pm 0.130$ & $-0.0445 \pm 0.0140$ & $0.2867 \pm 0.0020$ & $-5.0261 \pm 0.0220$ \\
1325.57255 & 900  & $1919.23 \pm 2.76$ & 33.38 & $6.8011 \pm 0.0055$ & $65.649 \pm 0.053$ & $-0.0349 \pm 0.0055$ & $0.2922 \pm 0.0008$ & $-4.9420 \pm 0.0050$ \\
1327.60286 & 900  & $1904.58 \pm 2.63$ & 35.03 & $6.8059 \pm 0.0053$ & $65.471 \pm 0.051$ & $-0.0331 \pm 0.0053$ & $0.2806 \pm 0.0006$ & $-4.9688 \pm 0.0050$ \\
1330.64971 & 900  & $1885.54 \pm 4.57$ & 22.64 & $6.7735 \pm 0.0091$ & $64.891 \pm 0.088$ & $-0.0220 \pm 0.0091$ & $0.2899 \pm 0.0009$ & $-4.9321 \pm 0.0100$ \\
1332.69745 & 900  & $1920.97 \pm 8.59$ & 16.94 & $6.8092 \pm 0.0170$ & $63.700 \pm 0.160$ & $-0.0124 \pm 0.0170$ & $0.2881 \pm 0.0010$ & $-5.0089 \pm 0.0210$ \\
1333.68693 & 900  & $1892.35 \pm 4.79$ & 22.08 & $6.7855 \pm 0.0096$ & $65.299 \pm 0.092$ & $-0.0361 \pm 0.0096$ & $0.2818 \pm 0.0009$ & $-5.0114 \pm 0.0140$ \\
1338.49270 & 900  & $1910.67 \pm 3.79$ & 26.65 & $6.8375 \pm 0.0076$ & $65.713 \pm 0.073$ & $-0.0430 \pm 0.0076$ & $0.2873 \pm 0.0010$ & $-4.9980 \pm 0.0092$ \\
1342.57678 & 900  & $1915.99 \pm 7.92$ & 18.23 & $6.9301 \pm 0.0160$ & $64.591 \pm 0.150$ & $-0.0329 \pm 0.0160$ & $0.2850 \pm 0.0008$ & $-5.0077 \pm 0.0190$ \\
1343.55042 & 900  & $1900510 \pm 7.90$ & 17.99 & $6.8917 \pm 0.0160$ & $64.886 \pm 0.150$ & $-0.0408 \pm 0.0160$ & $0.2788 \pm 0.0009$ & $-5.0079 \pm 0.0180$ \\
1344.54888 & 900  & $1916.01 \pm 5.29$ & 24.44 & $6.9385 \pm 0.0110$ & $64.573 \pm 0.098$ & $-0.0213 \pm 0.0110$ & $0.2802 \pm 0.0006$ & $-4.8694 \pm 0.0078$ \\
1345.59635 & 900  & $1930.04 \pm 5.93$ & 23.05 & $6.9158 \pm 0.0120$ & $64.610 \pm 0.110$ & $-0.0408 \pm 0.0120$ & $0.2855 \pm 0.0006$ & $-4.9650 \pm 0.0120$ \\
1360.55960 & 900  & $1905.83 \pm 6.28$ & 18.12 & $6.7700 \pm 0.0130$ & $64.342 \pm 0.120$ & $-0.0339 \pm 0.0130$ & $0.2760 \pm 0.0010$ & $-4.9936 \pm 0.0180$ \\
1362.55955 & 900  & $1911.21 \pm 3.35$ & 28.79 & $6.8077 \pm 0.0067$ & $65.373 \pm 0.064$ & $-0.0521 \pm 0.0067$ & $0.2795 \pm 0.0006$ & $-4.9666 \pm 0.0074$ \\
1363.48678 & 900  & $1890.17 \pm 2.67$ & 34.62 & $6.8182 \pm 0.0053$ & $65.570 \pm 0.051$ & $-0.0260 \pm 0.0053$ & $0.2828 \pm 0.0006$ & $-4.9242 \pm 0.0047$ \\
1365.55694 & 900  & $1918.92 \pm 3.19$ & 30.50 & $6.8296 \pm 0.0064$ & $65.516 \pm 0.061$ & $-0.0304 \pm 0.0064$ & $0.2874 \pm 0.0008$ & $-4.9350 \pm 0.0063$ \\
1366.55753 & 900  & $1900150 \pm 3.15$ & 30.27 & $6.8245 \pm 0.0063$ & $65.661 \pm 0.061$ & $-0.0357 \pm 0.0063$ & $0.2920 \pm 0.0006$ & $-4.9219 \pm 0.0064$ \\
1371.41258 & 900  & $1911.21 \pm 5.05$ & 21.50 & $6.8601 \pm 0.0100$ & $65.549 \pm 0.097$ & $-0.0302 \pm 0.0100$ & $0.2951 \pm 0.0010$ & $-4.9584 \pm 0.0130$ \\
1377.49536 & 900  & $1903.76 \pm 3.81$ & 26.60 & $6.8634 \pm 0.0076$ & $65.633 \pm 0.073$ & $-0.0272 \pm 0.0076$ & $0.3025 \pm 0.0009$ & $-4.9193 \pm 0.0078$ \\
1379.43056 & 900  & $1911.88 \pm 3.28$ & 29.88 & $6.8630 \pm 0.0066$ & $65.428 \pm 0.063$ & $-0.0303 \pm 0.0066$ & $0.2985 \pm 0.0008$ & $-4.8556 \pm 0.0052$ \\
1387.47776 & 900  & $1903.37 \pm 3.36$ & 29.24 & $6.7942 \pm 0.0067$ & $65.595 \pm 0.065$ & $-0.0234 \pm 0.0067$ & $0.2771 \pm 0.0008$ & $-4.9679 \pm 0.0073$ \\
1388.47407 & 900  & $1909.51 \pm 2.32$ & 38.84 & $6.8054 \pm 0.0046$ & $65.503 \pm 0.045$ & $-0.0230 \pm 0.0046$ & $0.2794 \pm 0.0006$ & $-4.9342 \pm 0.0038$ \\
1390.45648 & 900  & $1880.95 \pm 2.70$ & 34.30 & $6.8021 \pm 0.0054$ & $65.271 \pm 0.052$ & $-0.0290 \pm 0.0054$ & $0.2892 \pm 0.0007$ & $-4.9180 \pm 0.0046$ \\
1391.46178 & 1200 & $1909.57 \pm 3.08$ & 30.67 & $6.7945 \pm 0.0062$ & $65.362 \pm 0.059$ & $-0.0221 \pm 0.0062$ & $0.2854 \pm 0.0007$ & $-4.9385 \pm 0.0062$ \\
1413.44785 & 900  & $1865.76 \pm 11.1$ & 11.83 & $6.8222 \pm 0.0220$ & $65.369 \pm 0.210$ & $-0.0269 \pm 0.0220$ & $0.2925 \pm 0.0020$ & $-5.1563 \pm 0.0570$ \\
1416.42606 & 900  & $1879.46 \pm 2.73$ & 34.17 & $6.8046 \pm 0.0055$ & $65.447 \pm 0.053$ & $-0.0139 \pm 0.0055$ & $0.2857 \pm 0.0006$ & $-4.8865 \pm 0.0047$ \\
1417.40817 & 900  & $1891.00 \pm 2.37$ & 38.01 & $6.8069 \pm 0.0047$ & $65.551 \pm 0.046$ & $-0.0231 \pm 0.0047$ & $0.2908 \pm 0.0005$ & $-4.9282 \pm 0.0040$ \\
1428.41714 & 900  & $1893.48 \pm 8.48$ & 14.78 & $6.8543 \pm 0.0170$ & $65.593 \pm 0.160$ & $ 0.0027 \pm 0.0170$ & $0.2972 \pm 0.0020$ & $-5.0021 \pm 0.0280$ \\
1430.40073 & 900  & $1894.45 \pm 4.88$ & 22.45 & $6.8236 \pm 0.0098$ & $65.742 \pm 0.094$ & $-0.0169 \pm 0.0098$ & $0.2892 \pm 0.0010$ & $-4.9842 \pm 0.0130$ \\
1431.39303 & 900  & $1891.97 \pm 5.49$ & 20.53 & $6.8325 \pm 0.0110$ & $65.625 \pm 0.110$ & $-0.0078 \pm 0.0110$ & $0.2910 \pm 0.0010$ & $-5.0049 \pm 0.0160$ \\
1443.36698 & 900  & $1887.36 \pm 7.36$ & 16.67 & $6.8117 \pm 0.0150$ & $64.773 \pm 0.140$ & $-0.0278 \pm 0.0150$ & $0.3042 \pm 0.0020$ & $-4.8952 \pm 0.0180$ \\
1445.37791 & 900  & $1914.70 \pm 4.34$ & 24.77 & $6.8117 \pm 0.0087$ & $65.428 \pm 0.083$ & $-0.0499 \pm 0.0087$ & $0.2971 \pm 0.0010$ & $-4.9129 \pm 0.0090$ \\
1447.38125 & 900  & $1908.40 \pm 4.63$ & 23.62 & $6.8627 \pm 0.0093$ & $64.882 \pm 0.088$ & $-0.0520 \pm 0.0093$ & $0.2958 \pm 0.0010$ & $-4.9527 \pm 0.0120$ \\
1448.36871 & 900  & $1917.23 \pm 2.90$ & 33.61 & $6.8624 \pm 0.0058$ & $65.045 \pm 0.055$ & $-0.0366 \pm 0.0058$ & $0.2950 \pm 0.0008$ & $-4.8858 \pm 0.0049$ \\
1601.72295 & 900  & $1892.66 \pm 2.74$ & 36.38 & $6.8214 \pm 0.0055$ & $65.678 \pm 0.053$ & $-0.0274 \pm 0.0055$ & $0.3002 \pm 0.0006$ & $-4.8730 \pm 0.0049$ \\
1627.68590 & 1800 & $1930.06 \pm 2.87$ & 35.47 & $6.8820 \pm 0.0057$ & $64.967 \pm 0.054$ & $-0.0247 \pm 0.0057$ & $0.3055 \pm 0.0008$ & $-4.8484 \pm 0.0050$ \\
1681.67501 & 900  & $1888.91 \pm 4.99$ & 23.39 & $6.7894 \pm 0.0100$ & $65.671 \pm 0.097$ & $-0.0405 \pm 0.0100$ & $0.3047 \pm 0.0010$ & $-4.8015 \pm 0.0100$ \\
1682.56077 & 900  & $1904.34 \pm 2.99$ & 34.10 & $6.7953 \pm 0.0060$ & $65.722 \pm 0.058$ & $-0.0518 \pm 0.0060$ & $0.2880 \pm 0.0007$ & $-4.9017 \pm 0.0059$ \\
1683.51917 & 900  & $1900620 \pm 5.36$ & 22.22 & $6.8081 \pm 0.0110$ & $65.400 \pm 0.100$ & $-0.0353 \pm 0.0110$ & $0.2906 \pm 0.0010$ & $-4.9225 \pm 0.0140$ \\
1690.62973 & 900  & $1918.08 \pm 4.77$ & 23.74 & $6.7416 \pm 0.0095$ & $65.719 \pm 0.093$ & $-0.0379 \pm 0.0095$ & $0.2785 \pm 0.0010$ & $-4.9454 \pm 0.0130$ \\
1706.59083 & 900  & $1909.84 \pm 4.47$ & 25.03 & $6.7620 \pm 0.0089$ & $65.928 \pm 0.087$ & $-0.0348 \pm 0.0089$ & $0.2860 \pm 0.0010$ & $-4.9667 \pm 0.0130$ \\
1709.50521 & 900  & $1906.78 \pm 5.77$ & 20.77 & $6.7821 \pm 0.0120$ & $65.674 \pm 0.110$ & $-0.0271 \pm 0.0120$ & $0.2812 \pm 0.0010$ & $-4.8505 \pm 0.0140$ \\
1714.61125 & 900  & $1892.06 \pm 4.26$ & 25.28 & $6.7559 \pm 0.0085$ & $65.427 \pm 0.082$ & $-0.0168 \pm 0.0085$ & $0.2774 \pm 0.0010$ & $-4.9687 \pm 0.0110$ \\
1716.56264 & 900  & $1906.26 \pm 2.51$ & 37.40 & $6.7623 \pm 0.0050$ & $65.835 \pm 0.049$ & $-0.0463 \pm 0.0050$ & $0.2800 \pm 0.0008$ & $-4.9431 \pm 0.0048$ \\
1739.51240 & 900  & $1905.58 \pm 4.19$ & 26.12 & $6.7683 \pm 0.0084$ & $65.965 \pm 0.082$ & $-0.0353 \pm 0.0084$ & $0.2789 \pm 0.0010$ & $-4.9506 \pm 0.0110$ \\
1747.52990 & 900  & $1924.37 \pm 5.06$ & 22.94 & $6.8347 \pm 0.0100$ & $65.298 \pm 0.097$ & $-0.0337 \pm 0.0100$ & $0.2844 \pm 0.0010$ & $-5.0482 \pm 0.0180$ \\
1768.44447 & 900  & $1904.75 \pm 4.60$ & 23.72 & $6.8080 \pm 0.0092$ & $65.985 \pm 0.089$ & $-0.0310 \pm 0.0092$ & $0.2892 \pm 0.0010$ & $-5.0218 \pm 0.0150$ \\
1771.42417 & 1200 & $1895.38 \pm 4.00$ & 27.07 & $6.7911 \pm 0.0080$ & $65.607 \pm 0.077$ & $-0.0491 \pm 0.0080$ & $0.2848 \pm 0.0009$ & $-4.9018 \pm 0.0090$ \\
1789.38669 & 900  & $1900100 \pm 4.23$ & 25.72 & $6.7621 \pm 0.0085$ & $66.010 \pm 0.083$ & $-0.0179 \pm 0.0085$ & $0.2831 \pm 0.0010$ & $-5.0078 \pm 0.0130$ \\
1803.38264 & 900  & $1899.96 \pm 3.17$ & 32.26 & $6.7731 \pm 0.0063$ & $65.555 \pm 0.061$ & $-0.0287 \pm 0.0063$ & $0.2856 \pm 0.0008$ & $-4.9508 \pm 0.0073$ \\
1805.39783 & 900  & $1884.95 \pm 5.54$ & 21.25 & $6.7488 \pm 0.0110$ & $66.188 \pm 0.110$ & $-0.0108 \pm 0.0110$ & $0.2906 \pm 0.0020$ & $-5.0938 \pm 0.0240$ \\
\end{longtable}
}

\tiny{
\begin{longtable}{crccccccc}
\caption{TOI-1694 HARPS-N RV data points and activity indices.}
\label{tab:RV_TOI-1694} \\
\hline\hline \\ [-8pt] %
BJD$_{\rm TDB}$ & $T_{\rm exp}$ & RV & S/N & FWHM & Contrast & BIS & H$\alpha$ & $\log{R^{\prime}_{\rm HK}}$  \\
$-2458000$ & (s) & (m\,s$^{-1}$) & & (m\,s$^{-1}$) & (m\,s$^{-1}$) &  (m\,s$^{-1}$)\\ [2 pt]
\hline \\ [-8pt] %
\endfirsthead
\caption{continued.}\\
\hline \\ [-8pt] %
BJD$_{\rm TDB}$ & $T_{\rm exp}$ & RV & S/N & FWHM & Contrast & BIS & H$\alpha$ & $\log{R^{\prime}_{\rm HK}}$  \\
$-2458000$ & (s) & (m\,s$^{-1}$) & & (km\,s$^{-1}$) & (km\,s$^{-1}$) &  (km\,s$^{-1}$)\\ [2 pt]
\hline \\ [-8pt] %
\endhead
\hline
\endfoot
1125.63868 & 900  & $-22518.75 \pm 4.55$ & 21.59 & $6.4392 \pm 0.0091$ & $65.642 \pm 0.093$ & $-0.0422 \pm 0.0091$ & $0.2443 \pm 0.0009$ & $-4.8395 \pm 0.0226$ \\
1126.63353 & 1200 & $-22531.65 \pm 2.82$ & 31.43 & $6.4224 \pm 0.0056$ & $66.224 \pm 0.058$ & $-0.0459 \pm 0.0056$ & $0.2314 \pm 0.0006$ & $-4.9651 \pm 0.0158$ \\
1127.68306 & 900  & $-22540.64 \pm 2.52$ & 33.9  & $6.4288 \pm 0.0050$ & $66.312 \pm 0.052$ & $-0.0685 \pm 0.0050$ & $0.2405 \pm 0.0005$ & $-4.9412 \pm 0.0127$ \\
1130.72313 & 900  & $-22531.88 \pm 8.38$ & 14.01 & $6.4251 \pm 0.0170$ & $65.560 \pm 0.170$ & $-0.0465 \pm 0.0170$ & $0.2277 \pm 0.0020$ & $-4.9945 \pm 0.0664$ \\
1133.69840 & 900  & $-22524.65 \pm 2.69$ & 32.07 & $6.4302 \pm 0.0054$ & $66.588 \pm 0.056$ & $-0.0418 \pm 0.0054$ & $0.2436 \pm 0.0005$ & $-4.9625 \pm 0.0151$ \\
1134.74196 & 900  & $-22539.16 \pm 4.97$ & 20.53 & $6.4153 \pm 0.0099$ & $66.441 \pm 0.100$ & $-0.0515 \pm 0.0099$ & $0.2330 \pm 0.0010$ & $-4.8272 \pm 0.0256$ \\
1156.58972 & 900  & $-22537.45 \pm 2.61$ & 33.51 & $6.4427 \pm 0.0052$ & $66.255 \pm 0.054$ & $-0.0558 \pm 0.0052$ & $0.2375 \pm 0.0005$ & $-4.8972 \pm 0.0121$ \\
1170.69599 & 900  & $-22534.31 \pm 2.98$ & 30.56 & $6.4219 \pm 0.0060$ & $66.502 \pm 0.062$ & $-0.0497 \pm 0.0060$ & $0.2364 \pm 0.0007$ & $-4.9677 \pm 0.0172$ \\
1171.66186 & 900  & $-22541.06 \pm 3.08$ & 29.61 & $6.4291 \pm 0.0062$ & $66.491 \pm 0.064$ & $-0.0598 \pm 0.0062$ & $0.2372 \pm 0.0007$ & $-4.9370 \pm 0.0168$ \\
1172.59335 & 900  & $-22554.04 \pm 3.14$ & 28.85 & $6.4368 \pm 0.0063$ & $66.536 \pm 0.065$ & $-0.0404 \pm 0.0063$ & $0.2406 \pm 0.0006$ & $-5.0036 \pm 0.0204$ \\
1190.62323 & 900  & $-22552.28 \pm 3.46$ & 26.84 & $6.4229 \pm 0.0069$ & $66.620 \pm 0.072$ & $-0.0470 \pm 0.0069$ & $0.2440 \pm 0.0010$ & $-4.9153 \pm 0.0192$ \\
1212.52229 & 900  & $-22551.34 \pm 3.99$ & 23.70 & $6.4089 \pm 0.0080$ & $65.970 \pm 0.082$ & $-0.0514 \pm 0.0080$ & $0.2400 \pm 0.0008$ & $-5.0252 \pm 0.0299$ \\
1215.63654 & 900  & $-22551.20 \pm 2.84$ & 30.71 & $6.4201 \pm 0.0057$ & $66.255 \pm 0.059$ & $-0.0473 \pm 0.0057$ & $0.2386 \pm 0.0006$ & $-4.9999 \pm 0.0180$ \\
1216.65346 & 900  & $-22561.06 \pm 4.18$ & 23.18 & $6.4265 \pm 0.0084$ & $66.264 \pm 0.086$ & $-0.0430 \pm 0.0084$ & $0.2315 \pm 0.0009$ & $-5.0653 \pm 0.0365$ \\
1229.61471 & 900  & $-22581.60 \pm 7.10$ & 15.58 & $6.4223 \pm 0.0140$ & $66.789 \pm 0.150$ & $-0.0684 \pm 0.0140$ & $0.2428 \pm 0.0020$ & $-5.2120 \pm 0.1030$ \\
1235.56127 & 900  & $-22570.83 \pm 2.79$ & 31.28 & $6.4231 \pm 0.0056$ & $66.595 \pm 0.058$ & $-0.0491 \pm 0.0056$ & $0.2383 \pm 0.0006$ & $-4.9914 \pm 0.0174$ \\
1236.61849 & 900  & $-22588.23 \pm 3.61$ & 26.44 & $6.442  \pm 0.0072$ & $66.501 \pm 0.075$ & $-0.0570 \pm 0.0072$ & $0.2346 \pm 0.0008$ & $-4.9780 \pm 0.0246$ \\
1237.57139 & 1200 & $-22570.80 \pm 3.15$ & 28.89 & $6.4071 \pm 0.0063$ & $66.455 \pm 0.065$ & $-0.0532 \pm 0.0063$ & $0.2357 \pm 0.0006$ & $-4.9714 \pm 0.0203$ \\
1239.52536 & 900  & $-22570.68 \pm 1.90$ & 42.88 & $6.4306 \pm 0.0038$ & $66.410 \pm 0.039$ & $-0.0565 \pm 0.0038$ & $0.2396 \pm 0.0004$ & $-4.9496 \pm 0.0084$ \\
1240.53770 & 1200 & $-22585.35 \pm 3.04$ & 29.08 & $6.4212 \pm 0.0061$ & $65.854 \pm 0.062$ & $-0.0382 \pm 0.0061$ & $0.2343 \pm 0.0006$ & $-4.9148 \pm 0.0167$ \\
1244.54372 & 900  & $-22587.32 \pm 2.74$ & 32.66 & $6.4341 \pm 0.0055$ & $66.109 \pm 0.056$ & $-0.0629 \pm 0.0055$ & $0.2374 \pm 0.0006$ & $-4.8573 \pm 0.0124$ \\
1244.57931 & 900  & $-22584.54 \pm 1.82$ & 44.94 & $6.4374 \pm 0.0036$ & $66.369 \pm 0.038$ & $-0.0487 \pm 0.0036$ & $0.2407 \pm 0.0004$ & $-4.9417 \pm 0.0077$ \\
1245.49308 & 1200 & $-22567.64 \pm 2.24$ & 37.90 & $6.4300 \pm 0.0045$ & $66.292 \pm 0.046$ & $-0.0497 \pm 0.0045$ & $0.2387 \pm 0.0005$ & $-4.9941 \pm 0.0123$ \\
1246.56854 & 1200 & $-22570.00 \pm 2.47$ & 35.20 & $6.4400 \pm 0.0049$ & $66.300 \pm 0.051$ & $-0.0504 \pm 0.0049$ & $0.2401 \pm 0.0006$ & $-4.8395 \pm 0.0100$ \\
1253.52033 & 900  & $-22564.71 \pm 3.21$ & 27.89 & $6.4471 \pm 0.0064$ & $66.497 \pm 0.066$ & $-0.0568 \pm 0.0064$ & $0.2388 \pm 0.0006$ & $-5.1226 \pm 0.0293$ \\
1254.48504 & 900  & $-22573.50 \pm 2.18$ & 37.59 & $6.4289 \pm 0.0044$ & $66.665 \pm 0.045$ & $-0.0679 \pm 0.0044$ & $0.2404 \pm 0.0004$ & $-4.9150 \pm 0.0097$ \\
1255.56441 & 900  & $-22588.94 \pm 2.29$ & 37.32 & $6.4430 \pm 0.0046$ & $66.408 \pm 0.047$ & $-0.0557 \pm 0.0046$ & $0.2372 \pm 0.0005$ & $-4.9626 \pm 0.0118$ \\
1256.49587 & 900  & $-22581.96 \pm 4.16$ & 23.53 & $6.4281 \pm 0.0083$ & $66.538 \pm 0.086$ & $-0.0566 \pm 0.0083$ & $0.2376 \pm 0.0009$ & $-5.0685 \pm 0.0368$ \\
1272.51273 & 900  & $-22559.08 \pm 2.51$ & 34.77 & $6.4608 \pm 0.0050$ & $66.220 \pm 0.052$ & $-0.0606 \pm 0.0050$ & $0.2324 \pm 0.0006$ & $-4.8736 \pm 0.0107$ \\
1322.36656 & 900  & $-22567.48 \pm 4.62$ & 21.77 & $6.4334 \pm 0.0092$ & $66.440 \pm 0.095$ & $-0.0479 \pm 0.0092$ & $0.2440 \pm 0.0010$ & $-5.0544 \pm 0.0403$ \\
1323.36644 & 900  & $-22589.75 \pm 4.85$ & 21.04 & $6.4378 \pm 0.0097$ & $66.565 \pm 0.100$ & $-0.0413 \pm 0.0097$ & $0.2507 \pm 0.0010$ & $-4.9464 \pm 0.0327$ \\
1325.35738 & 900  & $-22555.13 \pm 2.30$ & 37.08 & $6.4421 \pm 0.0046$ & $66.431 \pm 0.047$ & $-0.0461 \pm 0.0046$ & $0.2444 \pm 0.0005$ & $-5.0561 \pm 0.0150$ \\
1327.40273 & 900  & $-22582.70 \pm 2.10$ & 40.10 & $6.4405 \pm 0.0042$ & $66.526 \pm 0.043$ & $-0.0575 \pm 0.0042$ & $0.2395 \pm 0.0005$ & $-5.0060 \pm 0.0118$ \\
1328.36954 & 900  & $-22550.96 \pm 3.64$ & 26.04 & $6.4508 \pm 0.0073$ & $66.213 \pm 0.075$ & $-0.0495 \pm 0.0073$ & $0.2374 \pm 0.0009$ & $-5.0134 \pm 0.0275$ \\
1513.77459 & 900  & $-22513.19 \pm 3.99$ & 26.23 & $6.4287 \pm 0.0080$ & $65.763 \pm 0.082$ & $-0.0476 \pm 0.0080$ & $0.2580 \pm 0.0010$ & $-5.0031 \pm 0.0321$ \\
1516.59128 & 900  & $-22527.90 \pm 3.55$ & 28.88 & $6.4202 \pm 0.0071$ & $66.677 \pm 0.074$ & $-0.0576 \pm 0.0071$ & $0.2470 \pm 0.0009$ & $-5.1071 \pm 0.0372$ \\
1560.71238 & 900  & $-22558.94 \pm 2.57$ & 37.02 & $6.4148 \pm 0.0051$ & $66.705 \pm 0.053$ & $-0.0604 \pm 0.0051$ & $0.2453 \pm 0.0006$ & $-4.8294 \pm 0.0122$ \\
1561.47011 & 900  & $-22549.14 \pm 2.83$ & 33.99 & $6.4080 \pm 0.0057$ & $66.681 \pm 0.059$ & $-0.0550 \pm 0.0057$ & $0.2405 \pm 0.0009$ & $-4.9323 \pm 0.0173$ \\
1565.47774 & 900  & $-22550.42 \pm 2.78$ & 34.92 & $6.4179 \pm 0.0056$ & $66.387 \pm 0.058$ & $-0.0488 \pm 0.0056$ & $0.2491 \pm 0.0007$ & $-4.8822 \pm 0.0153$ \\
1566.49186 & 900  & $-22537.14 \pm 2.23$ & 41.71 & $6.4234 \pm 0.0045$ & $66.399 \pm 0.046$ & $-0.0519 \pm 0.0045$ & $0.2478 \pm 0.0006$ & $-4.8229 \pm 0.0095$ \\
1567.74708 & 900  & $-22551.55 \pm 6.64$ & 18.43 & $6.4560 \pm 0.0130$ & $64.702 \pm 0.130$ & $-0.0527 \pm 0.0130$ & $0.2441 \pm 0.0020$ & $-4.7312 \pm 0.0342$ \\
1575.47779 & 900  & $-22561.19 \pm 1.92$ & 46.47 & $6.4362 \pm 0.0038$ & $66.635 \pm 0.040$ & $-0.0460 \pm 0.0038$ & $0.2423 \pm 0.0005$ & $-4.8951 \pm 0.0089$ \\
1579.62839 & 900  & $-22571.01 \pm 3.22$ & 31.10 & $6.4217 \pm 0.0064$ & $66.668 \pm 0.067$ & $-0.0411 \pm 0.0064$ & $0.2503 \pm 0.0008$ & $-4.8918 \pm 0.0200$ \\
1580.62230 & 900  & $-22556.87 \pm 6.43$ & 18.20 & $6.4073 \pm 0.0130$ & $66.617 \pm 0.130$ & $-0.0468 \pm 0.0130$ & $0.2366 \pm 0.0020$ & $-5.5249 \pm 0.2060$ \\
1581.61568 & 900  & $-22545.44 \pm 4.16$ & 25.68 & $6.4211 \pm 0.0083$ & $66.534 \pm 0.086$ & $-0.0299 \pm 0.0083$ & $0.2421 \pm 0.0010$ & $-4.7029 \pm 0.0185$ \\
1582.70292 & 900  & $-22553.06 \pm 6.25$ & 18.90 & $6.4244 \pm 0.0130$ & $66.707 \pm 0.130$ & $-0.0531 \pm 0.0130$ & $0.2313 \pm 0.0020$ & $-5.5481 \pm 0.2360$ \\
1583.58366 & 1400 & $-22570.09 \pm 7.25$ & 16.76 & $6.4062 \pm 0.0140$ & $66.871 \pm 0.150$ & $-0.0540 \pm 0.0140$ & $0.2255 \pm 0.0020$ & $-4.7991 \pm 0.0480$ \\
1587.60640 & 1800 & $-22569.53 \pm 3.29$ & 30.36 & $6.4211 \pm 0.0066$ & $66.624 \pm 0.068$ & $-0.0553 \pm 0.0066$ & $0.2427 \pm 0.0008$ & $-4.8871 \pm 0.0212$ \\
1601.62066 & 900  & $-22561.72 \pm 2.75$ & 34.95 & $6.4341 \pm 0.0055$ & $66.635 \pm 0.057$ & $-0.0538 \pm 0.0055$ & $0.2478 \pm 0.0006$ & $-4.8384 \pm 0.0140$ \\
1615.56873 & 900  & $-22556.78 \pm 3.11$ & 31.35 & $6.4328 \pm 0.0062$ & $66.633 \pm 0.064$ & $-0.0529 \pm 0.0062$ & $0.2443 \pm 0.0010$ & $-5.0300 \pm 0.0249$ \\
1616.57616 & 900  & $-22574.21 \pm 3.98$ & 26.37 & $6.4225 \pm 0.0080$ & $66.698 \pm 0.083$ & $-0.0657 \pm 0.0080$ & $0.2423 \pm 0.0010$ & $-4.9129 \pm 0.0296$ \\
1617.59017 & 1800 & $-22581.87 \pm 7.41$ & 16.39 & $6.4601 \pm 0.0150$ & $66.772 \pm 0.150$ & $-0.0593 \pm 0.0150$ & $0.2452 \pm 0.0020$ & $-4.8734 \pm 0.0628$ \\
1618.54846 & 900  & $-22563.85 \pm16.81$ & 9.002 & $6.4470 \pm 0.0340$ & $67.081 \pm 0.350$ & $-0.0410 \pm 0.0340$ & $0.2494 \pm 0.0040$ & $-4.4536 \pm 0.0662$ \\
1624.50452 & 900  & $-22579.77 \pm 2.01$ & 44.52 & $6.4271 \pm 0.0040$ & $66.588 \pm 0.042$ & $-0.0505 \pm 0.0040$ & $0.2476 \pm 0.0005$ & $-4.8719 \pm 0.0090$ \\
1625.54519 & 900  & $-22580.29 \pm 2.62$ & 36.27 & $6.4298 \pm 0.0052$ & $66.416 \pm 0.054$ & $-0.0462 \pm 0.0052$ & $0.2348 \pm 0.0008$ & $-4.8689 \pm 0.0132$ \\
1626.54701 & 900  & $-22553.10 \pm 4.98$ & 22.38 & $6.4229 \pm 0.0100$ & $65.902 \pm 0.100$ & $-0.0118 \pm 0.0100$ & $0.2487 \pm 0.0010$ & $-4.8626 \pm 0.0334$ \\
1627.52596 & 1800 & $-22517.27 \pm 6.51$ & 18.44 & $6.5281 \pm 0.0130$ & $64.802 \pm 0.130$ & $ 0.0067 \pm 0.0130$ & $0.2380 \pm 0.0020$ & $-5.4836 \pm 0.1850$ \\
1646.38218 & 900  & $-22574.15 \pm 2.22$ & 41.17 & $6.4278 \pm 0.0044$ & $66.678 \pm 0.046$ & $-0.0534 \pm 0.0044$ & $0.2524 \pm 0.0006$ & $-4.7184 \pm 0.0076$ \\
1647.39660 & 900  & $-22591.55 \pm 3.27$ & 30.92 & $6.4349 \pm 0.0065$ & $66.587 \pm 0.068$ & $-0.0458 \pm 0.0065$ & $0.2536 \pm 0.0009$ & $-4.7904 \pm 0.0150$ \\
1648.40645 & 900  & $-22580.02 \pm 2.96$ & 32.68 & $6.4306 \pm 0.0059$ & $66.575 \pm 0.061$ & $-0.0549 \pm 0.0059$ & $0.2509 \pm 0.0006$ & $-4.8450 \pm 0.0161$ \\
1649.39566 & 900  & $-22568.15 \pm 2.39$ & 38.71 & $6.4313 \pm 0.0048$ & $66.532 \pm 0.049$ & $-0.0502 \pm 0.0048$ & $0.2479 \pm 0.0005$ & $-4.7884 \pm 0.0098$ \\
1650.40605 & 900  & $-22578.82 \pm 2.44$ & 37.98 & $6.4410 \pm 0.0049$ & $66.613 \pm 0.051$ & $-0.0443 \pm 0.0049$ & $0.2620 \pm 0.0009$ & $-4.8225 \pm 0.0108$ \\
1651.53944 & 900  & $-22587.69 \pm 2.31$ & 40.17 & $6.4393 \pm 0.0046$ & $66.551 \pm 0.048$ & $-0.0365 \pm 0.0046$ & $0.2415 \pm 0.0004$ & $-4.8137 \pm 0.0106$ \\
1656.39900 & 1200 & $-22571.21 \pm 3.28$ & 30.45 & $6.4339 \pm 0.0066$ & $66.524 \pm 0.068$ & $-0.0547 \pm 0.0066$ & $0.2473 \pm 0.0008$ & $-4.7988 \pm 0.0169$ \\
1659.37521 & 900  & $-22595.27 \pm 2.05$ & 43.52 & $6.4260 \pm 0.0041$ & $66.741 \pm 0.043$ & $-0.0538 \pm 0.0041$ & $0.2523 \pm 0.0007$ & $-4.8276 \pm 0.0080$ \\
1660.37693 & 900  & $-22572.57 \pm 4.34$ & 24.85 & $6.4309 \pm 0.0087$ & $66.609 \pm 0.090$ & $-0.0577 \pm 0.0087$ & $0.2544 \pm 0.0010$ & $-4.8192 \pm 0.0249$ \\
1681.38517 & 900  & $-22586.54 \pm 3.09$ & 31.83 & $6.4313 \pm 0.0062$ & $66.633 \pm 0.064$ & $-0.0612 \pm 0.0062$ & $0.2538 \pm 0.0008$ & $-4.8896 \pm 0.0188$ \\
1682.38445 & 900  & $-22575.07 \pm 2.80$ & 34.21 & $6.4568 \pm 0.0056$ & $66.396 \pm 0.058$ & $-0.0576 \pm 0.0056$ & $0.2494 \pm 0.0006$ & $-4.7049 \pm 0.0106$ \\
1683.36443 & 900  & $-22545.29 \pm 5.76$ & 20.12 & $6.4137 \pm 0.0120$ & $66.582 \pm 0.120$ & $-0.0411 \pm 0.0120$ & $0.2459 \pm 0.0010$ &   \dots                 \\
1684.36011 & 900  & $-22565.95 \pm 6.36$ & 18.80 & $6.4796 \pm 0.0130$ & $66.139 \pm 0.130$ & $-0.0329 \pm 0.0130$ & $0.2560 \pm 0.0020$ & $-4.8540 \pm 0.0452$ \\
1685.35769 & 900  & $-22574.47 \pm 6.75$ & 18.02 & $6.4859 \pm 0.0140$ & $65.496 \pm 0.140$ & $-0.0427 \pm 0.0140$ & $0.2534 \pm 0.0020$ & $-4.9399 \pm 0.0561$ \\
1686.34985 & 900  & $-22530.58 \pm 3.91$ & 27.06 & $6.5098 \pm 0.0078$ & $65.428 \pm 0.079$ & $-0.0266 \pm 0.0078$ & $0.2530 \pm 0.0010$ & $-4.8473 \pm 0.0215$ \\
1821.73085 & 900  & $-22516.42 \pm 5.61$ & 19.84 & $6.4532 \pm 0.0110$ & $66.878 \pm 0.120$ & $-0.0410 \pm 0.0110$ & $0.2528 \pm 0.0020$ & $-4.8068 \pm 0.0362$ \\
1829.74920 & 900  & $-22504.80 \pm 2.93$ & 32.38 & $6.4422 \pm 0.0059$ & $66.393 \pm 0.060$ & $-0.0512 \pm 0.0059$ & $0.2499 \pm 0.0007$ & $-4.7998 \pm 0.0131$ \\
1835.74717 & 900  & $-22523.93 \pm 2.37$ & 38.22 & $6.4412 \pm 0.0047$ & $66.536 \pm 0.049$ & $-0.0553 \pm 0.0047$ & $0.2522 \pm 0.0006$ & $-4.7622 \pm 0.0088$ \\
1837.75609 & 900  & $-22500.34 \pm 2.24$ & 40.26 & $6.4511 \pm 0.0045$ & $66.006 \pm 0.046$ & $-0.0558 \pm 0.0045$ & $0.2576 \pm 0.0005$ & $-4.8133 \pm 0.0087$ \\
1843.74584 & 900  & $-22530.56 \pm 3.21$ & 30.25 & $6.4520 \pm 0.0064$ & $66.653 \pm 0.066$ & $-0.0525 \pm 0.0064$ & $0.2504 \pm 0.0007$ & $-4.8414 \pm 0.0170$ \\
1844.69836 & 900  & $-22508.38 \pm 4.31$ & 24.32 & $6.4518 \pm 0.0086$ & $66.741 \pm 0.089$ & $-0.0534 \pm 0.0086$ & $0.2550 \pm 0.0010$ & $-4.8730 \pm 0.0286$ \\
1845.73103 & 900  & $-22506.55 \pm 3.17$ & 30.65 & $6.4570 \pm 0.0063$ & $66.624 \pm 0.065$ & $-0.0469 \pm 0.0063$ & $0.2494 \pm 0.0008$ & $-4.7299 \pm 0.0128$ \\
1859.64096 & 900  & $-22515.28 \pm 3.02$ & 31.67 & $6.4349 \pm 0.0060$ & $66.583 \pm 0.063$ & $-0.0611 \pm 0.0060$ & $0.2577 \pm 0.0009$ & $-4.9486 \pm 0.0196$ \\
1860.66050 & 900  & $-22506.49 \pm 3.20$ & 29.69 & $6.4483 \pm 0.0064$ & $66.454 \pm 0.066$ & $-0.0493 \pm 0.0064$ & $0.2568 \pm 0.0010$ & $-4.8564 \pm 0.0170$ \\
1861.63368 & 900  & $-22526.10 \pm 4.06$ & 25.56 & $6.4618 \pm 0.0081$ & $66.036 \pm 0.083$ & $-0.0393 \pm 0.0081$ & $0.2592 \pm 0.0010$ & $-4.9299 \pm 0.0281$ \\
1864.68187 & 900  & $-22512.52 \pm 2.47$ & 36.96 & $6.4372 \pm 0.0049$ & $66.528 \pm 0.051$ & $-0.0543 \pm 0.0049$ & $0.2520 \pm 0.0007$ & $-4.9033 \pm 0.0128$ \\
1865.66798 & 900  & $-22527.63 \pm 3.34$ & 29.37 & $6.4548 \pm 0.0067$ & $66.543 \pm 0.069$ & $-0.0512 \pm 0.0067$ & $0.2457 \pm 0.0010$ & $-5.0109 \pm 0.0262$ \\
1868.68978 & 900  & $-22511.63 \pm 3.04$ & 31.72 & $6.4632 \pm 0.0061$ & $66.566 \pm 0.063$ & $-0.0399 \pm 0.0061$ & $0.2472 \pm 0.0008$ & $-4.9403 \pm 0.0195$ \\
1869.75485 & 900  & $-22539.65 \pm 4.74$ & 22.66 & $6.4404 \pm 0.0095$ & $66.526 \pm 0.098$ & $-0.0601 \pm 0.0095$ & $0.2469 \pm 0.0010$ & $-4.9545 \pm 0.0387$ \\
1871.67243 & 900  & $-22508.44 \pm 4.18$ & 24.70 & $6.4534 \pm 0.0084$ & $66.694 \pm 0.087$ & $-0.0402 \pm 0.0084$ & $0.2491 \pm 0.0010$ & $-4.7778 \pm 0.0210$ \\
1872.67749 & 900  & $-22520.14 \pm 3.42$ & 28.81 & $6.4648 \pm 0.0068$ & $66.543 \pm 0.070$ & $-0.0505 \pm 0.0068$ & $0.2525 \pm 0.0010$ & $-4.7855 \pm 0.0154$ \\
1890.61739 & 900  & $-22508.12 \pm 2.55$ & 36.48 & $6.4512 \pm 0.0051$ & $66.350 \pm 0.052$ & $-0.0473 \pm 0.0051$ & $0.2533 \pm 0.0007$ & $-4.7582 \pm 0.0096$ \\
\end{longtable}
}
\begin{table}
\caption{Priors and best-fitting values for the GP regression analysis of the HIRES RV dataset.}
\label{tab:rvgphires}
\centering
\begin{tabular}{lcc}
\hline \hline
\noalign{\smallskip}
Parameter & Prior & Best-fit value$^{(a)}$  \\
\noalign{\smallskip}
\hline
\noalign{\smallskip}
Stellar activity GP term\\
\noalign{\smallskip}
$\theta$ [d]& $\mathcal{U}(0,35)$ & $26.05\pm0.20$ \\
\noalign{\smallskip}
$\lambda$ [d] & $\mathcal{U}(0,1000)$ & $66.65^{+25.37}_{-20.02}$ \\
\noalign{\smallskip}
$h$ [m\,s$^{-1}$] & $\mathcal{U}(0,50)$ & $9.17^{+1.68}_{-1.26}$ \\
\noalign{\smallskip}
$w$ & $\mathcal{U}(0,1)$  & $0.26\pm0.04$ \\
\noalign{\smallskip}
\hline
\noalign{\smallskip}
Planetary parameters\\
\noalign{\smallskip}
$K_{\rm b}$ [m\,s$^{-1}$] & $\mathcal{U}(0,20)$ & $11.80^{+0.72}_{-0.69}$ \\
\noalign{\smallskip}
$P_{\rm b}$ [d]~$^{(b)}$ & $\mathcal{N}(3.3159765,0.0000015)$ & $3.3159765\,(15)$ \\
\noalign{\smallskip}
$T_{\rm b,\:conj}$ [BJD-2\,450\,000]~$^{(b)}$ & $\mathcal{N}(8713.03098,0.00040)$ & $8713.03097\,(39)$  \\
\noalign{\smallskip}
$\sqrt{e_{\rm b}}\cos\omega_{\star,\, \rm b}$~$^{(c)}$ & $\mathcal{U}$(-1,1) & $-0.347^{+0.073}_{-0.065}$ \\
\noalign{\smallskip}
$\sqrt{e_{\rm b}}\sin\omega_{\star, \, \rm b}$ ~$^{(c)}$ & $\mathcal{U}$(-1,1) & $0.384^{+0.077}_{-0.089}$ \\
\noalign{\smallskip}
\hline
\noalign{\smallskip}
Derived planetary parameters\\
\noalign{\smallskip}
$e_{\rm b}$ & $\cdots$ & $0.27\pm0.05$ \\
\noalign{\smallskip}
$\omega_{\rm b}$ [deg] & $\cdots$ & 
$131.78^{+10.89}_{-10.31}$ \\ [2pt]
\hline
\noalign{\smallskip}
Instrument-related parameters\\
$\sigma_{\rm jit}$ [m\,s$^{-1}$] & $\mathcal{U}(0,20)$  & $1.84^{+0.79}_{-0.86}$  \\
\noalign{\smallskip}
offset $\gamma$ [m\,s$^{-1}$] &  $\mathcal{U}(-100,100)$ & $0.83^{+2.33}_{-2.34}$ \\
\noalign{\smallskip}
\hline
\end{tabular}
\tablefoot{
\tablefoottext{a}{Percentiles (16$^{\rm th}$, 50$^{\rm th}$, and 84$^{\rm th}$) of the posterior distributions.}
\tablefoottext{b}{The numbers in brackets represent the uncertainties in the preceding digits.}
\tablefoottext{c}{We adopted the parametrization $\sqrt{e_{b}}\cos\omega_{\star,\,b}$ and $\sqrt{e_{b}}\sin\omega_{\star,\,b}$ instead of using $e_{b}$ and $\omega_{\star,\,b}$ as free parameters.} 
}
\end{table}

\newpage

\section{Additional plots}
\clearpage

\begin{figure}
\centering
\includegraphics[width=16.0cm]{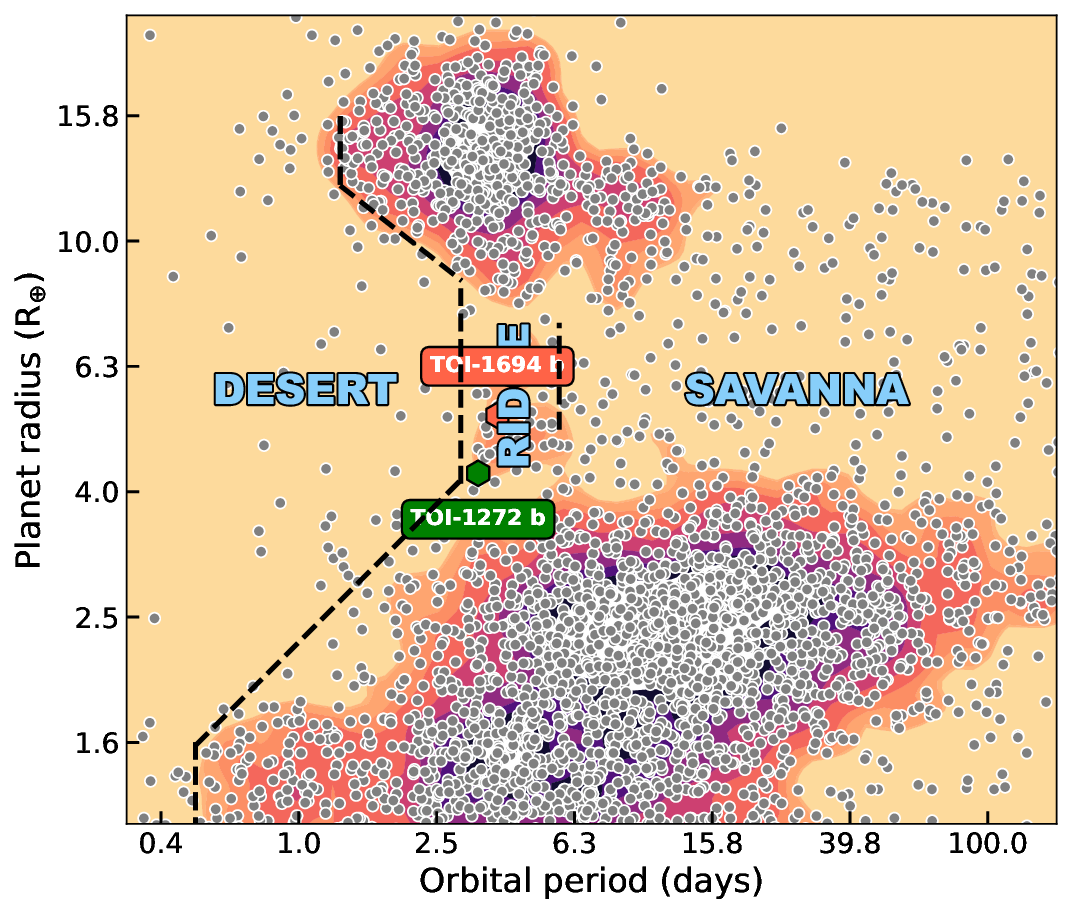}
\caption{Radius-period diagram of close-in exoplanets with mass and radius measured to an accuracy of at least $5\,\sigma$. The data were obtained from the NASA Exoplanet Archive on March 10, 2026. Error bars were suppressed for clarity. The positions of TOI-1272\,b and TOI-1694\,b (this work) are highlighted, together with the population-based boundaries of the Neptunian desert, ridge, and savanna, as derived by \citet{castro2024a}. This plot was generated with \texttt{nep-des} ({\url{github.com/castro-gzlz/nep-des}}).
} 
\label{fig:desert}
\end{figure}
\begin{figure*}[t!]
\centering
\includegraphics[width=9.0cm]{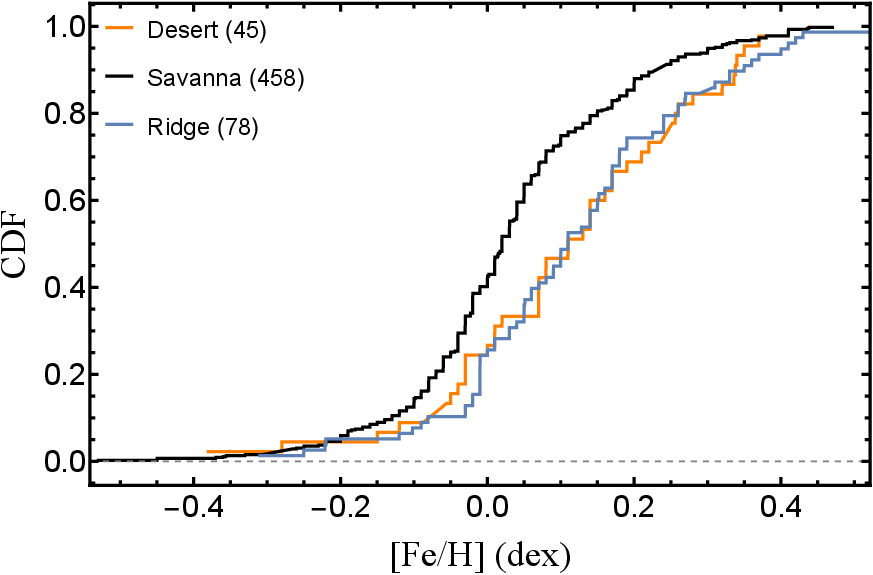}
\includegraphics[width=9.0cm]{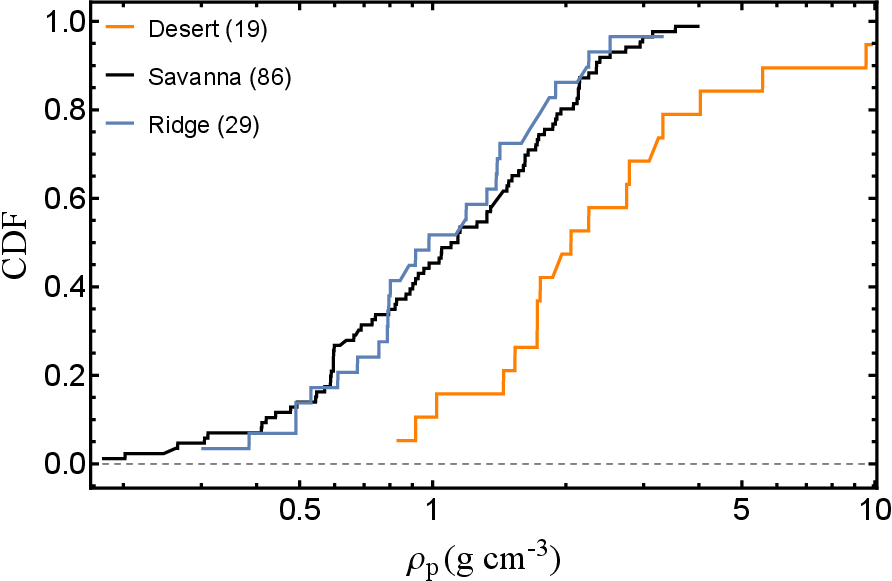}
\caption{{\it Left panel}: Comparison of host-star metallicity CDFs for different Neptune-sized-planet (3\,$R_{\oplus}$\,<\,$R_{\rm p}$\,<\,8.5\,$R_{\oplus}$) samples belonging to the desert (orange), ridge (blue) and savanna (black), whose boundaries are specified in the text. We considered only planets orbiting stars of known metallicity. {\it Right panel}: comparison of planetary-density CDFs for the same Neptune-sized-planet samples as in the left-hand panel. We only considered planets with a mean density measured with an accuracy to within $30\%$. Data taken from {\tt TEPCat}. 
} 
\label{fig:CDF}
\end{figure*}

\begin{figure*}[t!]
\centering
\includegraphics[width=9.0cm]{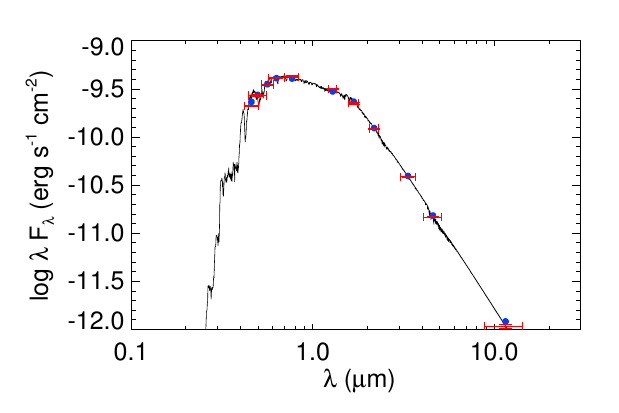}
\includegraphics[width=9.0cm]{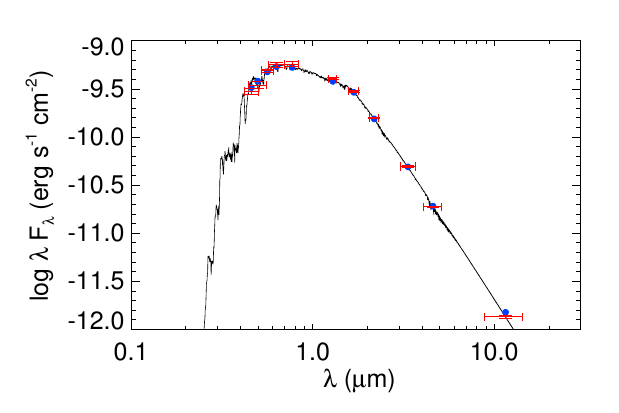} \\
\caption{SED of the host stars TOI-1272 (left panel) and TOI-1694 (right panel). The broadband measurements from the APASS Johnson and Sloan, 2MASS, and WISE magnitudes are displayed in red, and the corresponding theoretical values with blue circles. The solid black lines in both plots show the non-averaged best-fit models. The strong similarity of the SEDs is due to the almost identical stellar spectral types.} 
\label{fig:stellarSEDs}
\end{figure*}

\begin{figure}
\centering
\includegraphics[width=12.0cm]{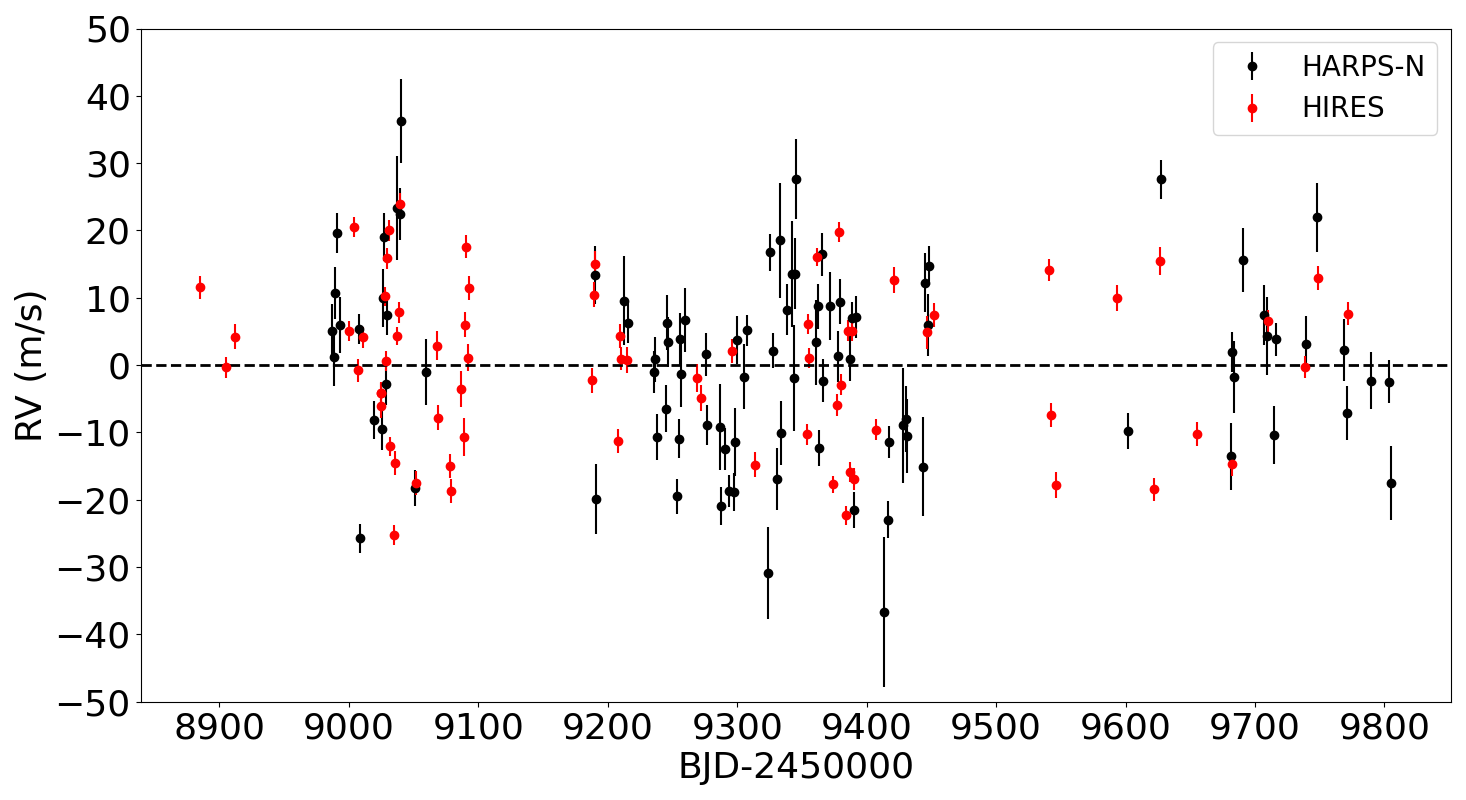}
\caption{TOI-1272 time series of the RVs extracted from HIRES \citep{polanski2024} and HARPS-N (this work) spectra.}
\label{fig:RVtimeseries}
\end{figure}


\begin{figure}
\centering
\includegraphics[width=12.0cm]{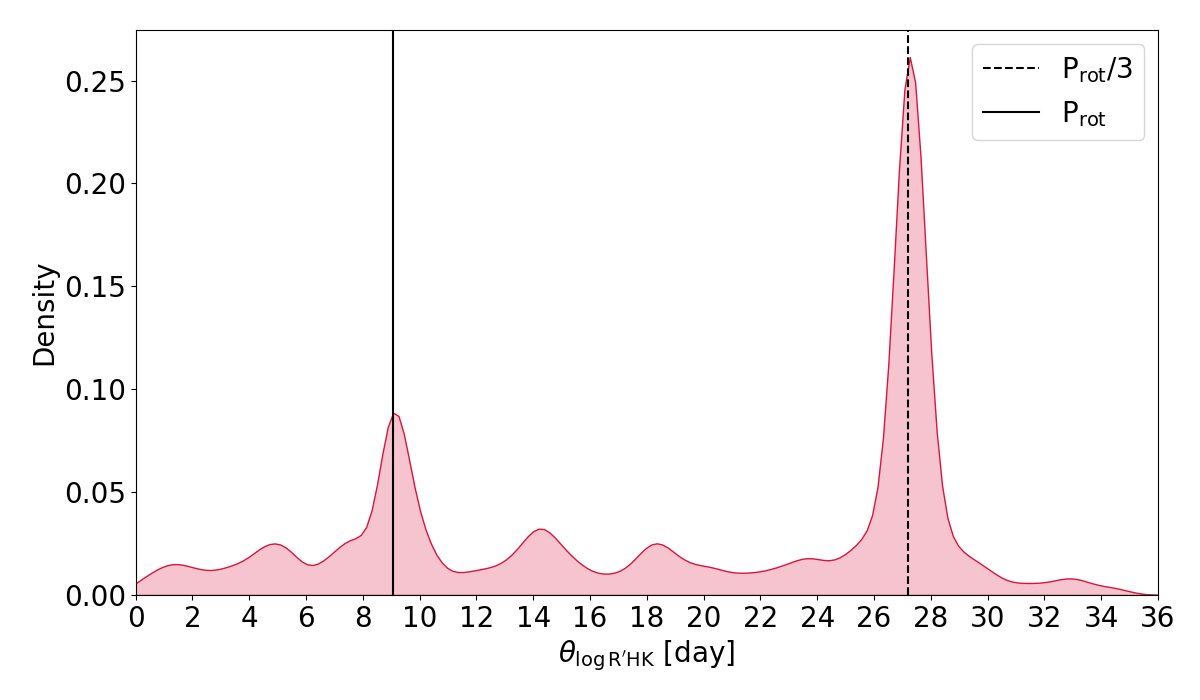}
\caption{Posterior distribution of the $\theta$ hyper-parameter (stellar $P_{\rm rot}$) obtained from a GP quasi-periodic regression of the $\log{R^{\prime}_{\rm HK}}$ time series measured from HARPS-N spectra. The rotation period and its second harmonic are indicated by a dashed and a solid vertical line, respectively.} 
\label{fig:posterior_prot}
\end{figure}

\begin{figure}
\centering
\includegraphics[width=12.0cm]{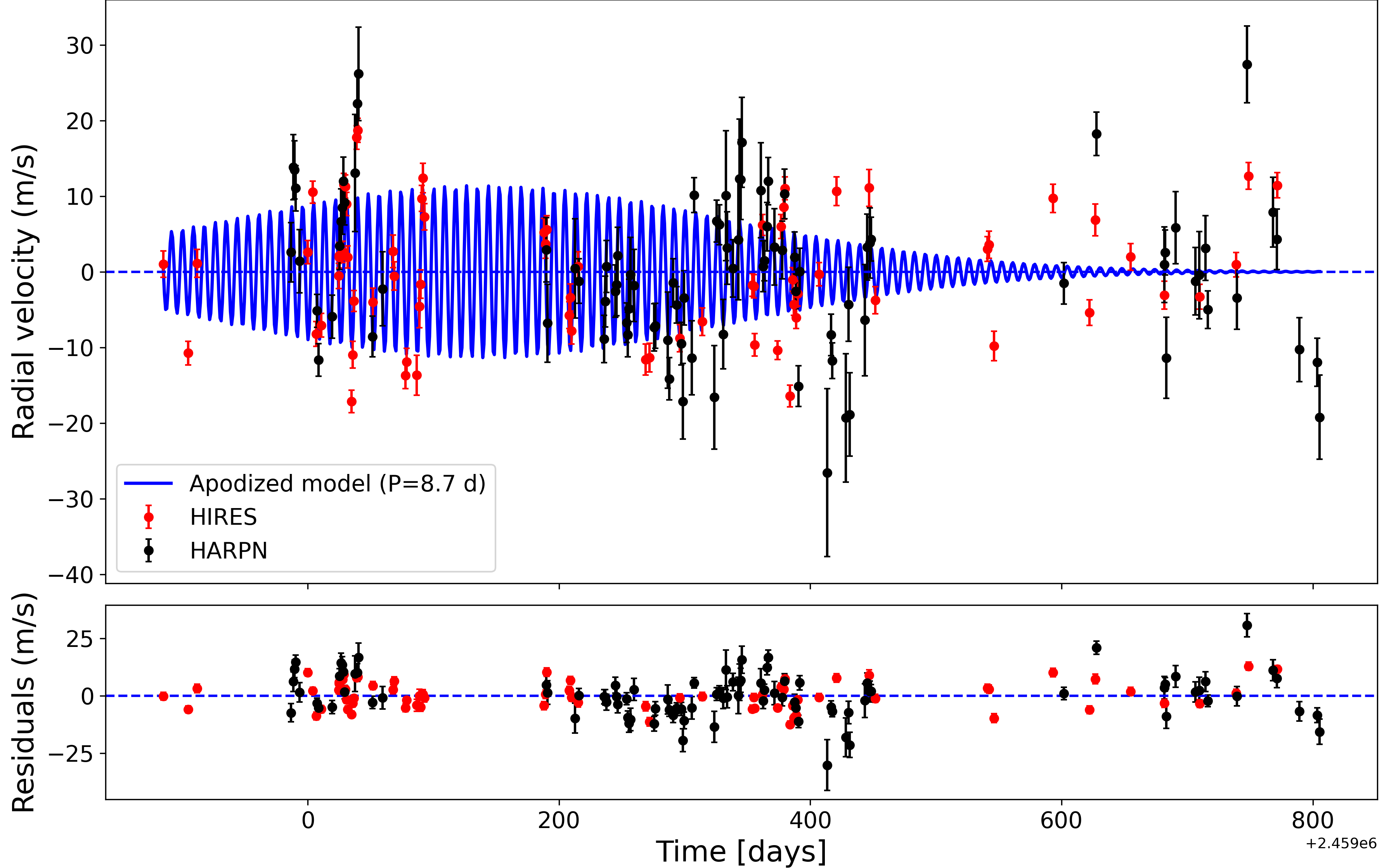}
\caption{Residual RVs after subtracting the signal of TOI-1272\,b overplotted with the apodised sine model identified at the period $P=8.7$\,d (in blue).} 
\label{fig:apodized}
\end{figure}

\end{appendix}
\end{document}